\documentclass{ws-ijmpe}

\renewcommand{\theequation}{\thesection.\arabic{equation}}
\renewcommand{\thefigure}{\thesection.\arabic{figure}}
\renewcommand{\thetable}{\thesection.\arabic{table}}

\begin{document}

\catchline{}{}{}{}{}

\title{Neutrino Masses, Lepton Flavor Mixing and Leptogenesis \\
in the Minimal Seesaw Model}

\author{\footnotesize {\bf Wan-lei Guo}\footnote{E-mail: guowl@itp.ac.cn}}

\address{ Institute of High Energy Physics, Chinese Academy of
Sciences, Beijing 100049, China
\\ and \\
Institute of Theoretical Physics, Chinese Academy of Sciences,
Beijing 100080, China }

\author{{\bf Zhi-zhong Xing}\footnote{E-mail :
xingzz@ihep.ac.cn} ~ and {\bf Shun Zhou}\footnote{E-mail:
zhoush@ihep.ac.cn}}

\address{Institute of High Energy Physics, Chinese Academy of
Sciences, Beijing 100049, China}

\maketitle

\begin{history}
\received{\today} \revised{\today}
\end{history}

\begin{abstract}
We present a review of neutrino phenomenology in the minimal
seesaw model (MSM), an economical and intriguing extension of the
Standard Model with only two heavy right-handed Majorana
neutrinos. Given current neutrino oscillation data, the MSM can
predict the neutrino mass spectrum and constrain the effective
masses of the tritium beta decay and the neutrinoless double-beta
decay. We outline five distinct schemes to parameterize the
neutrino Yukawa-coupling matrix of the MSM. The lepton flavor
mixing and baryogenesis via leptogenesis are investigated in some
detail by taking account of possible texture zeros of the Dirac
neutrino mass matrix. We derive an upper bound on the CP-violating
asymmetry in the decay of the lighter right-handed Majorana
neutrino. The effects of the renormalization-group evolution on
the neutrino mixing parameters are analyzed, and the correlation
between the CP-violating phenomena at low and high energies is
highlighted. We show that the observed matter-antimatter asymmetry
of the Universe can naturally be interpreted through the resonant
leptogenesis mechanism at the TeV scale. The
lepton-flavor-violating rare decays, such as $\mu \to e + \gamma$,
are also discussed in the supersymmetric extension of the MSM.
\end{abstract}

\tableofcontents

\markboth{Wan-lei Guo, Zhi-zhong Xing and Shun Zhou}{Neutrino
Masses, Lepton Flavor Mixing and Leptogenesis in the Minimal Seesaw
Model}

\section{Introduction}
\setcounter{equation}{0} \setcounter{figure}{0}

Recent solar,\cite{SNO} atmospheric,\cite{SK} reactor\cite{KM} and
accelerator\cite{K2K} neutrino oscillation experiments have
provided us with very robust evidence that neutrinos are massive
and lepton flavors are mixed. This great breakthrough opens a
novel window to new physics beyond the Standard Model (SM). In
order to generate neutrino masses, the most straightforward
extension of the SM is to preserve its $SU(2)^{}_{\rm L}\times
U(1)^{}_{\rm Y}$ gauge symmetry and introduce a right-handed
neutrino for each lepton family. Neutrinos can therefore acquire
masses via the Dirac mass term, which links the lepton doublets to
the right-handed singlets. If we adopt such a scenario and
confront the masses of Dirac neutrinos with current experimental
data, we have to give a reasonable explanation for the extremely
tiny neutrino Yukawa couplings. This unnaturalness can be
overcome, however, provided neutrinos are Majorana particles
instead of Dirac particles. In this case, it is also possible to
write out a lepton-number-violating mass term in terms of the
fields of right-handed Majorana neutrinos. Since the latter are
$SU(2)^{}_{\rm L}$ singlets, their masses are not subject to the
spontaneous gauge symmetry breaking. Given the Dirac neutrino mass
term of the same order as the electroweak scale $\Lambda^{}_{\rm
EW} \sim 10^2$ GeV, the small masses of left-handed Majorana
neutrinos can be generated by pushing the masses of right-handed
Majorana neutrinos up to a superhigh-energy scale close to the
scale of grand unified theories $\Lambda^{}_{\rm GUT} \sim
10^{16}~{\rm GeV}$. This is just the well-known seesaw
mechanism,\cite{SEESAW} which has been extensively discussed in
the literature. There are of course some other ways to make
neutrinos massive. For instance, one may extend the SM with the
scalar singlets or triplets which couple to two lepton doublets
and form a gauge invariant mass term.\cite{Li} Neutrinos can then
gain the Majorana masses after the relevant scalars gain their
vacuum expectation values. But why is the seesaw mechanism so
attractive? An immediate answer to this question is that the
seesaw mechanism can not only account for the smallness of
neutrino masses in a natural way, but also provide a natural
possibility to interpret the observed matter-antimatter asymmetry
of the Universe.

The cosmological baryon-antibaryon asymmetry is a long-standing
problem in particle physics and cosmology. To dynamically generate
a net baryon number asymmetry in the Universe, three Sakharov
conditions have to be satisfied:\cite{Sakharov} (1) baryon number
non-conservation; (2) C and CP violation; (3) a departure from
thermal equilibrium. Fortunately, both $B$- and $L$-violating
anomalous interactions exist in the SM and can be in thermal
equilibrium when the temperature is much higher than the
electroweak scale. Fukugita and Yanagida have pointed out that it
is possible to understand baryogenesis by means of the mechanism
of leptogenesis,\cite{LEP} in which a net lepton number asymmetry
is generated from the CP-violating and out-of-equilibrium decays
of heavy right-handed Majorana neutrinos. This lepton number
asymmetry is partially converted into the baryon number asymmetry
via the $(B-L)$-conserving sphaleron interaction,\cite{Kuzmin}
such that the matter-antimatter asymmetry comes into being in the
Universe.

The fact of neutrino oscillations and the elegance of leptogenesis
convince us of the rationality of the seesaw mechanism. However,
the seesaw models are usually pestered with too many parameters.
In the framework of the SM extended with three right-handed
Majorana neutrinos, for instance, there are fifteen free
parameters in the Dirac Yukawa couplings as well as three unknown
mass eigenvalues of heavy Majorana neutrinos. But the effective
neutrino mass matrix resulting from the seesaw relation contains
only nine physical parameters. That is to say, specific
assumptions have to be made for the model so as to get some
testable predictions for the neutrino mass spectrum, neutrino
mixing angles and CP violation. Among many realistic seesaw models
existing in the literature, the most economical one is the
so-called minimal seesaw model (MSM) proposed by Frampton, Glashow
and Yanagida.\cite{FGY} The MSM contains only two right-handed
Majorana neutrinos,\footnote{One may in principle introduce a
single right-handed Majorana neutrino into the SM to realize the
seesaw mechanism. In this case, two left-handed Majorana neutrinos
turn out to be massless, in conflict with the solar and
atmospheric neutrino oscillation data.} hence the number of its
free parameters is eleven instead of eighteen. Motivated by the
simplicity and predictability of the MSM, a number of authors have
explored its phenomenology. In particular, the following topics
have been investigated: (a) the neutrino mass spectrum and its
implication on the tritium beta decay and the neutrinoless
double-beta decay; (b) specific neutrino mass matrices and their
consequences on lepton flavor mixing and CP violation in neutrino
oscillations; (c) radiative corrections to the neutrino mass and
mixing parameters from the seesaw scale to the electroweak scale;
(d) baryogenesis via leptogenesis at a superhigh-energy scale or
via resonant leptogenesis\cite{Pilaftsis} at the TeV scale; (e)
lepton-flavor-violating processes (e.g., $\mu \to e + \gamma$) in
the minimal supersymmetric extension of the SM. The purpose of
this article is just to review a variety of works on these topics
in the framework of the MSM.

The remaining parts of this review are organized as follows. In
Sec. 2, we first describe the main features of the MSM and its
minimal supersymmetric extension, and then discuss the neutrino
mass spectrum and the lepton flavor mixing pattern. Stringent
constraints are obtained on the effective masses of the tritium
$\beta$ decay and the neutrinoless double-$\beta$ decay. Sec. 3 is
devoted to a summary of five distinct parameterizations of the
Dirac neutrino Yukawa couplings. They will be helpful for us to
gain some insight into physics at high energies, when the relevant
parameters are measured or constrained at low energies. In Sec. 4,
we present a phenomenological analysis of the MSM with specific
texture zeros in its Dirac Yukawa coupling matrix. Neutrino
masses, lepton flavor mixing angles and CP-violating phases are
carefully analyzed for the two-zero textures, in which the
renormalization-group running effects on the neutrino mixing
parameters are also calculated. Assuming the masses of two heavy
right-handed Majorana neutrinos to be hierarchical, we derive an
upper bound on the CP-violating asymmetry in the decay of the
lighter right-handed Majorana neutrino in Sec. 5. We present a
resonant leptogensis scenario at the TeV scale and a conventional
leptogenesis scenario at much higher energy scales to interpret
the cosmological baryon-antibaryon asymmetry. The correlation
between the CP-violating phenomena at high and low energies is
highlighted. For completeness, we also give some brief discussions
about the lepton-flavor-violating processes $l^{}_j \to l^{}_i +
\gamma$ in the supersymmetric MSM. In Sec. 6, we draw a number of
conclusions and remark the importance of the MSM as an instructive
example for model building in neutrino physics.

\section{The Minimal Seesaw Model (MSM)}
\setcounter{equation}{0} \setcounter{figure}{0}

\subsection{Salient Features of the MSM}

In the MSM, two heavy right-handed Majorana neutrinos $N^{~}_{i
\rm R}$ (for $i = 1, 2$) are introduced as the $SU(2)^{}_{\rm L}$
singlets. The Lagrangian relevant for lepton masses can be written
as\cite{FGY}
\begin{equation}
-{\cal L}^{}_{\rm lepton} \; =\; \overline{l^{~}_{\rm L}} Y^{~}_l
E^{~}_{\rm R} H + \overline{l^{~}_{\rm L}} Y^{~}_\nu N^{~}_{\rm R}
\tilde{H} + \frac{1}{2} \overline{N^{\rm c}_{\rm R}} M^{~}_{\rm R}
N^{~}_{\rm R} + {\rm h.c.} \; ,
\end{equation}
where $\tilde{H} \equiv i\sigma^{~}_2 H^*$ and $l_{\rm L}$ denotes
the left-handed lepton doublet, while $E^{~}_{\rm R}$ and
$N^{~}_{\rm R}$ stand respectively for the right-handed
charged-lepton and neutrino singlets. After the spontaneous gauge
symmetry breaking, one obtains the charged-lepton mass matrix
$M^{~}_l = v Y^{~}_l$ and the Dirac neutrino mass matrix
$M^{~}_{\rm D} = v Y^{~}_\nu$ with $v \simeq 174 ~ {\rm GeV}$
being the vacuum expectation value (vev) of the neutral component
of the Higgs doublet $H$. The heavy right-handed Majorana neutrino
mass matrix $M^{~}_{\rm R}$ is a $ 2 \times 2 $ symmetric matrix.
The overall lepton mass term turns out to be
\begin{equation}
-{\cal L}^{}_{\rm mass} \; =\; \overline{E^{~}_{\rm L}} M^{~}_l
E^{~}_{\rm R} + \frac{1}{2} \overline{(\nu^{~}_{\rm L}, ~N^{\rm
c}_{\rm R} )} \left ( \begin{matrix} {\bf 0} & M^{~}_{\rm D} \cr
M^T_{\rm D} & M^{~}_{\rm R} \cr \end{matrix} \right ) \left (
\begin{matrix} \nu^{\rm c}_{\rm L} \cr N^{~}_{\rm R} \cr
\end{matrix} \right ) + {\rm h.c.} \; ,
\end{equation}
where $E$, $\nu^{~}_{\rm L}$ and $N^{~}_{\rm R}$ represent the
column vectors of $(e, \mu, \tau)$, $(\nu^{~}_e, \nu^{~}_\mu,
\nu^{~}_\tau)^{~}_{\rm L}$ and $(N^{~}_1, N^{~}_2)^{~}_{\rm R}$
fields, respectively. Without loss of generality, we work in the
flavor basis where $M^{~}_l$ and $M^{~}_{\rm R}$ are both
diagonal, real and positive; i.e., $M^{~}_l = {\rm Diag}\{m^{~}_e,
m^{~}_\mu, m^{~}_\tau \}$ and $M^{~}_{\rm R} = {\rm
Diag}\{M^{~}_1, M^{~}_2\}$. The general form of $M^{~}_{\rm D}$ is
\begin{equation}
M^{~}_{\rm D} \; =\; \left ( \begin{matrix} a^{~}_1 & ~ b^{~}_1 \cr
a^{~}_2 & ~ b^{~}_2 \cr a^{~}_3 & ~ b^{~}_3 \cr \end{matrix} \right
) \; ,
\end{equation}
where $a^{~}_i$ and $b^{~}_i$ (for $i = 1, 2, 3$) are complex.
After diagonalizing the $5\times 5$ neutrino mass matrix in Eq.
(2.2), we obtain the effective mass matrix of three light
(left-handed) Majorana neutrinos:
\begin{eqnarray}
M^{~}_\nu = - M^{~}_{\rm D} M^{-1}_{\rm R} M^T_{\rm D} \; .
\end{eqnarray}
Note that this canonical seesaw relation holds up to the accuracy
of ${\cal O}(M^2_{\rm D}/M^2_{\rm R})$.\cite{XZ} Since the masses
of right-handed Majorana neutrinos are not subject to the
electroweak symmetry breaking, they can be much larger than $v$
and even close to $\Lambda^{~}_{\rm GUT} \sim 10^{16} ~ {\rm
GeV}$. Thus Eq. (2.4) provides an elegant explanation for the
smallness of three left-handed Majorana neutrino masses.

In the framework of the minimal supersymmetric standard model
(MSSM), one may similarly have the supersymmetric version of the
MSM with the following lepton mass term:
\begin{equation}
-{\cal L}^{}_{\rm lepton} \; =\; \overline{l^{~}_{\rm L}} Y^{~}_l
E^{~}_{\rm R} H^{}_1 + \overline{l^{~}_{\rm L}} Y^{~}_\nu
N^{~}_{\rm R} H^{}_2 + \frac{1}{2} \overline{N^{\rm c}_{\rm R}}
M^{~}_{\rm R} N^{~}_{\rm R} + {\rm h.c.} \; ,
\end{equation}
where $H^{}_1$ and $H^{}_2$ (with hypercharges $\pm 1/2$) are the
MSSM Higgs doublet superfields. In this case, the seesaw relation
in Eq. (2.4) remains valid, but $M^{}_{\rm D}$ is given by
$M^{}_{\rm D} = Y^{}_\nu v^{}_2$ with $v^{}_i$ being the vev of
the Higgs doublet $H^{}_i$ (for $i=1, 2$). The ratio of $v^{}_2$
to $v^{}_1$ is commonly defined as $\tan \beta \equiv
v^{}_2/v^{}_1$. Although $\tan\beta$ plays a crucial role in the
supersymmetric MSM, its value is unfortunately unknown.

Let us give some comments on the salient features of the MSM.
First of all, one of the light (left-handed) Majorana neutrinos
must be massless. This observation is actually straightforward:
since $M^{~}_{\rm R}$ is of rank 2, $M^{}_\nu$ is also a rank-2
matrix with $|{\rm Det}(M^{~}_\nu)| = m^{~}_1 m^{~}_2 m^{~}_3 =
0$, where $m^{~}_i$ (for $i=1, 2, 3$) are the masses of three
light neutrinos. It is therefore possible to fix the neutrino mass
spectrum by using current neutrino oscillation data (see Sec. 2.2
for a detailed analysis). Another merit of the MSM is that it has
fewer free parameters than other seesaw models. Hence the MSM is
not only realistic but also predictive in the phenomenological
study of neutrino masses and leptogenesis. Furthermore, the MSM
can be regarded as a special example of the conventional seesaw
model with three right-handed Majorana neutrinos, if one of the
following conditions or limits is satisfied: (1) one column of the
$3\times 3$ Dirac neutrino Yukawa coupling matrix is vanishing or
vanishingly small; (2) one of the right-handed Majorana neutrino
masses is extremely larger than the other two, such that this
heaviest neutrino essentially decouples from the model at low
energies and almost has nothing to do with neutrino phenomenology.

\subsection{Neutrino Masses and Mixing}

As for three neutrino masses $m^{}_i$ (for $i=1, 2, 3$), the solar
neutrino oscillation data have set $m^{~}_2
> m^{~}_1$.\cite{SNO} Now that the lightest neutrino in the MSM must
be massless, we are then left with either $m^{~}_1=0$ (normal mass
hierarchy) or $m^{~}_3 =0$ (inverted mass hierarchy). After a
redefinition of the phases of three charged-lepton fields, the
effective neutrino mass matrix $M^{~}_\nu$ can in general be
expressed as
\begin{equation}
M^{~}_\nu \; =\; V  \left ( \begin{matrix} m^{~}_1 & 0 & 0 \cr 0 &
m^{~}_2 & 0 \cr 0 & 0 & m^{~}_3 \cr \end{matrix} \right ) V^T \;
\end{equation}
in the above-chosen flavor basis, where
\begin{equation}
V \; = \; \left ( \begin{matrix} c^{~}_x c^{~}_z & s^{~}_x c^{~}_z
& s^{~}_z \cr - c^{~}_x s^{~}_y s^{~}_z - s^{~}_x c^{~}_y
e^{-i\delta} & - s^{~}_x s^{~}_y s^{~}_z + c^{~}_x c^{~}_y
e^{-i\delta} & s^{~}_y c^{~}_z \cr - c^{~}_x c^{~}_y s^{~}_z +
s^{~}_x s^{~}_y e^{-i\delta} & - s^{~}_x c^{~}_y s^{~}_z - c^{~}_x
s^{~}_y e^{-i\delta} & c^{~}_y c^{~}_z \cr \end{matrix} \right )
\left ( \begin{matrix} 1 & 0 & 0 \cr 0 & e^{i\sigma} & 0 \cr 0 & 0
& 1 \cr \end{matrix} \right ) \;
\end{equation}
is the Maki-Nakagawa-Sakata (MNS) lepton flavor mixing
matrix\cite{MNS} with $s^{~}_x \equiv \sin \theta^{~}_x$, $c^{~}_x
\equiv \cos \theta^{~}_x$ and so on\footnote{The flavor mixing
angles in our parametrization are equivalent to those in the
``standard" parametrization:\cite{PDG} $\theta^{}_x =
\theta^{}_{12}$, $\theta^{}_y = \theta^{}_{23}$ and $\theta^{}_z =
\theta^{}_{13}$.}. It is worth remarking that there is only a
single nontrivial Majorana CP-violating phase ($\sigma$) in the
MSM, as a straightforward consequence of $m^{}_1 =0$ or $m^{}_3
=0$.

A global analysis of current neutrino oscillation data\cite{FIT}
yields
\begin{eqnarray}
30^\circ  \leq & \theta^{~}_x & \leq \; 38^\circ \; ,
\nonumber \\
36^\circ \leq & \theta^{~}_y &  \leq \; 54^\circ \; ,
\nonumber \\
0^\circ \leq & \theta^{~}_z &  < \; 10^\circ \; ,
\end{eqnarray}
at the $99\%$ confidence level (the best-fit values: $\theta^{}_x
= 34^\circ$, $\theta^{}_y = 45^\circ$ and $\theta^{}_z =
0^\circ$). The mass-squared differences of solar and atmospheric
neutrino oscillations are defined respectively as $\Delta m^2_{\rm
sun} \equiv m^2_2 - m^2_1$ and $\Delta m^2_{\rm atm} \equiv |m^2_3
- m^2_2|$. At the $99\%$ confidence level, we have\cite{FIT}
\begin{eqnarray}
7.2 \times 10^{-5} ~ {\rm eV^2} \leq & \Delta m^2_{\rm sun} & \leq
8.9 \times 10^{-5} ~ {\rm eV^2} \; ,
\nonumber \\
1.7 \times 10^{-3} ~ {\rm eV^2} \leq & \Delta m^2_{\rm atm} & \leq
3.3 \times 10^{-3} ~ {\rm eV^2} \; ,
\end{eqnarray}
together with the best-fit values $\Delta m^{2}_{\rm sun} =
8.0\times 10^{-5}~{\rm eV}^2$ and $\Delta m^{2}_{\rm atm} =
2.5\times 10^{-3}~{\rm eV}^2$. Whether $m^{~}_2 < m^{~}_3$ or
$m^{~}_2 > m^{~}_3$, corresponding to whether $m^{}_1 =0$ or
$m^{}_3 =0$ in the MSM, remains an open question. This ambiguity
has to be clarified by the future neutrino oscillation
experiments.

If $m^{~}_1 =0$ holds in the MSM, one can easily obtain
\begin{eqnarray}
m^{~}_2 & = & \sqrt{\Delta m^2_{\rm sun}} \;\; ,
\nonumber \\
m^{~}_3 & = & \sqrt{\Delta m^2_{\rm sun} + \Delta m^2_{\rm atm}}
\;\; .
\end{eqnarray}
On the other hand, $m^{~}_3 =0$ will lead to
\begin{eqnarray}
m^{~}_1 & = & \sqrt{\Delta m^2_{\rm atm} - \Delta m^2_{\rm sun}}
\;\; ,
\nonumber \\
m^{~}_2 & = & \sqrt{\Delta m^2_{\rm atm}} \;\; .
\end{eqnarray}
Taking account of Eq. (2.9), we are able to constrain the ranges
of $m^{~}_2$ and $m^{~}_3$ by using Eq. (2.10) or the ranges of
$m^{~}_1$ and $m^{~}_2$ by using Eq. (2.11). Our numerical results
are shown in Fig. 2.1(a) and Fig. 2.1(b), respectively. The
allowed ranges of two non-vanishing neutrino masses are
\begin{eqnarray}
0.00849 \; {\rm eV} \leq & m^{~}_2 & \leq 0.00943 \; {\rm eV} \; ,
\nonumber \\
0.0421 \; {\rm eV} \leq & m^{~}_3 & \leq 0.0582 \; {\rm eV}
\end{eqnarray}
for the normal neutrino mass hierarchy ($m^{~}_1 = 0$); and
\begin{eqnarray}
0.0401 \; {\rm eV} \leq & m^{~}_1 & \leq 0.0568 \; {\rm eV} \; ,
\nonumber \\
0.0412 \; {\rm eV} \leq & m^{~}_2 & \leq 0.0574 \; {\rm eV}
\end{eqnarray}
for the inverted neutrino mass hierarchy ($m^{~}_3 = 0$).
\begin{figure}[tbp]
\vspace{-0.5cm}
\begin{center}
\includegraphics[width=7.5cm,height=7.3cm,angle=0]{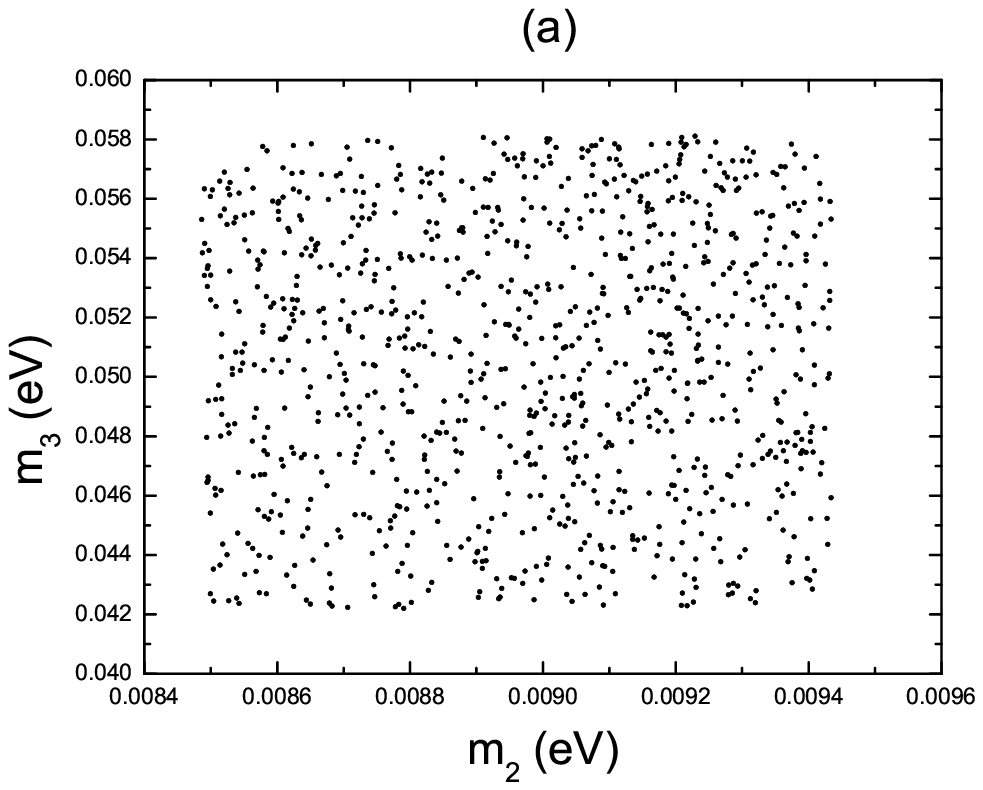}
\includegraphics[width=7.5cm,height=7.3cm,angle=0]{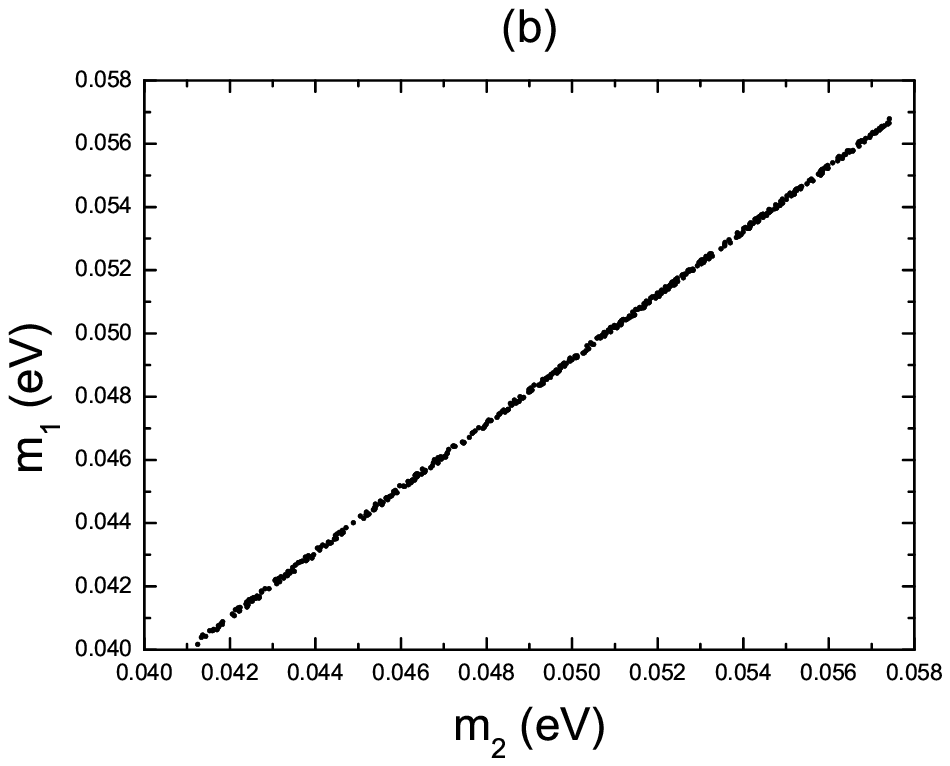}
\end{center} \vspace{-0.8cm}
\caption{Allowed region of (a) $m^{~}_2$ and $m^{~}_3$ for
$m^{~}_1 =0$ or (b) $m^{~}_1$ and $m^{~}_2$ for $m^{~}_3 =0$ in
the MSM.}
\end{figure}

\subsection{Tritium $\beta$ Decay and Neutrinoless Double-$\beta$ Decay}

If neutrinos are Majorana particles, the neutrinoless
double-$\beta$ decay may occur. The rate of this
lepton-number-violating process depends both on an effective
neutrino mass term $\langle m \rangle_{ee}$ and on the associated
nuclear matrix element. The latter can be calculated, but it
involves some uncertainties.\cite{Bilenky} Here we aim to explore
possible consequences of the MSM on the tritium $\beta$ decay
($^3_1{\rm H} \rightarrow$ $^3_2{\rm He} + e^- +
\overline{\nu}^{~}_e$) and the neutrinoless double-$\beta$ decay
($^A_Z X \rightarrow$ $^{~~~A}_{Z+2} X + 2e^-$), whose effective
mass terms are
\begin{equation}
\langle m\rangle^{~}_e \; \equiv \; \sqrt{ m^2_1 |V^{~}_{e1}|^2 +
m^2_2 |V^{~}_{e2}|^2 + m^2_3 |V^{~}_{e3}|^2} \;
\end{equation}
and
\begin{equation}
\langle m\rangle^{~}_{ee} \; \equiv \; \left | m^{~}_1 V_{e1}^2 +
m^{~}_2 V_{e2}^2 + m^{~}_3 V_{e3}^2 \right | \; , ~~~~~~~
\end{equation}
respectively,\cite{XING03} where $V^{~}_{ei}$ (for $i = 1, 2, 3$)
are the elements of the MNS matrix $V$. While $\langle
m\rangle^{~}_{ee} \neq 0$ must imply that neutrinos are Majorana
particles, $\langle m\rangle^{~}_{ee} = 0$ does not {\it
necessarily} ensure that neutrinos are Dirac particles. The reason
is simply that the Majorana phases hidden in $V$ may lead to
significant cancellations in $ \langle m \rangle^{~}_{ee}$, making
$\langle m\rangle^{~}_{ee}$ vanishing or too small to be
detectable.\cite{Bilenky2,Xing03a} But we are going to show that
$\langle m\rangle^{}_{ee} =0$ is actually impossible in the MSM.

Now let us calculate the effective mass terms $\langle
m\rangle^{~}_e$ and $\langle m\rangle^{~}_{ee}$. With the help of
Eqs. (2.7), (2.10), (2.11) and (2.14), we obtain\cite{GX}
\begin{equation}
\langle m\rangle^{~}_e \; =\; \left \{
\begin{array}{l}
\displaystyle \sqrt{\Delta m^2_{\rm sun} s^2_x c^2_z + \left (
\Delta m^2_{\rm sun} + \Delta m^2_{\rm atm} \right ) s^2_z} \;\; ,
~~~~ (m^{~}_1 = 0) \; ,
\cr\cr
\displaystyle \sqrt{\left ( \Delta
m^2_{\rm atm} - \Delta m^2_{\rm sun} c^2_x \right ) c^2_z} \;\; ,
~~~~~~~~~~~~~~~~~~~~ (m^{~}_3 = 0) \; .
\end{array}
\right . ~~~~~~~~~~~~~~~~~~
\end{equation}
On the other hand, we get the expression of $\langle
m\rangle^{}_{ee}$ by combining Eqs. (2.7), (2.10), (2.11) and
(2.15):\cite{GX}
\begin{equation}
\langle m\rangle^{~}_{ee} \; =\; \left \{
\begin{array}{l}
\displaystyle \sqrt{\Delta m^2_{\rm sun} s^4_x c^4_z + \left (
\Delta m^2_{\rm sun} + \Delta m^2_{\rm atm} \right ) s^4_z +
T^{~}_1 \cos 2\sigma} \;\; , ~~~ (m^{~}_1 = 0) \; ,
\cr\cr
\displaystyle \sqrt{\Delta m^2_{\rm atm} s^4_x c^4_z + \left (
\Delta m^2_{\rm atm} - \Delta m^2_{\rm sun} \right ) c^4_x c^4_z +
T^{~}_3 \cos 2\sigma} \;\; ,  (m^{~}_3 = 0) \; ,
\end{array}
\right .
\end{equation}
where
\begin{eqnarray}
T^{}_1 & = & 2 \sqrt{\Delta m^2_{\rm sun} \left ( \Delta m^2_{\rm
sun} + \Delta m^2_{\rm atm} \right )} ~ s^2_x c^2_z s^2_z \; ,
\nonumber \\
T^{}_3 & = & 2 \sqrt{\Delta m^2_{\rm atm} \left ( \Delta m^2_{\rm
atm} - \Delta m^2_{\rm sun} \right )} ~ c^2_x s^2_x c^4_z \; .
\end{eqnarray}
Just as expected, $\langle m\rangle^{~}_{ee}$ depends on the
Majorana CP-violating phase $\sigma$. This phase parameter does
not affect CP violation in neutrino-neutrino and
antineutrino-antineutrino oscillations, but it may play a
significant role in the scenarios of leptogenesis\cite{LEP} due to
the lepton-number-violating and CP-violating decays of two heavy
right-handed Majorana neutrinos.

With the help of current experimental data listed in Eqs. (2.8)
and (2.9), we can obtain the numerical predictions for $\langle
m\rangle^{~}_e$ and $\langle m\rangle^{~}_{ee}$ by using Eqs.
(2.16) and (2.17). The results are shown in Fig. 2.2 for two
different neutrino mass spectra. It is then straightforward to
arrive at
\begin{eqnarray}
0.00424 \; {\rm eV} \leq & \langle m\rangle^{~}_e & \leq 0.0116 \;
{\rm eV} \; ,
\nonumber \\
0.00031 \; {\rm eV} \leq & \langle m\rangle^{~}_{ee} & \leq 0.0052
\; {\rm eV}
\end{eqnarray}
for $m_1 = 0$; and
\begin{eqnarray}
0.0398 \; {\rm eV} \leq & \langle m\rangle^{~}_e & \leq 0.0571 \;
{\rm eV} \; ,
\nonumber \\
0.0090 \; {\rm eV} \leq & \langle m\rangle^{~}_{ee} & \leq 0.0571 \;
{\rm eV}
\end{eqnarray}
for $m^{~}_3 = 0$. Two comments are in order:
\begin{figure}[tbp]
\vspace{-0.5cm}
\begin{center}
\includegraphics[width=7.5cm,height=7.3cm,angle=0]{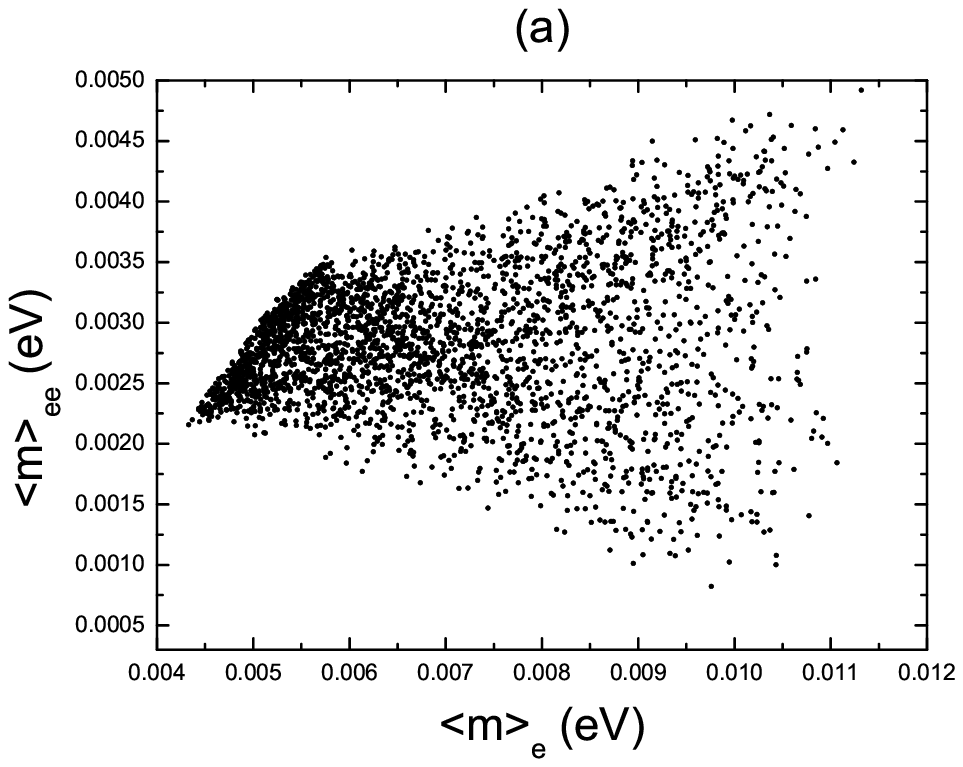}
\includegraphics[width=7.5cm,height=7.3cm,angle=0]{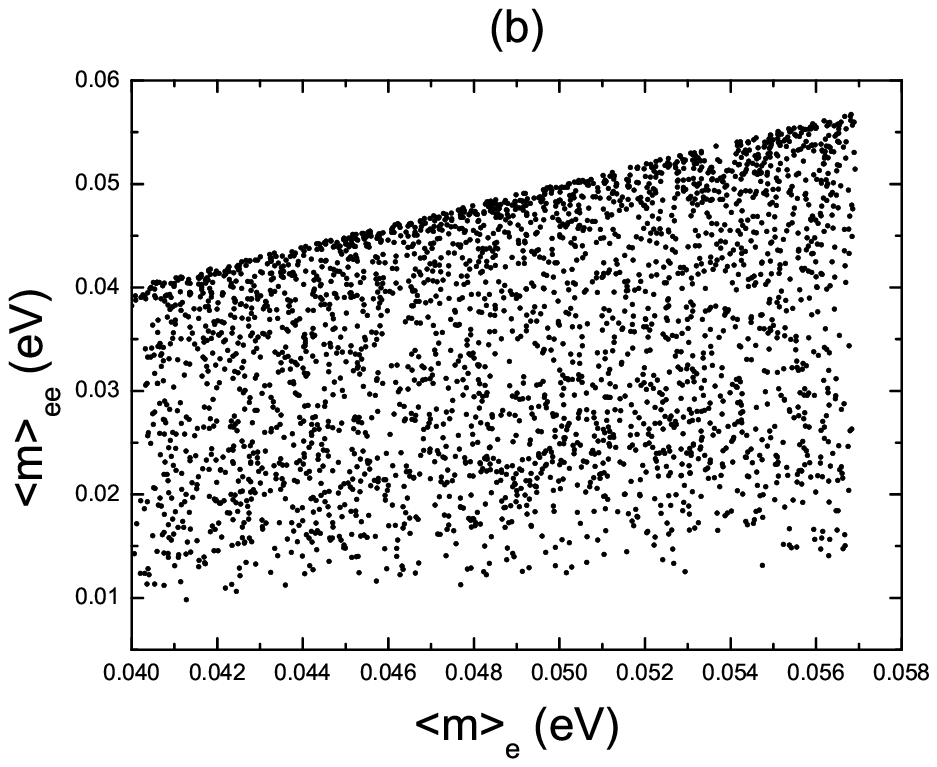}
\end{center}\vspace{-0.8cm}
\caption{Allowed region of $\langle m\rangle^{}_e$ and $\langle
m\rangle^{}_{ee}$ in the MSM: (a) $m^{}_1 =0$ and (b) $m^{}_3 =0$.
}
\end{figure}

(a) Whether $\langle m\rangle^{~}_e$ and $\langle
m\rangle^{~}_{ee}$ can be measured remains an open question. The
present experimental upper bounds are $\langle m\rangle^{~}_e < 2
~ {\rm eV}$ and $\langle m\rangle^{~}_{ee} < 0.35 ~ {\rm eV}$ at
the $90\%$ confidence level.\cite{PDG,HM} They are much larger
than our predictions for the upper bounds of $\langle
m\rangle^{~}_e$ and $\langle m\rangle^{~}_{ee}$ in the MSM. The
proposed KATRIN experiment is possible to reach the sensitivity
$\langle m\rangle^{~}_e \sim 0.3 ~ {\rm eV}$.\cite{KATRIN} If a
signal of $\langle m\rangle^{~}_e \sim 0.1 ~ {\rm eV}$ is seen,
the MSM will definitely be ruled out. On the other hand, a number
of the next-generation experiments for the neutrinoless
double-$\beta$ decay\cite{Bilenky} are possible to probe $\langle
m\rangle^{~}_{ee}$ at the level of 10 meV to 50 meV. Such
experiments are expected to test our prediction for $\langle
m\rangle^{~}_{ee}$ given in Eq. (2.20); i.e., in the case of
$m^{~}_3 =0$.

(b) Now that the magnitude of $\langle m\rangle^{~}_{ee}$ in the
case of $m^{~}_3 =0$ is experimentally accessible in the future,
its sensitivity to the unknown parameters $\theta^{~}_z$ and
$\sigma$ is worthy of some discussions. Eq. (2.17) shows that
$\langle m\rangle^{~}_{ee}$ depends only on $c^{~}_z$ for $m^{~}_3
=0$. Hence we conclude that $\langle m\rangle^{~}_{ee}$ is
insensitive to the change of $\theta^{~}_z$ in its allowed range
(i.e., $0^\circ \leq \theta^{~}_z < 10^\circ$).\cite{FIT} The
dependence of $\langle m\rangle^{~}_{ee}$ on the Majorana
CP-violating phase $\sigma$ is illustrated in Fig. 2.3.\cite{GX}
We observe that $\langle m\rangle^{~}_{ee}$ is significantly
sensitive to $\sigma$. Thus a measurement of $\langle
m\rangle^{~}_{ee}$ will allow us to determine or constrain this
important phase parameter in the MSM.
\begin{figure}[tbp]
\vspace{-0.4cm}
\begin{center}
\includegraphics[width=7.5cm,height=7.3cm,angle=0]{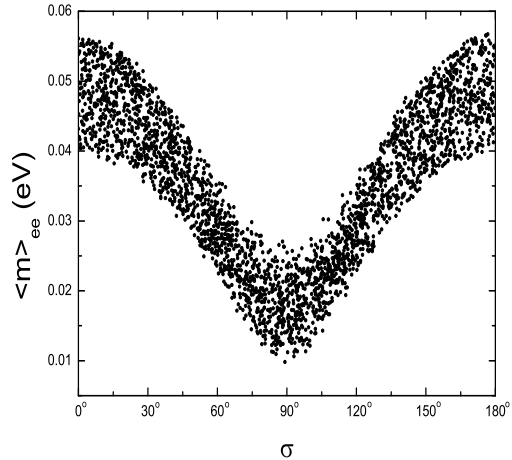}
\end{center}
\vspace{-0.9cm} \caption{Dependence of $\langle m\rangle^{~}_{ee}$
on the Majorana CP-violating phase $\sigma$ for $m^{~}_3 =0$ in
the MSM.}
\end{figure}

\section{How to Describe the MSM}
\setcounter{equation}{0} \setcounter{figure}{0}

In the flavor basis where $M^{}_l$ and $M^{}_{\rm R}$ are both
taken to be diagonal, it is easy to count the number of free
parameters in the MSM: two heavy Majorana neutrino masses
($M^{~}_1$, $M^{~}_2$) and nine real parameters in the Dirac
neutrino mass matrix $M^{~}_{\rm D}$. Note that three trivial
phases in the Dirac Yukawa couplings can be rotated away by
rephasing the charged-lepton fields. On the other hand, the
effective neutrino mass matrix $M^{~}_\nu$ contains seven
parameters: two non-vanishing neutrino masses, three flavor mixing
angles and two nontrivial CP-violating phases (the Dirac phase
$\delta$ and the Majorana phase $\sigma$), as one can easily see
from Eqs. (2.6) and (2.7). Since $M^{}_\nu$ is related to
$M^{}_{\rm D}$ and $M^{}_{\rm R}$ via the seesaw relation given in
Eq. (2.4), the parameters of $M^{}_\nu$ are therefore dependent on
those of $M^{}_{\rm D}$ and $M^{}_{\rm R}$. In principle, the
light Majorana neutrino masses, flavor mixing angles and
CP-violating phases may all be measured at low energies. Hence it
is possible to reconstruct the Dirac Yukawa coupling matrix
$Y^{}_\nu$ (or equivalently $M^{}_{\rm D}$) by means of two heavy
Majorana neutrino masses, seven low-energy observables and two
extra real parameters.

A few distinct parametrization schemes have been proposed to
describe the MSM by using different combinations of eleven
parameters. This kind of attempt is by no means trivial, because
some intriguing phenomena (e.g., leptogenesis and the
lepton-flavor-violating rare decays) are closely related to the
Dirac Yukawa couplings. A brief summary of the existing schemes
for the reconstruction of the MSM will be presented below,
together with some comments on their respective advantages in the
study of neutrino phenomenology.

\subsection{Casas-Ibarra-Ross Parametrization}

Ibarra and Ross\cite{ROSS} have advocated a useful parametrization
of the Dirac neutrino mass matrix:
\begin{eqnarray}
M^{~}_{\rm D} \; = \; i  V  \sqrt{m} ~ R \sqrt{M^{~}_{\rm R}} \;\;
, ~~
\end{eqnarray}
where $V$ is the MNS matrix, $m \equiv {\rm Diag}\{m^{~}_1,
m^{~}_2, m^{~}_3\}$ with either $m^{}_1 =0$ or $m^{}_3 =0$, and
$R$ is a $3 \times 2$ complex matrix which satisfies the
normalization relation $R R^T = {\rm Diag} \{0, 1, 1 \} $ for the
$m^{~}_1 = 0$ case or $R R^T = {\rm Diag} \{1, 1, 0 \} $ for the
$m^{~}_3 = 0$ case. Given $m^{}_1 =0$, $R$ can in general be
parameterized as
\begin{equation}
R \; =\; \left ( \begin{matrix} 0 &  0 \cr \cos z &  -\sin z \cr \pm
\sin z & \pm \cos z \cr \end{matrix} \right ) \; ,
\end{equation}
where $z$ is a complex number. Given $m^{}_3 =0$, $R$ is of the
form
\begin{equation}
R \; =\; \left ( \begin{matrix} \cos z &  -\sin z \cr \pm \sin z &
\pm \cos z \cr  0 &  0 \cr \end{matrix} \right ) \; .
\end{equation}
To be more explicit, $z$ can be written as $z = \alpha^{~}_z + i
\beta^{~}_z$. Taking the normal neutrino mass hierarchy for
example, we obtain
\begin{equation}
R \; =\; \left ( \begin{matrix} 0 &  0 \cr \cosh \beta^{~}_z &  - i
\sinh \beta^{~}_z \cr \pm i \sinh \beta^{~}_z &  \pm \cosh
\beta^{~}_z \cr \end{matrix} \right ) \left ( \begin{matrix} \cos
\alpha^{~}_z & -\sin \alpha^{~}_z \cr \sin \alpha^{~}_z & \cos
\alpha^{~}_z \end{matrix} \right )\; .
\end{equation}
Without loss of generality, one may take $-\pi \leq \alpha^{~}_z
\leq \pi$ and leave $\beta^{~}_z$ unconstrained. With the help of
Eqs. (3.1), (3.2) and (3.3), six elements of $M^{}_{\rm D}$ can
then be expressed as
\begin{eqnarray}
\left(M^{}_{\rm D}\right)^{~}_{\alpha 1} = \left\{
\begin{array}{cc}
\displaystyle i \sqrt{M^{~}_1} \left ( V^{~}_{\alpha 2}
\sqrt{m^{~}_2} \cos z \pm
V^{~}_{\alpha 3} \sqrt{m^{~}_3} \sin z \right ) & ~~~~~~ (m^{~}_1 = 0) \\
\displaystyle i \sqrt{M^{~}_1} \left (V^{~}_{\alpha 1}
\sqrt{m^{~}_1} \cos z \pm V^{~}_{\alpha 2} \sqrt{m^{~}_2} \sin z
\right ) & ~~~~~~ (m^{~}_3 = 0)
\end{array} \right. \; ,
\end{eqnarray}
and
\begin{eqnarray}
\left(M^{}_{\rm D}\right)^{~}_{\alpha 2} = \left\{
\begin{array}{cc}
\displaystyle - i \sqrt{M^{~}_2} \left (V^{~}_{\alpha 2}
\sqrt{m^{~}_2} \sin z \mp V^{~}_{\alpha 3} \sqrt{m^{~}_3}
\cos z \right ) & ~~~~ (m^{~}_1 = 0) \\
\displaystyle - i \sqrt{M^{~}_2} \left (V^{~}_{\alpha 1}
\sqrt{m^{~}_1} \sin z \mp V^{~}_{\alpha 2} \sqrt{m^{~}_2} \cos z
\right ) & ~~~~ (m^{~}_3 = 0)
\end{array} \right. \; ,
\end{eqnarray}
where the subscript $\alpha$ runs over $e$, $\mu$ and $\tau$. It
is straightforward to verify that $V^\dagger M^{~}_\nu V^* = m$
holds, where $M^{~}_\nu$ is determined by the seesaw formula.

Indeed, such a parametrization scheme was first proposed by Casas
and Ibarra to describe the seesaw model with three right-handed
Majorana neutrinos.\cite{Casas} It has proved to be particularly
useful to understand the generic features of different models in
dealing with thermal leptogenesis.\cite{DiBari,RL1,Turzynski}

\subsection{Bi-unitary Parametrization}

Endoch {\it et al} have pointed out a different way to
parameterize $M^{}_{\rm D}$, which is here referred to as the
bi-unitary parametrization.\cite{T1} Given $m^{}_1 =0$, $M^{}_{\rm
D}$ can in general be written as
\begin{equation}
M^{}_{\rm D} \; =\; V^{}_{\rm L} \left ( \begin{matrix} 0 & ~ 0 \cr
d^{}_2 & ~ 0 \cr 0 & ~ d^{}_3 \cr \end{matrix} \right ) U^{}_{\rm R}
\; ,
\end{equation}
where $d^{}_2$ and $d^{}_3$ are real and positive; $V^{}_{\rm L}$
and $U^{}_{\rm R}$ are the $3\times 3$ and $2\times 2$ unitary
matrices, respectively. An explicit parametrization of $V^{}_{\rm
L}$ is
\begin{eqnarray}
V^{}_{\rm L} = \left ( \begin{matrix} c^{~}_1 c^{~}_3 & s^{~}_1
c^{~}_3 & s^{~}_3 \cr - c^{~}_1 s^{~}_2 s^{~}_3 - s^{~}_1 c^{~}_2
e^{-i\delta^{}_{\rm L}} & - s^{~}_1 s^{~}_2 s^{~}_3 + c^{~}_1
c^{~}_2 e^{-i\delta^{}_{\rm L}} & s^{~}_2 c^{~}_3 \cr - c^{~}_1
c^{~}_2 s^{~}_3 + s^{~}_1 s^{~}_2 e^{-i\delta^{}_{\rm L}} & -
s^{~}_1 c^{~}_2 s^{~}_3 - c^{~}_1 s^{~}_2 e^{-i\delta^{}_{\rm L}}
& c^{~}_2 c^{~}_3 \cr \end{matrix} \right ) \left ( \begin{matrix}
1 & 0 & 0 \cr 0 & e^{- i\gamma^{}_{\rm L}} & 0 \cr 0 & 0 & e^{ +
i\gamma^{}_{\rm L}} \cr \end{matrix} \right ) \; , ~~~~~
\end{eqnarray}
while $U^{}_{\rm R}$ can be parameterized as
\begin{eqnarray}
U^{}_{\rm R} & = & \left ( \begin{matrix} c^{}_{\rm R} & ~ s^{}_{\rm
R} \cr -s^{}_{\rm R} & ~ c^{}_{\rm R} \cr\end{matrix} \right ) \left
( \begin{matrix} e^{- i {\gamma^{}_{\rm R}}} & ~ 0 \cr 0 & ~ e^{ + i
{\gamma^{}_{\rm R}}} \cr \end{matrix} \right ) \; ,
\end{eqnarray}
where $c^{}_i \equiv \cos \theta^{}_i$ and $s^{}_i \equiv \sin
\theta^{}_i$ (for $i=1, 2, 3$) as well as $c^{}_{\rm R} \equiv
\cos \theta^{}_{\rm R}$ and $s^{}_{\rm R} \equiv \sin
\theta^{}_{\rm R}$. In this scheme, the eleven parameters of the
MSM are $M^{}_1$, $M^{}_2$, $d^{}_2$, $d^{}_3$, $\theta^{}_1$,
$\theta^{}_2$, $\theta^{}_3$, $\theta^{}_{\rm R}$, $\delta^{}_{\rm
L}$, $\gamma^{}_{\rm L}$ and $\gamma^{}_{\rm R}$. Note that
$V^{}_{\rm L}$ itself is not the MNS matrix. Note also that a
parametrization of $M^{}_{\rm D}$ in the $m^{}_3 =0$ case can be
considered in a similar way.\footnote{In this case, we have
$\displaystyle M^{}_{\rm D} = V^{}_{\rm L} \left ( \begin{matrix}
d^{}_1 & ~ 0 \cr 0 & ~ d^{}_2 \cr 0 & ~ 0 \cr \end{matrix} \right
) U^{}_{\rm R}$, where $d^{}_1$ and $d^{}_2$ are real and
positive.}

As far as leptogenesis is concerned in the MSM, it is convenient
to define two effective neutrino masses
\begin{equation}
\tilde{m}^{}_i \equiv \frac{({{M^\dagger_{\rm D}} M^{}_{\rm
D}})_{ii}}{M^{}_i} = 8 \pi \Gamma^{}_i \left ( \frac{v}{M^{}_i}
\right )^2 \; ,
\end{equation}
where $\Gamma^{}_i \equiv  M^{}_i \left( Y^\dagger_\nu Y^{}_\nu
\right)^{}_{ii}/(8\pi)$ is the tree-level decay width of the heavy
Majorana neutrino $N^{}_i$ (for $i=1, 2$). The magnitude of
$\tilde{m}^{}_i$ will be crucial in evaluating the washout effects
associated with the out-of-equilibrium decays of $N^{}_i$. Note
that ($\theta^{}_{\rm R}$, $\gamma^{}_{\rm R}$) and ($d^{}_2$,
$d^{}_3$) can also be expressed in terms of a new set of
parameters\cite{T1}
\begin{eqnarray}
\cos 4 \gamma^{}_{\rm R} & = & \frac{m^2_2 + m^2_3 - \tilde{m}^2_1
- \tilde{m}^2_2}{2 \left (\tilde{m}^{}_1 \tilde{m}^{}_2 - m^{}_2
m^{}_3 \right )} \; ,
\end{eqnarray}
and
\begin{eqnarray}
(c^{}_{\rm R}, s^{}_{\rm R}) & = & \left( \sqrt{\frac{\rho +
\sigma^{}_{-}}{2 \rho}} , -\sqrt{\frac{\rho - \sigma^{}_{-}}{2
\rho}} \right ) \; , \nonumber \\
(d^2_2, \; d^2_3) & = & \sqrt{M^{}_1 M^{}_2} \left (\sigma^{}_{+}
- \rho, \sigma^{}_{+} + \rho \right ) \; ,
\end{eqnarray}
where $\sigma^{}_{\pm} = (\tilde{m}^{}_2 \pm \tilde{m}^{}_1 \zeta)
/(2 \sqrt{\zeta})$, $\rho = \sqrt{(\tilde{m}^{}_1 \tilde{m}^{}_2 -
m_2 m_3) + \sigma^2_{-}}$ and $\zeta \equiv M_1 / M_2$. The
CP-violating phase $\gamma^{}_{\rm R}$ plays a special role in
this parametrization scheme, as it shows up in both the high- and
low-scale phenomena of CP violation. In comparison, the
CP-violating phase $\delta$ of $V$ in the Casas-Ibarra-Ross
parametrization scheme has nothing to do with leptogenesis.

\subsection{Natural Reconstruction}

Barger {\it et al} have pointed out a more natural way to
reconstruct the MSM.\cite{HE} From Eqs. (2.3) and (2.4), one may
directly obtain
\begin{equation}
M^{}_\nu \; =\; - \left ( \begin{matrix} \displaystyle
\frac{a_1^2}{M_1} + \frac{b_1^2}{M_2} &  \displaystyle \frac{a_1
a_2}{M_1} + \frac{b_1 b_2}{M_2} & \displaystyle \frac{a_1 a_3}{M_1}
+ \frac{b_1 b_3}{M_2} \cr\cr \displaystyle \frac{a_1 a_2}{M_1} +
\frac{b_1 b_2}{M_2} & \displaystyle \frac{a_2^2}{M_1} +
\frac{b_2^2}{M_2} & \displaystyle \frac{a_2 a_3}{M_1} + \frac{b_2
b_3}{M_2} \cr\cr \displaystyle \frac{a_1 a_3}{M_1} + \frac{b_1
b_3}{M_2} & \displaystyle \frac{a_2 a_3}{M_1} + \frac{b_2 b_3}{M_2}
& \displaystyle \frac{a_3^2}{M_1} + \frac{b_3^2}{M_2} \cr
\end{matrix} \right ) \; .
\end{equation}
Given $M^{}_1$, $M^{}_2$, $(M^{}_\nu)^{}_{11}$ and $a^{}_1$ (or
$b^{}_1$), the parameter $b^{}_1$ (or $a^{}_1$) reads
\begin{equation}
b^{}_1  =  \pm  \sqrt{- \left(M^{}_\nu\right)^{}_{11} M^{}_2 -
\frac{M^{}_2}{M^{}_1}a^2_1} \; ,
\end{equation}
or
\begin{equation}
a^{}_1 =  \pm \sqrt{ - \left(M^{}_\nu\right)^{}_{11} M_1 - b_1^2
\frac{M_1}{M_2}} \; .
\end{equation}
Then the remaining five elements of $M^{}_{\rm D}$ can be
expressed in terms of $M^{}_1$, $M^{}_2$, $(M^{}_\nu )^{}_{ij}$
and $a^{}_1$ (or $b^{}_1$) as follows:
\begin{eqnarray}
a^{}_i & = & \frac{1}{\left(M^{}_\nu\right)^{}_{11}} \left\{
a^{}_1 \left(M^{}_\nu\right)^{}_{1i} + \xi^{}_i b^{}_1
\sqrt{\frac{M^{}_1}{M^{}_2}} \sqrt{\left(M^{}_\nu\right)^{}_{11}
\left(M^{}_\nu\right)^{}_{ii} -\left(M^{}_\nu\right)^2_{1i}}
\right \} \; , \nonumber \\
b^{}_i & = & \frac{1}{\left(M^{}_\nu\right)^{}_{11}} \left\{
b^{}_1 \left(M^{}_\nu\right)^{}_{1i} - \xi^{}_i a^{}_1
\sqrt{\frac{M^{}_2}{M^{}_1}} \sqrt{\left(M^{}_\nu\right)^{}_{11}
\left(M^{}_\nu\right)_{ii} - \left(M^{}_\nu\right)^2_{1i}}
\right\} \; ,
\end{eqnarray}
where $i = 2$ or 3, and $\xi^{}_i$ takes either $+1$ or $-1$. Note
that we have assumed $(M^{}_\nu )^{}_{11}$ to be nonzero in the
calculation. It is worth remarking that Eqs. (3.14), (3.15) and
(3.16) are valid for both $m^{}_1 =0$ and $m^{}_3 =0$ cases.

Since Eq. (3.13) is invariant under the permutations $a^{}_1
\leftrightarrow a^{}_2$, $b^{}_1 \leftrightarrow b^{}_2$,
$(M^{}_\nu )^{}_{11} \leftrightarrow (M^{}_\nu )^{}_{22}$ and
$(M^{}_\nu )^{}_{13} \leftrightarrow (M^{}_\nu )^{}_{23}$, we may
also express $a^{}_i$ and $b^{}_i$ in terms of $a^{}_2$ or
$b^{}_2$ (for $(M^{}_\nu )^{}_{22} \neq 0$). The case of
$(M^{}_\nu )^{}_{33} \neq 0$ can be similarly treated. It is easy
to count the number of model parameters in this natural
parametrization: two right-handed Majorana neutrino masses from
$M^{}_{\rm R}$; two non-vanishing left-handed Majorana neutrino
masses, three flavor mixing angles and two CP-violating phases
from $M^{}_\nu$, together with the real and imaginary parts of one
free complex parameter (e.g., $a^{}_1$ or $b^{}_1$) from
$M^{}_{\rm D}$.

\subsection{Modified Casas-Ibarra-Ross Scheme}

Ibarra has also proposed an interesting parameterization scheme
for the MSM,\cite{IBARRA} in which all eleven model parameters can
in principle be measured. This scheme is actually a modified
version of the Casas-Ibarra-Ross scheme. Defining the Hermitian
matrix
\begin{eqnarray}
P \equiv M^{}_{\rm D}M^\dagger_{\rm D}= V \sqrt{m} R M^{}_{\rm R}
R^\dagger \sqrt{m} V^\dagger \; ,
\end{eqnarray}
where Eq. (3.1) has been used, we immediately get $(V^\dagger
P)^{}_{1i} = 0$. As a result,
\begin{eqnarray}
P^{}_{11} & = & - \frac{P^*_{12} V^*_{21} + P^*_{13}
V^*_{31}}{V^*_{11}} \; , \nonumber \\
P^{}_{22} & = & - \frac{P^{}_{12} V^*_{11} + P^*_{23}
V^*_{31}}{V^*_{21}} \; , \nonumber \\
P^{}_{33} & = & - \frac{P^{}_{13} V^*_{11} + P^{}_{23}
V^*_{21}}{V^*_{31}} \; .
\end{eqnarray}
Since the diagonal elements of $P$ are real and positive, it is
easy to derive the phases of $P^{}_{13}$ and $P^{}_{23}$ from the
first and second relations in Eq. (3.18):
\begin{eqnarray}
e^{i \phi^{}_{13}} & = & \frac{-i \; {\rm Im} (P^{~}_{12}
V^{~}_{21} V^*_{11}) \pm \sqrt{|P^{}_{13}|^2 |V^{}_{11}|^2
|V^{}_{31}|^2 - \left[{\rm Im} (P^{~}_{12} V^{~}_{21}
V^*_{11})\right]^2} }{|P^{~}_{13}| V^{~}_{31} V^*_{11}} \; ,
\nonumber \\
e^{i \phi^{}_{23}} & = & \frac{+i \; {\rm Im} \left(P^{~}_{12}
V^{~}_{21} V^*_{11}\right) \pm \sqrt{|P^{}_{23}|^2 |V^{}_{21}|^2
|V^{}_{31}|^2 - \left[{\rm Im} (P^{~}_{12} V^{~}_{21}
V^*_{11})\right]^2} }{|P^{~}_{23}| V^{~}_{31} V^*_{21}} \; , ~~~~~
\end{eqnarray}
where $\phi^{}_{13} \equiv \arg (P^{}_{13})$ and $\phi^{}_{23}
\equiv \arg (P^{}_{23})$. The above analysis shows that only
$P^{}_{12}$, $|P^{}_{13}|$ and $|P^{}_{23}|$ are the independent
parameters of $P$.

Now let us define the Hermitian matrix $Q \equiv V^\dagger P V$.
Its elements $Q^{}_{22}$, $Q^{}_{23}$, $Q^{}_{33}$ can be
expressed in terms of $M^{}_1$, $M^{}_2$ and $z$. It is then
possible to use Eq. (3.17) to inversely derive the exact
expressions for these three parameters:
\begin{eqnarray}
&& M^{}_1 = \frac{1}{2} \left [
\sqrt{\left(\frac{Q^{}_{33}}{m^{}_3} +
\frac{Q^{}_{22}}{m^{}_2}\right)^2 - 4 \frac{({\rm
Im}Q^{}_{23})^2}{m^{}_2 m^{}_3} } -
\sqrt{\left(\frac{Q^{}_{33}}{m^{}_3}
-\frac{Q^{}_{22}}{m^{}_2}\right)^2 + 4 \frac{({\rm Re}
Q^{}_{23})^2}{m^{}_2 m^{}_3} } \right ] \; ,
\nonumber \\
&& M^{}_2 = \frac{1}{2} \left [
\sqrt{\left(\frac{Q^{}_{33}}{m^{}_3} +
\frac{Q^{}_{22}}{m^{}_2}\right)^2 - 4 \frac{({\rm
Im}Q^{}_{23})^2}{m^{}_2 m^{}_3} } +
\sqrt{\left(\frac{Q^{}_{33}}{m^{}_3}
-\frac{Q^{}_{22}}{m^{}_2}\right)^2 + 4 \frac{({\rm Re}
Q^{}_{23})^2}{m^{}_2 m^{}_3} } \right ] \; ,
\nonumber \\
&& \cos 2z = \frac{1}{M^2_1 -M^2_2} \left( \frac{Q^2_{22}}{m^2_2}
- \frac{Q^2_{33}}{m^2_3} + 4i \frac{{\rm Re} Q^{}_{23} {\rm Im}
Q^{}_{23}}{m^{}_2 m^{}_3} \right) \; .
\end{eqnarray}
It is worth remarking that $|P^{}_{12}|$, $|P^{}_{13}|$ and
$|P^{}_{23}|$ could be measured through the
lepton-flavor-violating rare decays $l^{}_j \to l^{}_i + \gamma$
in the supersymmetric case.\cite{Casas,LFV} The only phase
appearing in $P^{}_{12}$ might be determined from a measurement of
the electric dipole moment of the electron, on which the present
experimental upper bound is $d_e < 1.6 \times 10^{-27}$ e
cm.\cite{Dipole} Of course, $\theta^{}_{x}, \theta^{}_{y},
\theta^{}_{z}, \delta, \sigma, m^{}_2$ and $m^{}_3$ are seven
low-energy observables. Thus all the eleven independent parameters
of the MSM are in principle measurable in this parameterization
scheme. Although the above discussion has been restricted to the
$m^{}_1 = 0$ case, it can easily be extended to the $m^{}_3 =0$
case.

\subsection{Vector Representation}

Fujihara {\it et al} have parameterized the Dirac neutrino mass
matrix as\cite{T2}
\begin{equation}
M^{}_{\rm D} \; =\; \left ( \begin{matrix} a^{}_1 & b^{}_1 \cr
a^{}_2 & b^{}_2 \cr a^{}_3 & b^{}_3 \cr \end{matrix} \right ) =
({\bf a}, {\bf b}) \left ( \begin{matrix} D^{}_1 & 0 \cr 0 &
D^{}_2 \cr \end{matrix} \right ) \; ,
\end{equation}
where ${\bf a} = (a^{}_e, a^{}_\mu, a^{}_\tau)^{T}$ and ${\bf b} =
(b^{}_e, b^{}_\mu, b^{}_\tau)^{T}$ are two unit vectors (i.e.,
${\bf a^\dagger \cdot a} = 1$ and ${\bf b^\dagger \cdot b} = 1$).
Both $D^{}_1$ and $D^{}_2$ are real and positive parameters.
Without loss of generality, we take ${\bf a}$ and ${\bf b}$ to be
real and complex, respectively. In this case, all low-energy
parameters can be expressed in terms of ${\bf a}$, ${\bf b} $,
$D^{}_1$, $D^{}_2$, $M^{}_1$ and $M^{}_2$. By using the seesaw
relation and solving the eigenvalue equation ${\rm
Det}\left(M^{}_\nu M^\dagger_\nu - n^2 \right) = 0$, we obtain
\begin{eqnarray}
n^2_{\pm} & = &\frac{X^2_1+X^2_2+ 2 X^{}_1 X^{}_2 {\rm Re} [{\bf
a^\dagger \cdot b}]^2}{2} \nonumber \\
&& \pm \frac{\sqrt{\left(X^2_1+X^2_2+ 2 X^{~}_1 X^{~}_2 {\rm Re}
[{\bf a^\dagger \cdot b}]^2\right)^2 - 4 X^2_1 X^2_2\left(1-|{\bf
a^{\dagger} \cdot b}|^2\right)^2}}{2} \; ,
\end{eqnarray}
where $X^{}_i = D^2_i / M^{}_i$ (for $i = 1, 2 $). For the normal
neutrino mass hierarchy (i.e., $m^{}_1 =0$), the non-vanishing
neutrino masses read
\begin{equation}
m^2_2 = n^2_{-} \; , ~~~~ m^2_3 = n^2_{+} \; ;
\end{equation}
and  for the inverted neutrino mass hierarchy (i.e., $m^{}_3 =0$),
the result is
\begin{equation}
m^2_1 = n^2_{-} \; , ~~~~ m^2_2 = n^2_{+} \; .
\end{equation}
Meanwhile, one may decompose the MNS matrix $V$ into a product of
unitary matrices. For simplicity, here we only concentrate on the
$m^{}_1 =0$ case. We express $V$ as $V=UK$, where
\begin{eqnarray}
U & = &\left( \begin{matrix}
 \displaystyle \frac{b^*_\mu a^*_\tau -b^*_\tau a^*_\mu}
 {\displaystyle \sqrt{1-|\bf a^{\dagger} \cdot b|^2}}
  &\frac{ \displaystyle b^{}_e - a^{}_e \bf a^{\dagger} \cdot b}
  {\sqrt{\displaystyle 1-|\bf a^{\dagger} \cdot b|^2}}
   & a^{}_e \cr
  \frac{
 \displaystyle b^*_\tau a^*_e-b^*_e a^*_\tau
  }{\displaystyle \sqrt{1-|\bf a^{\dagger} \cdot b|^2}} &
  \frac{\displaystyle b^{}_\mu-a^{}_\mu \bf a^{\dagger} \cdot b}
  {\displaystyle \sqrt{1-|\bf a^{\dagger} \cdot b|^2}}
  & a^{}_\mu \cr
  \frac{\displaystyle b^*_e a^*_\mu-b^*_\mu a^*_e
  }{\displaystyle \sqrt{1-|\bf a^{\dagger} \cdot b|^2}}
  & \frac{\displaystyle b^{}_\tau - a^{}_\tau \bf a^{\dagger} \cdot b}
  {\displaystyle \sqrt{1-|\bf a^{\dagger} \cdot b|^2}}
  &  a^{}_\tau \cr \end{matrix} \right) \; ,
\end{eqnarray}
and
\begin{eqnarray}
K &=& \left( \begin{matrix} 1 & 0 & 0  \cr 0 & \cos\theta^{}_{\rm N}
& \sin\theta^{}_{\rm N} e^{-i \phi^{}_{\rm N}} \cr 0 & -\sin
\theta^{}_{\rm N} e^{i \phi^{}_{\rm N}} & \cos \theta^{}_{\rm N} \cr
\end{matrix} \right) \left( \begin{matrix} 1 & 0 & 0 \cr 0 & e^{i \alpha^{}_{\rm
N}}& 0 \cr 0 & 0  & e^{-i \alpha^{}_{\rm N}} \cr \end{matrix}
\right) \; .
\end{eqnarray}
The parameters $\theta^{}_{\rm N}$, $\phi^{}_{\rm N}$ and
$\alpha^{}_{\rm N}$ in Eq. (3.26) are given by
\begin{eqnarray}
\tan 2 \theta^{}_{\rm N} &=&\frac{2 X^{}_2 \sqrt{1-|{\bf
a^{\dagger} \cdot b}|^2} \left|{X^{}_1 \bf (a^{\dagger} \cdot
b)^{\ast}}+ X_2 \left ({\bf a^{\dagger} \cdot b} \right ) \right|}
{X^2_1 + X^2_2 \left(2 \left|{\bf a^{\dagger} \cdot
b}\right|^2-1\right) + 2 X_1 X_2 {\rm Re}\left[{\bf a^{\dagger}
\cdot b}\right]^2} \; ,
\end{eqnarray}
and
\begin{eqnarray}
\phi^{}_{\rm N} &=& \arg \left [X_1 \left ({\bf a^{\dagger} \cdot
b} \right )^{\ast} + X_2 \left ({\bf a^{\dagger} \cdot b} \right
)\right ] \; ,
\nonumber \\
\alpha^{}_{\rm N} & = & \frac{1}{2} \arg \left [ \left (Z^{}_{\rm
N} \right )^{}_{22} \cos^2 \theta^{}_{\rm N} + \left (Z^{}_{\rm N}
\right )^{}_{33} \sin^2 \theta^{}_{\rm N} e^{- 2 i \phi^{}_{\rm
N}}- \left (Z^{}_{\rm N} \right )^{}_{23} \sin 2 \theta^{}_{\rm N}
e^{-i \phi^{}_{\rm N}} \right ] \; ~~~~~~~~
\end{eqnarray}
with
\begin{eqnarray}
(Z^{}_{\rm N})^{}_{22} &=& -X^{}_2 \left(1 - \left |{\bf
a^{\dagger} \cdot b} \right |^2 \right) \; ,
\nonumber \\
(Z^{}_{\rm N})^{}_{33} &=& -\left[ X_1 + X_2 \left ({\bf
a^{\dagger} \cdot b} \right )^2 \right ] \; ,
\nonumber \\
(Z^{}_{\rm N})^{}_{23} &=& -X_2 \sqrt{1-|{\bf a^{\dagger} \cdot
b}|^2} \left({\bf a^{\dagger} \cdot b} \right) \; .
\end{eqnarray}
The $m^{}_3 =0$ case can be discussed in a similar way.

In such a vector representation of the MSM, the eleven model
parameters are $M^{}_1$, $M^{}_2$ and nine real parameters from
$(M^{}_{\rm D})^{}_{ij}$; or equivalently $D^{}_1$, $D^{}_2$,
$X^{}_1$, $X^{}_2$ and seven real parameters from ${\bf a}$ and
${\bf b}$. This parameterization scheme has been applied to the
analysis of baryogenesis via leptogenesis by taking into account
the contribution from individual lepton flavors.\cite{T2}

To summarize, we have outlined the main features of five typical
parameterization schemes for the MSM. Each of them has its own
advantage and disadvantage in the analysis of neutrino
phenomenology. A ``hybrid" parameterization scheme,\cite{Para6}
which is more or less similar to one of the representations
discussed above, has also been proposed. These generic
descriptions of the MSM are instructive, but specific assumptions
have to be made on the texture of $M^{}_{\rm D}$ in order to
achieve specific predictions for the neutrino mixing angles,
CP-violating phases and leptogenesis.

\section{Texture Zeros in the MSM}
\setcounter{equation}{0} \setcounter{figure}{0}

Among eleven independent parameters of the MSM, only seven of them
(two non-vanishing left-handed Majorana neutrino masses, three
flavor mixing angles and two CP-violating phases) are possible to
be measured in some low-energy neutrino experiments. Hence the
predictability of the MSM depends on how its remaining four free
parameters can be constrained. To reduce the freedom in the MSM, a
phenomenologically popular and theoretically meaningful approach
is to introduce texture zeros\cite{FGY,RAIDAL,GUO1,Zero} or flavor
symmetries.\cite{FlavorR} It is worth mentioning that certain
texture zeros may be a natural consequence of a certain flavor
symmetry.\cite{RABY,MOHAPATRA} In this section, we concentrate on
possible texture zeros in the MSM and investigate their
implications on neutrino mixing and CP violation at low energies.

\subsection{One-zero Textures}

If the Dirac neutrino mass matrix $M^{}_{\rm D}$ has one vanishing
element,\cite{HE,KANG,BRANCO} two free real parameters can then be
eliminated from the model. There are totally six one-zero textures
for $M^{}_{\rm D}$. Here let us take $b^{}_1 = 0$ in Eq. (2.3) for
example. By adopting the Casas-Ibarra-Ross
parametrization\cite{ROSS} and using the expression of $V$ in Eq.
(2.7), we get
\begin{eqnarray}
- s^{}_x c^{}_z e^{i \sigma} \sqrt{m^{}_2} \sin z \pm  s^{}_z
\sqrt{m^{}_3} \cos z = 0 \;
\end{eqnarray}
from $b^{}_1 =0$ in the $m^{}_1 =0$ case. This relation implies
that it is now possible to fix the free parameter $z$:
\begin{eqnarray}
\tan z = \pm \frac{s^{}_z e^{- i \sigma}}{s^{}_x c^{}_z
\sqrt{r^{}_{23}}} \;\; ,
\end{eqnarray}
where $r^{}_{23} \equiv m^{}_2/m^{}_3$. Similarly, one may
determine $z$ from $b^{}_1 =0$ in the $m^{}_3 =0$ case. If the
scheme of natural reconstruction\cite{HE} is used, the texture
zero $b^{}_1 = 0$ can help us to compute the other five elements
of $M^{}_{\rm D}$ through Eqs. (3.15) and (3.16). Namely,
\begin{eqnarray}
a^{}_1 = \pm i \sqrt{(M^{}_\nu)^{}_{11} M^{}_1} \;\; ,
\end{eqnarray}
and
\begin{eqnarray}
a^{}_i & = & a^{}_1 \frac{(M^{}_\nu)^{}_{1i}}{(M^{}_\nu)^{}_{11}}
\; , \nonumber \\
b^{}_i & = &  - \xi^{}_i \frac{a^{}_1}{(M^{}_\nu)^{}_{11}}
\sqrt{\frac{M^{}_2}{M^{}_1}} \sqrt{(M^{}_\nu)^{~}_{11}
(M^{}_\nu)^{~}_{ii} - (M^{}_\nu)^2_{1i}} \;\; ,
\end{eqnarray}
where $i = 2, 3$ and $\xi^{}_i = \pm 1$. More detailed discussions
about the one-zero textures of $M^{}_{\rm D}$ in the MSM can be
found in Refs. 41 and 42.

Similar to the one-zero hypothesis for the texture of $M^{}_{\rm
D}$, the equality between two elements of $M^{}_{\rm D}$ can also
be assumed. As pointed out by Barger {\it et al},\cite{HE} there
are fifteen possibilities to set the equality, which is horizontal
(e.g., $a^{}_1 = b^{}_1$), vertical (e.g., $a^{}_1 = a^{}_2 $) or
crossed (e.g., $a^{}_1= b^{}_2$). This kind of equality might come
from an underlying flavor symmetry in much more concrete scenarios
of the MSM.\cite{Moha}

\subsection{Two-zero Textures}

If $M^{}_{\rm D}$ involves two texture zeros, the MSM will have
some testable predictions for neutrino phenomenology. There are
totally fifteen two-zero textures of $M^{}_{\rm D}$, among which
only five can coincide with current neutrino oscillation
data.\cite{ROSS} One of these five viable textures is referred to
as the FGY ansatz, since it was first proposed and discussed by
Frampton, Glashow and Yanagida (FGY).\cite{FGY} We shall reveal a
very striking feature of the FGY ansatz: its nontrivial
CP-violating phases can be calculated in terms of three neutrino
mixing angles $(\theta^{}_x, \theta^{}_y, \theta^{}_z)$ and the
ratio of two neutrino mass-squared differences $\Delta m^2_{\rm
sun}$ and $\Delta m^2_{\rm atm}$.

In the FGY ansatz, $M^{}_{\rm D}$ is of the form
\begin{equation}
M^{}_{\rm D} = \left ( \begin{matrix} a^{}_1 & ~ {\bf 0} \cr
a^{}_2 & ~ b^{}_2 \cr {\bf 0} & ~ b^{}_3 \cr \end{matrix} \right )
\; .
\end{equation}
Two texture zeros in $M^{}_{\rm D}$ may arise from a horizontal
flavor symmetry.\cite{RABY,MOHAPATRA} With the help of Eq. (2.4),
we immediately obtain
\begin{equation}
M^{}_\nu \; =\; - \left ( \begin{matrix} \displaystyle
\frac{a^2_1}{M^{}_1} & \displaystyle \frac{a^{}_1 a^{}_2}{M^{}_1} &
{\bf 0} \cr \displaystyle \frac{a^{}_1 a^{}_2}{M^{}_1} &
\displaystyle \frac{a^2_2}{M^{}_1} + \frac{b^2_2}{M^{}_2} &
\displaystyle \frac{b^{}_2 b^{}_3}{M^{}_2} \cr {\bf 0} &
\displaystyle \frac{b^{}_2 b^{}_3}{M^{}_2} & \displaystyle
\frac{b^2_3}{M^{}_2} \cr
\end{matrix} \right ) \; .
\end{equation}
Without loss of generality, we can always redefine the phases of
left-handed lepton fields to make $a^{}_1$, $b^{}_2$ and $b^{}_3$
real and positive. In this basis, only $a^{}_2$ is complex and its
phase $\phi \equiv \arg(a^{}_2)$ is the sole source of CP
violation in the model under consideration. Because $a^{}_1$,
$b^{}_2$ and $b^{}_3$ of $M^{}_{\rm D}$ have been taken to be real
and positive, $M^{}_\nu$ may not be diagonalized as in Eq. (2.6).
In this phase convention, a more general way to express $M^{}_\nu$
is
\begin{equation}
M^{}_\nu \; =\; \left (P^{}_l V \right ) \left ( \begin{matrix}
m^{}_1 & 0 & 0 \cr 0 & m^{}_2 & 0 \cr 0 & 0 & m^{}_3 \cr
\end{matrix} \right ) \left (P^{}_l V \right )^T ,
\end{equation}
where $P^{}_l = i {\rm Diag}\{e^{i\alpha}, e^{i\beta},
e^{i\gamma}\}$ is a phase matrix, and $V$ is just the MNS matrix
parameterized as in Eq. (2.7).

For the normal neutrino mass hierarchy ($m^{}_1 = 0$), six
independent elements of $M^{}_\nu$ can be written as\cite{GUO1}
\begin{eqnarray}
\left(M^{}_\nu\right)^{}_{11} & = & - e^{2i\alpha} \left [m^{}_2
s^2_x c^2_z e^{2i\sigma} + m^{}_3 s^2_z \right ] \; ,
\nonumber  \\
(M^{}_\nu)^{}_{22} & = & - e^{2i\beta} \left [m^{}_2 \left (-s^{}_x
s^{}_y s^{}_z + c^{}_x c^{}_y e^{-i\delta} \right )^2 e^{2i\sigma} +
m^{}_3 s^2_y c^2_z \right ] \; ,
\nonumber \\
(M^{}_\nu)^{}_{33} & = & - e^{2i\gamma} \left [m^{}_2 \left (s^{}_x
c^{}_y s^{}_z + c^{}_x s^{}_y e^{-i\delta} \right )^2 e^{2i\sigma} +
m^{}_3 c^2_y c^2_z \right ] \; ;
\end{eqnarray}
and
\begin{eqnarray}
(M^{}_\nu)^{}_{12} & = & - e^{i(\alpha + \beta)} \left [ m^{}_2
s^{}_x c^{}_z \left (-s^{}_x s^{}_y s^{}_z + c^{}_x c^{}_y
e^{-i\delta} \right ) e^{2i\sigma} + m^{}_3 s^{}_y s^{}_z c^{}_z
\right ] \; ,
\nonumber  \\
(M^{}_\nu)^{}_{13} & = & - e^{i(\alpha+\gamma)} \left [ -m^{}_2
s^{}_x c^{}_z \left (s^{}_x c^{}_y s^{}_z + c^{}_x s^{}_y
e^{-i\delta} \right ) e^{2i\sigma} + m^{}_3 c^{}_y s^{}_z c^{}_z
\right ] \; ,
\nonumber \\
(M^{}_\nu)^{}_{23} & = & - e^{i(\beta+\gamma)} \left [ m^{}_2 \left
(s^{}_x c^{}_y s^{}_z + c^{}_x s^{}_y e^{-i\delta} \right ) \left
(s^{}_x s^{}_y s^{}_z - c^{}_x c^{}_y e^{-i\delta} \right )
e^{2i\sigma} + m^{}_3 s^{}_y c^{}_y c^2_z \right ] \; . \nonumber \\
\end{eqnarray}
Because of $(M^{}_\nu)^{}_{13}=0$ as shown in Eq. (4.6), we
immediately arrive at
\begin{eqnarray}
\delta & = & \pm \arccos \left [ \frac{c^2_y s^2_z - r^2_{23} s^2_x
\left (c^2_x s^2_y + s^2_x c^2_y s^2_z \right )}{2 r^2_{23} s^3_x
c^{}_x s^{}_y c^{}_y s^{}_z} \right ] \; ,
\nonumber \\
\sigma & = & \frac{1}{2} \arctan \left [\frac{c^{}_x s^{}_y
\sin\delta} {s^{}_x c^{}_y s^{}_z + c^{}_x s^{}_y \cos\delta} \right
] \; ,
\end{eqnarray}
where $r^{}_{23} \equiv m^{}_2/m^{}_3 \approx 0.18$ obtained from
Eqs. (2.10) and (2.12). This result implies that both $\delta$ and
$\sigma$ can definitely be determined, if and only if the smallest
mixing angle $\theta^{}_z$ is measured. To establish the
relationship between $\phi$ and $\delta$, we need to figure out
$\alpha$, $\beta$ and $\gamma$. As $a^{}_1$, $b^{}_2$ and $b^{}_3$
are real and positive, $(M^{}_\nu)^{}_{11}$, $(M^{}_\nu)^{}_{23}$
and $(M^{}_\nu)^{}_{33}$ must be real and negative. Then $\alpha$,
$\beta$ and $\gamma$ can be derived from Eqs. (4.9) and (4.10):
\begin{eqnarray}
\alpha & = & -\frac{1}{2} \arctan \left [ \frac{r^{2}_{23} s^2_x
c^2_z \sin 2\sigma}{s^2_z + r^2_{23} s^2_x c^2_z \cos 2\sigma}
\right ] \; ,
\nonumber \\
\beta & = & -\gamma - \arctan \left [ \frac{c^{}_x c^{}_y s^{}_z
\sin\delta} {s^{}_x s^{}_y - c^{}_x c^{}_y s^{}_z \cos\delta} \right
] \; ,
\nonumber \\
\gamma & = & +\frac{1}{2} \arctan \left [ \frac{s^2_z \sin
2\sigma}{r^2_{23} s^2_x c^2_z + s^2_z \cos 2\sigma} \right ] \; .
\end{eqnarray}
The overall phase of $-(M^{}_\nu)^{}_{12}$, which is equal to the
phase of $a^{}_2$, is given by
\begin{equation}
~~~ \phi \; =\; \alpha + \beta - \arctan \left [ \frac{s^{}_x c^{}_y
s^{}_z \sin\delta}{c^{}_x s^{}_y + s^{}_x c^{}_y s^{}_z \cos\delta}
\right ] \; .
\end{equation}
Eqs. (4.10), (4.11) and (4.12) show that all six phase parameters
($\delta$, $\sigma$, $\phi$, $\alpha$, $\beta$ and $\gamma$) can
be determined in terms of $r^{}_{23}$, $\theta^{}_x$,
$\theta^{}_y$ and $\theta^{}_z$. Similar results can also be
obtained for the inverted neutrino mass hierarchy ($m^{}_3
=0$),\cite{GUO1} but we do not elaborate on them here.

A measurement of the unknown neutrino mixing angle $\theta^{}_z$
is certainly crucial to test the FGY ansatz. Because $|\cos
\delta| \leq 1$ must hold, Eq. (4.10) allows us to constrain the
magnitude of $\theta^{}_z$. Taking the best-fit values of $\Delta
m^2_{\rm sun}$, $\Delta m^2_{\rm atm}$, $\theta^{}_x$ and
$\theta^{}_y$ as our typical inputs, we find that $\theta^{}_z$ is
restricted to a very narrow range $4.4^\circ \leq \theta^{}_z \leq
4.9^\circ$ (i.e., $0.077 \leq s^{}_z \leq 0.086$). This result
implies that the FGY ansatz with $m^{}_1 =0$ is highly sensitive
to $\theta^{}_z$ and can easily be ruled out if the experimental
value of $\theta^{}_z$ does not lie in the predicted region. We
illustrate the numerical dependence of six phase parameters
$(\delta, \sigma, \phi, \alpha, \beta, \gamma)$ on the smallest
mixing angle $\theta_z$ in Fig. 4.1. To a good degree of accuracy,
we obtain $\delta \approx 2\sigma$, $\phi \approx \alpha \approx
-\sigma$, $\beta \approx -\gamma$ and $\gamma \approx 0$. These
instructive relations can essentially be observed from Eqs.
(4.10), (4.11) and (4.12), because of $s^{}_z \ll 1$. Note that we
have only shown the dependence of $\delta$ on $\theta_z$ in the
range $0< \delta <\pi$. The reason is simply that only this range
may lead to the correct sign for the cosmological baryon number
asymmetry $Y^{}_{\rm B}$, when the mechanism of baryogenesis via
leptogenesis is taken into account.\cite{GUO1} As a by-product,
the Jarlskog invariant of CP violation\cite{J} and the effective
mass of the neutrinoless double-$\beta$ decay are found to be $0 <
J^{}_{\rm CP} \leq 0.019$ and $2.6 ~ {\rm meV} \leq \langle m
\rangle^{}_{ee} \leq 3.1$ meV in the $m^{}_1=0$ case. It is
possible to measure $|J^{}_{\rm CP}| \sim {\cal O}(10^{-2})$ in
the future long-baseline neutrino oscillation experiments. The
interesting correlation between $Y^{}_{\rm B}$ and $J^{}_{\rm CP}$
will be illustrated in Sec. 5.4.
\begin{figure}[tbp]
\begin{center}
\includegraphics[width=7.2cm,height=7.2cm,angle=0]{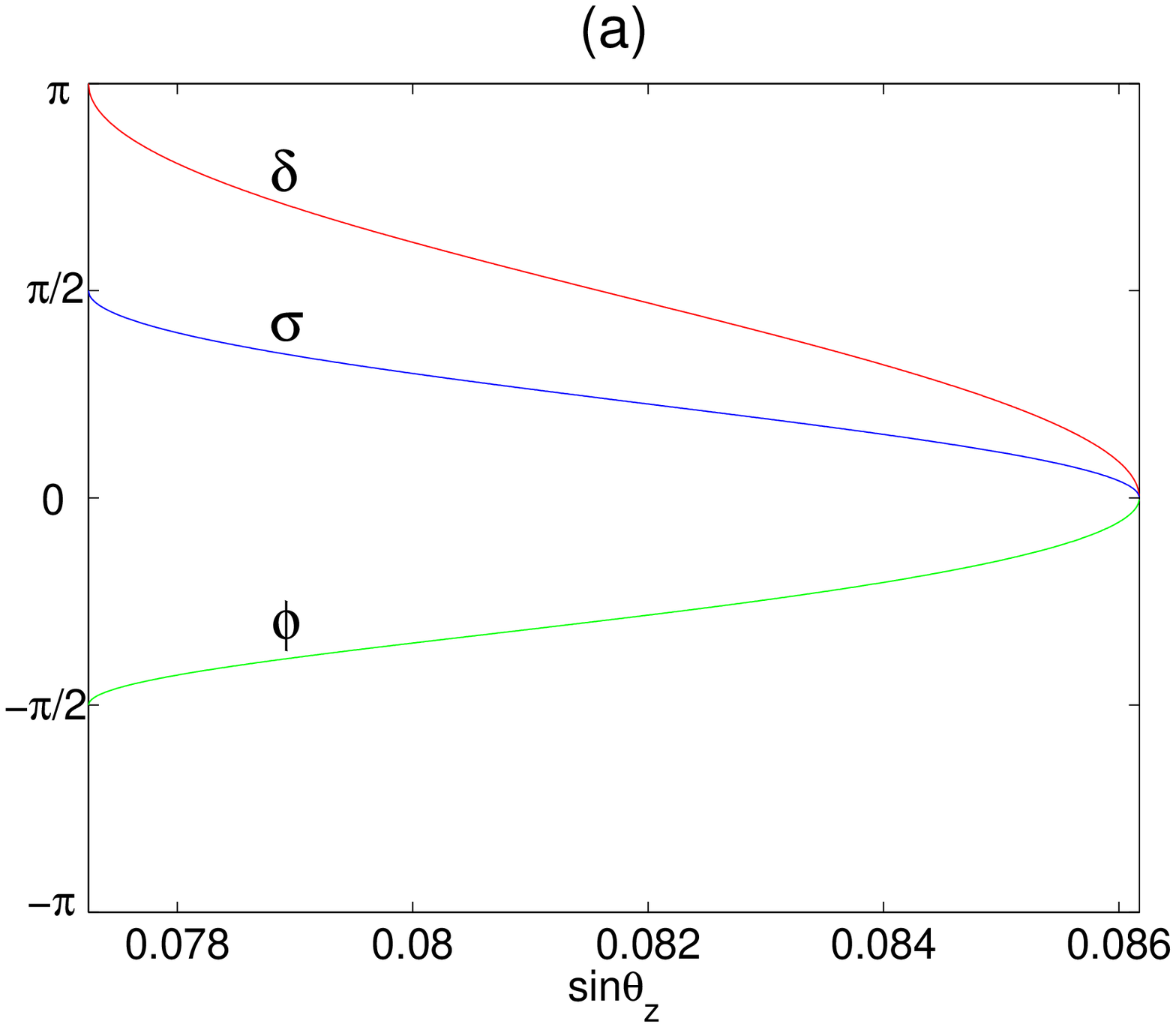}
\includegraphics[width=7.2cm,height=7.2cm,angle=0]{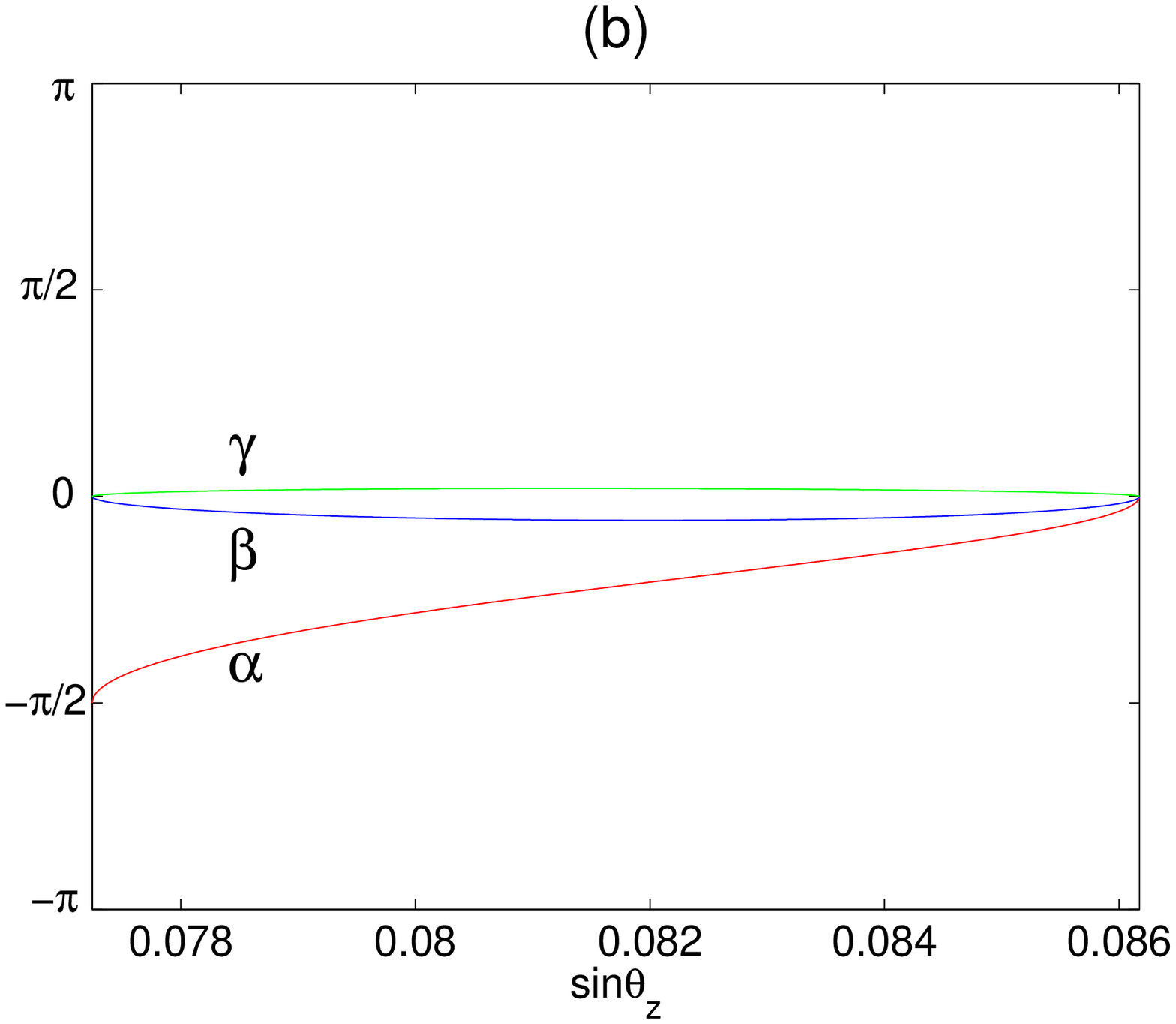}
\end{center}
\vspace{-0.3cm} \caption{Numerical results for the FGY ansatz with
$m^{}_1=0$: (a) dependence of $\delta$, $\sigma$ and $\phi$ on
$\sin\theta^{}_z$; (b) dependence of $\alpha$, $\beta$ and
$\gamma$ on $\sin\theta^{}_z$.}
\end{figure}

Finally let us take a look at another two-zero texture of
$M^{}_{\rm D}$, in which $a^{}_2 = 0$ and $b^{}_1 = 0$ hold. The
resultant neutrino mass matrix $M^{}_\nu$ has a vanishing
entry:\footnote{Note that two texture zeros in $M^{}_{\rm D}$
naturally lead to one texture zero in $M^{}_\nu$. A systematic
analysis of the one-zero textures of $M^{}_\nu$ in the MSM has
been done in Ref. 44.} $(M^{}_\nu)^{}_{12} =0$. In this case, one
may choose $a^{}_3$ to be complex. The relevant phase parameters
can then be calculated by setting $(M^{}_\nu)^{}_{12} =0$ in Eq.
(4.9). We find that the simple replacements $\delta \rightarrow
\delta - \pi$ and $\theta^{}_y \rightarrow \pi/2 - \theta^{}_y$
allow us to directly write out the expressions of $\sigma$,
$\phi$, $\alpha$, $\beta$ and $\gamma$ in the $(M^{}_\nu)^{}_{12}
=0$ case from Eqs. (4.10), (4.11) and (4.12). It turns out that
the numerical results of $\sigma$, $\phi$ and $\alpha$ are
essentially unchanged, but those of $\beta$, $\gamma$ and
$J^{}_{\rm CP}$ require the replacements $\beta \leftrightarrow
\gamma$ and $J^{}_{\rm CP} \rightarrow -J^{}_{\rm CP}$.

\subsection{More Texture Zeros}

It is straightforward to consider more texture zeros in $M^{}_{\rm
D}$. If $n$ (for $n=1, 2, \cdots, 6$) elements of $M^{}_{\rm D}$
are vanishing, there are totally
\begin{equation}
{\bf C}^n_6 = \frac{6 !}{n ! (6 - n) !}
\end{equation}
patterns of $M^{}_{\rm D}$. In the case of $n=3$, we are left with
$20$ distinct textures of $M^{}_{\rm D}$:
\begin{eqnarray}
{\rm Patterns ~ A1 ~ to ~ A4} &:& ~~ \left(\begin{matrix} {\bf 0}
& {\bf 0} \cr {\bf 0} & \times \cr \times & \times \cr
\end{matrix}\right) \; ,~ \left(\begin{matrix} {\bf 0} & {\bf 0}
\cr \times & {\bf 0} \cr \times & \times \cr
\end{matrix}\right) \; ,~ \left(\begin{matrix} {\bf 0} & {\bf 0} \cr
\times & \times \cr {\bf 0} & \times \cr
\end{matrix}\right) \; ,~ \left(\begin{matrix} {\bf 0} & {\bf 0} \cr
\times & \times \cr \times & {\bf 0} \cr
\end{matrix}\right) \; ; ~~~~~~~~
\end{eqnarray}
\begin{eqnarray}
{\rm Patterns ~ B1 ~ to ~ B4} &:& ~~ \left(\begin{matrix} {\bf 0}
& \times \cr {\bf 0} & {\bf 0} \cr \times & \times \cr
\end{matrix}\right) \; ,~ \left(\begin{matrix} \times & {\bf 0} \cr {\bf 0} & {\bf 0}
\cr \times & \times \cr
\end{matrix}\right) \; ,~ \left(\begin{matrix}
\times & \times \cr {\bf 0} & {\bf 0} \cr {\bf 0} & \times \cr
\end{matrix}\right) \; ,~ \left(\begin{matrix}
\times & \times \cr {\bf 0} & {\bf 0} \cr \times & {\bf 0} \cr
\end{matrix}\right) \; ; ~~~~~~~~
\end{eqnarray}
\begin{eqnarray}
{\rm Patterns ~ C1 ~ to ~ C4} &:& ~~ \left(\begin{matrix} {\bf 0}
& \times \cr \times & \times \cr {\bf 0} & {\bf 0} \cr
\end{matrix}\right) \; ,~ \left(\begin{matrix} \times & {\bf 0} \cr \times & \times
\cr {\bf 0} & {\bf 0} \cr
\end{matrix}\right) \; ,~ \left(\begin{matrix}
\times & \times \cr {\bf 0} & \times \cr {\bf 0} & {\bf 0} \cr
\end{matrix}\right) \; ,~ \left(\begin{matrix}
\times & \times \cr \times & {\bf 0} \cr {\bf 0} & {\bf 0} \cr
\end{matrix}\right) \; ; ~~~~~~~~
\end{eqnarray}
\begin{eqnarray}
{\rm Patterns ~ D1 ~ to ~ D4} &:& ~~ \left(\begin{matrix} {\bf 0}
& \times \cr {\bf 0} & \times \cr {\bf 0} & \times \cr
\end{matrix}\right) \; ,~ \left(\begin{matrix} \times & {\bf 0}
\cr {\bf 0} & \times \cr {\bf 0} & \times \cr
\end{matrix}\right) \; ,~ \left(\begin{matrix} {\bf 0} & \times \cr
\times & {\bf 0} \cr {\bf 0} & \times \cr
\end{matrix}\right) \; ,~ \left(\begin{matrix} {\bf 0} & \times \cr
{\bf 0} & \times \cr \times & {\bf 0} \cr
\end{matrix}\right) \; ; ~~~~~~~~
\end{eqnarray}
\begin{eqnarray}
{\rm Patterns ~ E1 ~ to ~ E4} &:& ~~~ \left(\begin{matrix} \times
& {\bf 0} \cr \times & {\bf 0} \cr \times & {\bf 0} \cr
\end{matrix}\right) \; ,~ \left(\begin{matrix} {\bf 0} & \times
\cr \times & {\bf 0} \cr \times & {\bf 0} \cr
\end{matrix}\right) \; ,~ \left(\begin{matrix} \times & {\bf 0} \cr
{\bf 0} & \times \cr \times & {\bf 0} \cr
\end{matrix}\right) \; ,~ \left(\begin{matrix} \times & {\bf 0} \cr
\times & {\bf 0} \cr {\bf 0} & \times \cr
\end{matrix}\right) \; , ~~~~~~~~~
\end{eqnarray}
in which ``$\times$" denotes an arbitrary non-vanishing matrix
element.

It is quite obvious that the textures of $M^{}_\nu$ resulting from
Category A, B or C of $M^{}_{\rm D}$ have been ruled out by
current experimental data, because they only have non-vanishing
entries in the (2,3), (3,1) or (1,2) block and cannot give rise to
the phenomenologically-favored bi-large neutrino mixing pattern.
Categories D and E of $M^{}_{\rm D}$ can be transformed into each
other by the exchange between $a^{}_i$ and $b^{}_i$ (for $i=1, 2,
3$). Hence let us examine the four patterns of $M^{}_{\rm D}$ in
Category D. Given three (or more) texture zeros in $M^{}_{\rm D}$,
its non-vanishing elements can all be chosen to be real by
redefining the phases of three charged lepton fields. Considering
Pattern D1, for example, we have
\begin{equation}
M^{}_\nu = \frac{1}{M^{}_2}\left(
\begin{matrix} b^2_1 & ~~ b^{}_1 b^{}_2 ~~ & b^{}_1
b^{}_3 \cr b^{}_2 b^{}_1 & b^2_2 & b^{}_2 b^{}_3 \cr b^{}_3 b^{}_1
& b^{}_3 b^{}_2 & b^2_3 \cr
\end{matrix}\right) \; ,
\end{equation}
which is actually of rank one and has two vanishing neutrino mass
eigenvalues. This result {\it does} conflict with the neutrino
oscillation data. As for Patterns D2, D3 and D4, the resultant
textures of $M^{}_\nu$ are
\begin{equation}
~~~~~~~\left( \begin{matrix} \times & {\bf 0} & {\bf 0} \cr {\bf
0} & \times & \times \cr {\bf 0} & \times & \times \cr
\end{matrix} \right) \; , ~~~~
\left( \begin{matrix} \times & {\bf 0} & \times \cr {\bf 0} &
\times & {\bf 0} \cr \times & {\bf 0} & \times \cr
\end{matrix} \right) \; , ~~~~
\left( \begin{matrix} \times & \times & {\bf 0} \cr \times &
\times & {\bf 0} \cr {\bf 0} & {\bf 0} & \times \cr
\end{matrix} \right) \; ,
\end{equation}
respectively. These three two-zero textures of $M^{}_\nu$ have
also been excluded by the present experimental
data.\cite{XING03,Zero} Therefore, we conclude that the patterns
of $M^{}_{\rm D}$ with three or more texture zeros are all
phenomenologically disfavored in the MSM.

\subsection{Radiative Corrections}

Now we discuss the possible renormalization-group running effects
on neutrino masses and lepton flavor mixing parameters between the
electroweak scale and the seesaw scale in the MSM. At energies far
below the mass of the lighter right-handed Majorana neutrino
$M^{}_1$, two right-handed Majorana neutrino fields can be
integrated out from the theory. Such a treatment will induce a
dimension-5 operator $\overline{l^{}_{\rm L}} \tilde{H} \kappa
\tilde{H}^T l^c_{\rm L}$ in the effective Lagrangian, whose
coupling matrix takes the canonical seesaw form at the scale $\mu
= M^{}_1$:
\begin{eqnarray}
\kappa^{~} (M^{}_1) \; = \; - Y^{~}_\nu M^{-1}_{\rm R} Y^{\rm T}_\nu
\; .
\end{eqnarray}
After the spontaneous gauge symmetry breaking, one may obtain the
effective mass matrix of three light (left-handed) Majorana
neutrinos $M_\nu = v^2 \kappa (M_Z)$ at the electroweak scale $\mu
= M^{}_Z$.

In the flavor basis where the charged-lepton and right-handed
Majorana neutrino mass matrices are both diagonal, one can
simplify the one-loop renormalization-group equations
(RGEs).\cite{RGE,MEI} The effective coupling matrix $\kappa$ will
receive radiative corrections when the energy scale runs from
$M^{}_1$ down to $M^{}_Z$. To be more explicit, $\kappa (M_Z)$ and
$\kappa (M_1)$ can be related to each other via
\begin{equation}
\kappa (M_Z) \; =\;  I^{}_\alpha \left ( \begin{matrix} I^{}_e & 0
& 0 \cr 0 & I^{}_\mu & 0 \cr 0 & 0 & I^{}_\tau \cr\end{matrix}
\right ) \kappa (M^{}_1) \left ( \begin{matrix} I^{}_e & 0 & 0 \cr
0 & I^{}_\mu & 0 \cr 0 & 0 & I^{}_\tau \cr \end{matrix} \right )
\; ,
\end{equation}
where $I^{}_\alpha$ and $I^{}_l$ ( for $l = e, \mu, \tau$ ) are
the RGE evolution functions.\cite{MEI} The overall factor
$I^{}_\alpha$ only affects the magnitudes of light neutrino
masses, while $I^{~}_l$ can modify the neutrino masses, flavor
mixing angles and CP-violating phases.\cite{Ellis} The strong mass
hierarchy of three charged leptons (i.e., $m^{}_e < m^{}_\mu <
m^{}_\tau$) implies that $I^{}_e < I^{}_\mu < I^{}_\tau$ holds
below the scale $\mu = M^{}_1$.\cite{MEI} Two comments are in
order.
\begin{enumerate}
\item The determinant of $\kappa$, which vanishes at $\mu =
M^{}_1$, keeps vanishing at $\mu = M^{}_Z$. This point can clearly
be seen from the relation
\begin{equation}
{\rm Det}[\kappa (M^{}_Z)] \; =\; I^3_\alpha I^2_e I^2_\mu
I^2_\tau ~ {\rm Det}[\kappa (M^{}_1)] \; .
\end{equation}
Taking account of $m^{}_1 = 0$ or $m^{}_3 = 0$, we have
$\left|{\rm Det}[\kappa (M^{}_Z)] \right| = m^{}_1 m^{}_2 m^{}_3 /
v^6 = 0$.

\item Comparing between Eqs. (4.21) and (4.22), we find that the
radiative correction to $\kappa$ can effectively be expressed as
the RGE running effects in the elements of $M^{}_{\rm D}$ (i.e.,
$a^{}_i$ and $b^{}_i$):
\begin{eqnarray}
a^{}_1 (M^{}_Z) & = & I^{}_e \sqrt{I^{}_\alpha} ~ a^{}_1 (M^{}_1)
\; ,
\nonumber \\
a^{}_2 (M^{}_Z) & = & I^{}_\mu \sqrt{I^{}_\alpha} ~ a^{}_2
(M^{}_1) \; ,
\nonumber \\
a^{}_3 (M^{}_Z) & = & I^{}_\tau \sqrt{I^{}_\alpha} ~ a^{}_3
(M^{}_1) \; ,
\end{eqnarray}
with the assumption that $M^{}_1$ keeps unchanged. The same
relations can be obtained for $b_i$ (for $i=1, 2 ,3$) at two
different energy scales. This observation indicates that possible
texture zeros of $\kappa$ at $\mu = M^{}_1$ remain there even at
$\mu = M^{}_Z$, at least at the one-loop level of the RGE
evolution. In other words, the texture zeros of $\kappa$ are
essentially stable against quantum corrections from $M^{}_1$ to
$M^{}_Z$.
\end{enumerate}
To illustrate, we typically take the top-quark mass $m^{}_t
(M^{}_Z) \approx 181~{\rm GeV}$ to calculate the evolution
functions $I^{}_\alpha$ and $I^{~}_l$ (for $l=e,\mu,\tau$). It
turns out that $I^{}_e \approx I^{}_\mu \approx 1$ is an excellent
approximation in the SM. Thus the RGE running of $\kappa$ is
mainly governed by $I^{}_\alpha$ and $I^{}_\tau$. The behaviors of
$I^{}_\alpha$ and $I^{}_\tau$ changing with $M^{}_1$ are shown in
Fig. 4.2. One can see that $I^{}_\tau \approx 1$ is also a good
approximation in the SM and in the MSSM with mild values of
$\tan\beta$. Hence the evolution of three light neutrino masses
are dominated by $I^{}_\alpha$, which may significantly deviate
from unity.
\begin{figure}
\vspace{2cm}
\epsfig{file=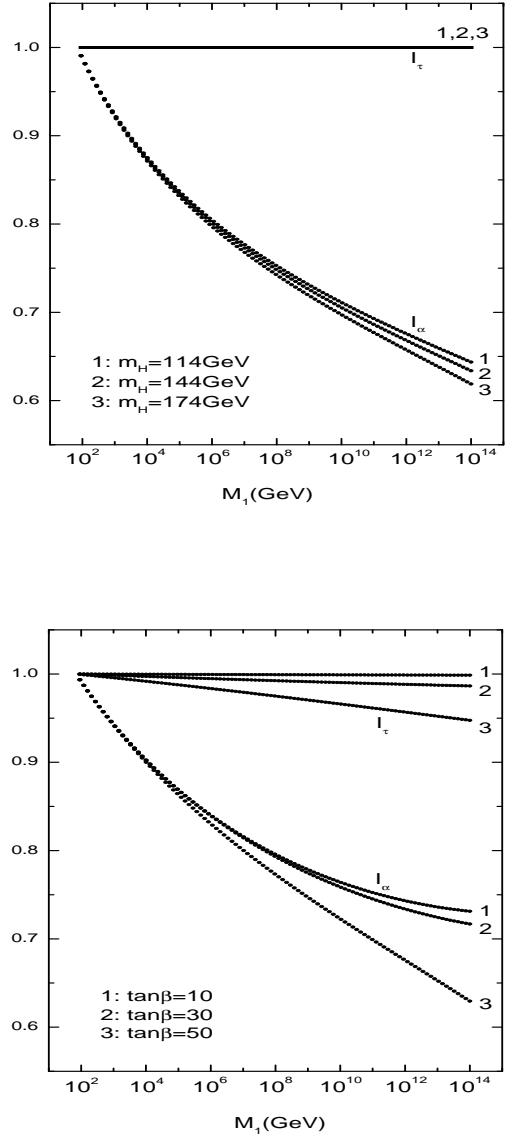,bbllx=2cm,bblly=2cm,bburx=21cm,bbury=24cm,
width=14.0cm,height=15cm,angle=0,clip=0} \vspace{-2.0cm}
\caption{Numerical illustration of the evolution functions
$I^{}_\alpha$ and $I^{}_\tau $ changing with $M^{}_1$ for
different values of the Higgs mass $m^{~}_H$ in the SM (up) or for
different values of $\tan\beta$ in the MSSM (down).}
\end{figure}

We proceed to discuss radiative corrections to three neutrino
masses. For simplicity, here we mainly consider the $m^{}_1 =0$
case. The RGE running of $m^{}_i$, $\dot{m}^{}_i \equiv {\rm d}
m^{}_i/{\rm d}t$ with $t = [\ln (\mu/M^{}_Z)]/(16\pi^2)$, is
proportional to $m^{}_i$ itself (for $i = 1,2,3$) at the one-loop
level.\cite{RGE} Explicitly,\cite{MEI}
\begin{eqnarray}
\dot{m}^{}_1 & = & 0 \; ,
\nonumber \\
\dot{m}^{}_2 & \approx & \frac{1}{16 \pi^2} \left( \alpha + 2 C
f_\tau^2 c_x^2 s_y^2 \right) m^{}_2 \; ,
\nonumber \\
\dot{m}^{}_3 & \approx & \frac{1}{16 \pi^2} \left( \alpha + 2 C
f_\tau^2 c_y^2 \right) m^{}_3 \; ,
\end{eqnarray}
where $C= -3/2$ (SM) or 1 (MSSM), $\alpha$ denotes the
contribution from both the gauge couplings and the top-quark
Yukawa coupling,\cite{RGE} and $f^{}_\tau$ is the tau-lepton
Yukawa coupling. It becomes clear that the running behaviors of
$m^{}_2$ and $m^{}_3$ are essentially identical.\cite{MEI} For
illustration, we show the ratio ${\rm R} \equiv
m^{}_2(M^{}_Z)/m^{}_2(M^{}_1)$ changing with the Higgs mass
$m^{~}_H$ (SM) or with $\tan \beta$ (MSSM) in Fig. 4.3, where
$M^{}_1 = 10^{14}$ GeV has typically been taken and the $m^{}_3
=0$ case is also included. One can see that ${\rm R}_{m^{}_1 =0}
\approx {\rm R}_{m^{}_3 =0} \approx I^{}_\alpha$ is an excellent
approximation in the SM or in the MSSM with mild values of
$\tan\beta$.
\begin{figure}
\vspace{2cm}
\epsfig{file=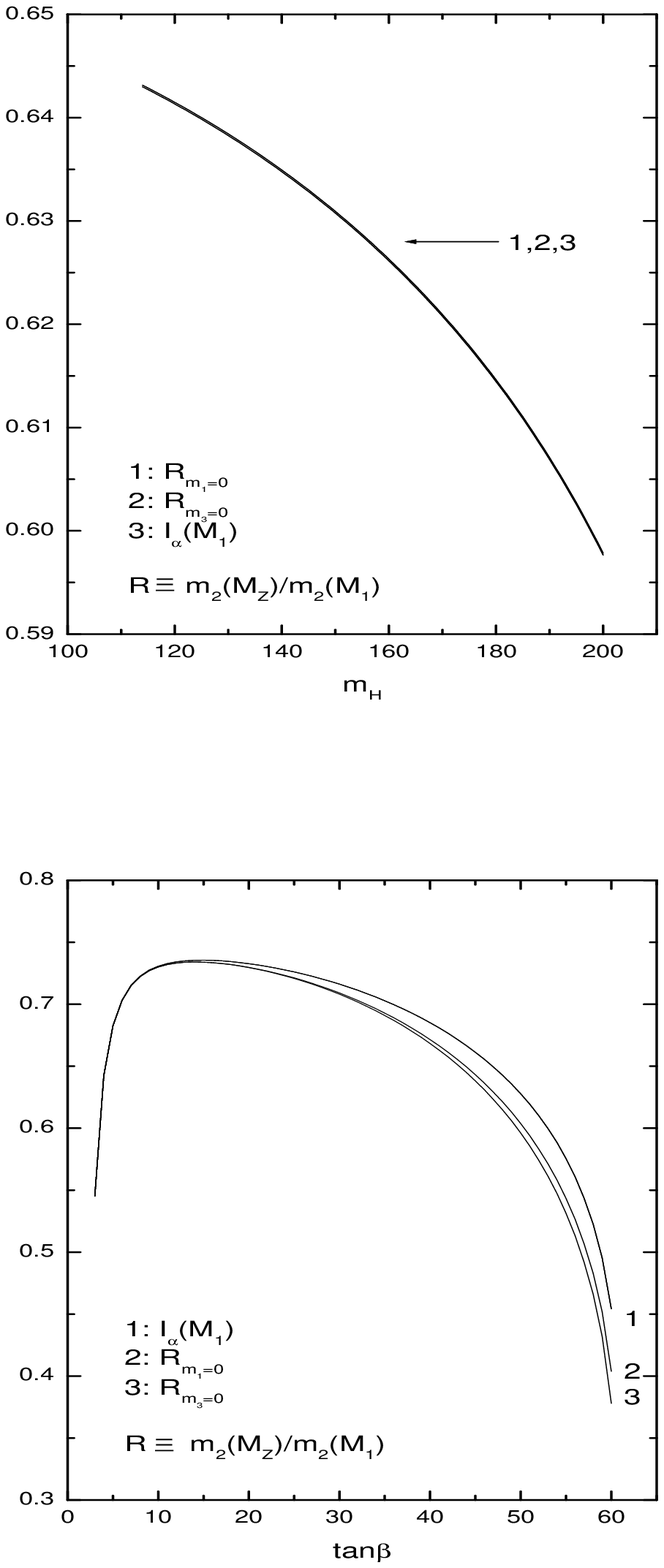,bbllx=2cm,bblly=2cm,bburx=21cm,bbury=24cm,
width=14.0cm,height=15cm,angle=0,clip=0} \vspace{-2.35cm}
\caption{Numerical illustration of the ratio $R\equiv m^{}_2
(M^{}_Z)/ m^{}_2 (M^{}_1)$ as a function of $m^{}_H$ in the SM
(up) or of $\tan\beta$ in the MSSM (down), where $M^{}_1 =
10^{14}$ GeV has typically been input for the MSM with either
$m^{}_1 =0$ or $m^{}_3 =0$.}
\end{figure}

The RGEs of three flavor mixing angles $(\theta^{}_x, \theta^{}_y,
\theta^{}_z)$ and two CP-violating phases $(\delta, \sigma)$ in
the $m^{}_1 =0$ case are approximately given by\cite{MEI}
\begin{eqnarray}
\dot{\theta}^{}_x & \approx & - \frac{C f_\tau^2}{16 \pi^2} s^{}_x
c^{}_x s_y^2 \; ,
\nonumber \\
\dot{\theta}^{}_y & \approx & - \frac{C f_\tau^2}{16 \pi^2} s^{}_y
c^{}_y \left (1 + 2 r^{}_{23} c_x^2 \cos \delta \right ) \; ,
\nonumber \\
\dot{\theta}^{}_z & \approx & - \frac{C f_\tau^2}{8 \pi^2}
r^{}_{23} s^{}_x c^{}_x s^{}_y c^{}_y \; ;
\end{eqnarray}
and
\begin{eqnarray}
\dot{\sigma} & \approx & \frac{C f_\tau^2}{8 \pi^2} \left (s^{}_x
c^{}_x s^{}_y c^{}_y \frac{r^{}_{23}}{s^{}_z} \right ) r^{}_{23}
\sin \delta \; ,
\nonumber \\
\dot{\delta} & \approx & \frac{C f_\tau^2}{8 \pi^2} \left [s^{}_x
c^{}_x s^{}_y c^{}_y \frac{r^{}_{23}}{s^{}_z} + c^2_x \left( c_y^2
- s_y^2 \right) \right] r^{}_{23} \sin\delta \; ,
\end{eqnarray}
where $r^{}_{23} \equiv m^{}_2/m^{}_3$ has been defined before. We
see that the running effects of these five parameters are all
governed by $f^2_\tau$. Because of $f^2_\tau \approx 10^{-4}$ in
the SM, the evolution of $(\theta^{}_x, \theta^{}_y, \theta^{}_z)$
and $(\sigma, \delta)$ is negligibly small. When $\tan\beta$ is
sufficiently large (e.g., $\tan\beta \sim 50$) in the MSSM,
however, $f^2_\tau \approx 10^{-4} /\cos^2\beta$ can be of ${\cal
O}(0.1)$ and even close to unity --- in this case, some small
variation of $(\theta^{}_x, \theta^{}_y, \theta^{}_z)$ and
$(\sigma, \delta)$ due to the RGE running from $M^{}_Z$ to
$M^{}_1$ will appear. A detailed analysis\cite{MEI} has shown that
the smallest neutrino mixing angle $\theta^{}_z$ is most sensitive
to radiative corrections, but its change from $\mu = M^{}_Z$ to
$\mu = M^{}_1$ is less than $10\%$ even if $M^{}_1 = 10^{14}$ GeV
and $\tan\beta = 50$ are taken. Thus we conclude that the RGE
effects on three flavor mixing angles and two CP-violating phases
are practically negligible in the MSM with $m^{}_1 =0$. As for the
$m^{}_3 =0$ case, it is found that the near degeneracy between
$m^{}_1$ and $m^{}_2$ may result in significant RGE running
effects on the mixing angle $\theta^{}_x$ in the MSSM, and the
evolution of two CP-violating phases can also be appreciable if
both $M^{}_1$ and $\tan\beta$ take sufficiently large
values.\cite{MEI}

\subsection{Non-Diagonal $M^{}_l$ and $M^{}_{\rm R}$}

So far we have been working in the flavor basis where both
$M^{}_l$ and $M^{}_{\rm R}$ are diagonal. In an arbitrary flavor
basis, however, $M^{}_l$ and $M^{}_{\rm R}$ need to be
diagonalized by using proper unitary transformations:
\begin{equation}
M^{}_l \; =\; U^{}_l  \left ( \begin{matrix} m^{}_e & 0 & 0 \cr 0 &
m^{}_\mu & 0 \cr 0 & 0 & m^{}_\tau \cr \end{matrix} \right ) \tilde
{U}^\dagger_l \; ,
\end{equation}
and
\begin{equation}
M^{}_{\rm R} \; = \; U^{}_{\rm R} \left ( \begin{matrix} M^{}_1 ~
& ~ 0 \cr 0 ~ & ~ M^{}_2 \cr  \end{matrix} \right ) U^T_{\rm R} \;
.
\end{equation}
When $M^{}_l$ is Hermitian or symmetric, we have $\tilde{U}^{}_l =
U^{}_l$ or $\tilde{U}^{}_l = U^*_l$. The MNS matrix is in general
given by $V^{}_{\rm MNS} = U^\dagger_l V$, where $V$ is the
unitary matrix used to diagonalize the effective neutrino mass
matrix $M^{}_\nu$ in Eq. (2.6). Without loss of generality,
$U^{}_l$ can be parameterized in terms of three rotation angles
and one phase, while $U^{}_{\rm R}$ can be parameterized in terms
of one rotation angle and one phase.

Let us make some brief comments on the texture of $M^{}_{\rm D}$
in the flavor basis where $M^{}_l$ and $M^{}_{\rm R}$ are not
diagonal. There are two possibilities:
\begin{enumerate}
\item      $M^{}_{\rm D}$ has no texture zeros. By redefining the
fields $l^{}_{\rm L}$, $E^{}_{\rm R}$ and $N^{}_{\rm R}$, we
transform $M^{}_l$ and $M^{}_{\rm R}$ into the diagonal mass
matrices $\tilde{M}^{}_l$ and $\tilde{M}^{}_{\rm R}$. Then
$M^{}_{\rm D}$ becomes $\tilde{M}^{}_{\rm D} = U_l^\dagger
M^{}_{\rm D} U_{\rm R}^*$ in the new basis. If $\tilde{M}^{}_{\rm
D}$ has no texture zeros, we cannot get any extra constraint on
the seesaw relation. Provided $\tilde{M}^{}_{\rm D}$ has texture
zeros, $(\tilde{M}_{\rm D})_{ij} = 0$, then we have
\begin{equation}
(\tilde{M}_{\rm D})^{}_{ij} \; = \left(i V^{}_{\rm MNS} \sqrt{m} R
\sqrt{M'_{\rm R}} \right )^{}_{ij} = 0 \;
\end{equation}
in the Casas-Ibarra-Ross parameterization, where $\sqrt{M'_{\rm
R}}= R' \sqrt{M^{}_{\rm R}} U^*_{\rm R}$ is a diagonal matrix with
$R'$ being a $2 \times 2$ orthogonal matrix.

\item      $M^{}_{\rm D}$ has texture zeros.  Then
$(M^{}_{\rm D})^{}_{ij} = 0$ means
\begin{equation}
(M^{}_{\rm D})^{}_{ij} \; =\; \left (i U^{}_l V^{}_{\rm MNS}
\sqrt{m} R \sqrt{M^{}_{\rm R}} \right )^{}_{ij} = 0
\end{equation}
in the Casas-Ibarra-Ross parameterization. After transforming
$M^{}_l$ and $M^{}_{\rm R}$ into $\tilde{M}^{}_l$ and
$\tilde{M}^{}_{\rm R}$, we get $\tilde{M}^{}_{\rm D} = U_l^\dagger
M^{}_{\rm D} U_{\rm R}^*$ in the new basis. If $\tilde{M}^{}_{\rm
D}$ has texture zeros, Eq. (4.30) will be applicable. Otherwise,
only Eq. (4.31) can impose some constraints on the
model.\cite{ROSS,MOHAPATRA}
\end{enumerate}

\subsection{Comments on Model Building}

To dynamically understand possible texture zeros in $M^{}_{\rm
D}$, one may incorporate a certain flavor symmetry in the
supersymmetric version of the MSM. For illustration, we first
consider the $SU(2)^{}_{\rm H}$ horizontal symmetry. In the
presence of a local $SU(2)^{}_{\rm H}$ horizontal symmetry under
which right-handed charged leptons transform nontrivially, freedom
from global anomalies requires that there be at least two
right-handed neutrinos with masses of order of the horizontal
symmetry breaking scale.\cite{MOHAPATRA} Taking account of the
quark-lepton symmetry, one may introduce an extra right-handed
neutrino, which is the $SU(2)^{}_{\rm H}$ singlet and too heavy to
couple to low-energy physics. In the leptonic sector, the
$SU(2)^{}_{\rm H}$ doublets include $(l^{}_{e \rm L}, l^{}_{\mu
\rm L})$, $(\mu^{}_{\rm R}, - e^{}_{\rm R})$ and $(\nu^{}_{\mu \rm
R}, - \nu^{}_{e \rm R})$, and the $SU(2)^{}_{\rm H}$ singlets are
$l^{}_{\tau \rm L}$, $\tau^{}_{\rm R}$ and $\nu^{}_{\tau \rm R}$.
In addition to the MSSM Higgs doublets $H^{}_1$ and $H^{}_2$, the
new Higgs doublets $\chi = (\chi^{}_1, \chi^{}_2)$ and $\bar{\chi}
= (-\bar{\chi}^{}_2, \bar{\chi}^{}_1)$ are assumed. The
gauge-invariant Yukawa couplings relevant for the Dirac neutrino
mass matrix is given by\cite{MOHAPATRA}
\begin{equation}
W^{}_{\rm Y} = h^{}_0 \left (l^{}_{e \rm L} H^{}_2 \nu^{}_{e \rm
R} + l^{}_{\mu \rm L} H^{}_2 \nu^{}_{\mu \rm R} \right ) + h^{}_1
l^{}_{\tau \rm L} \left (\nu^{}_{\mu \rm R} \chi^{}_2 + \nu^{}_{e
\rm R} \chi^{}_1 \right ) H^{}_2 / M \; ,
\end{equation}
where $M$ can be regarded as the scale of the horizontal symmetry
breaking. After the horizontal and gauge symmetries are
spontaneously broken down, the Higgs fields gain their vevs as
$\langle H^{}_i \rangle = v^{}_i$, $\langle \chi^{}_i \rangle =
u^{}_i$ (for $i = 1, 2$). Then we obtain the Dirac neutrino mass
matrix
\begin{equation}
M^{}_{\rm D} = \left( \begin{matrix} h^{}_0 v^{}_2 & {\bf 0} \cr
{\bf 0} & h^{}_0 v^{}_2 \cr h^{}_1 w^{}_1 & h^{}_1 w^{}_2 \cr
\end{matrix}\right) \; ,
\end{equation}
where $w^{}_i \equiv v^{}_2 u^{}_i / M$ (for $i=1, 2$). Note that
the mass matrices $M^{}_l$ and $M^{}_{\rm R}$ are in general not
diagonal. This scenario indicates that the MSM can be viewed as
the special case of a more generic seesaw model with three
right-handed Majorana neutrinos, when one of them is so heavy that
it essentially decouples from low-energy physics.

Another simple scenario, in which the MSM is incorporated with a
$SU(2)\times U(1)$ family symmetry, has also been
proposed.\cite{RABY} It can naturally result in the texture of
$M^{}_{\rm D}$ in Eq. (4.5). The superpotential relevant for
$M^{}_{\rm D}$ in this model is written as\cite{RABY}
\begin{equation}
W^{}_{\rm Y} = \frac{H}{M}  \left (L^{}_a  \phi^a  N^{}_1  +
L^{}_a \tilde {\phi}^a N^{}_2 + l^{}_3 \omega N^{}_2 \right )   +
\frac{1}{2} \left (S^{~}_1 N_1^2 + S^{~}_2 N_2^2 \right ) \; ,
\end{equation}
where $L^{}_a = (l^{}_{\rm L 1}, ~ l^{}_{\rm L 2} )^T$ is a
doublet of the $SU(2)$ family symmetry, while $l^{}_{\rm L 3}$ is
a singlet. In addition, two flavor (anti)-doublets ($\phi^a$ and
$\tilde{ \phi}^a$), four flavor singlets ($N^{}_1$, $N^{}_2$,
$S^{}_1$ and $S^{}_2$) and the SM Higgs doublet $H$ are
introduced. Note that $M$ is a superhigh mass scale in Eq. (4.34).
In the basis where $M^{}_l$ is diagonal, the $U(1)$ charge
assignments for the fields $\{ L^{}_a, l^{}_3, N^{}_1, N^{}_2,
\phi^a, \tilde {\phi}^a, \omega, S^{}_1, \ S^{}_2 \}$ are $\{ 1,
\xi, x, y, -(x+1), -(y + 1), -( \xi + y), -2 x, -2 y \}$ with $x
\neq y$. We assume that $\phi$ and $\tilde{\phi}$ can get vevs
$\langle \phi \rangle = (\phi^1, ~\phi^2)^T$ and $\langle
\tilde{\phi} \rangle = (0, ~\tilde{\phi}^2)^T$. The vevs $\langle
S_i \rangle = M_i$  (for $i = 1,2$) are also needed to give the
states $N^{}_i$ sufficiently large masses. These vevs can be
obtained via the suitable terms added to the above
superpotential.\cite{RABY} Then we obtain the texture of
$M^{}_{\rm D}$ as given in Eq. (4.5), where
\begin{eqnarray}
a^{}_1 & = & v \sin \beta \frac{\phi^1}{ \sqrt{2} M} \; ,
\nonumber \\
a^{}_2 & = & v \sin \beta \frac{\phi^2}{\sqrt{2} M} \; ,
\nonumber \\
b^{}_2 & = & v \sin \beta \frac{\tilde{\phi}^2}{\sqrt{2} M} \; ,
\nonumber \\
b^{}_3 & = & v \sin \beta \frac{\omega}{\sqrt{2} M} \; ,
\end{eqnarray}
together with $\langle H \rangle = (0, ~ v \sin \beta)^T$. These
vevs are in general complex.

Finally, it is worth mentioning that the FGY ansatz can also be
derived from certain extra-dimensional models.\cite{RAIDAL}
Another possibility to obtain the texture zeros in $M^{}_{\rm D}$
is to require the vanishing of certain CP-odd invariants together
with a reasonable assumption of no conspiracy among the parameters
of $M^{}_{\rm D}$ and $M^{}_{\rm R}$.\cite{BRANCO}

\section{Baryogenesis via Leptogenesis}
\setcounter{equation}{0} \setcounter{figure}{0}

The cosmological baryon number asymmetry is one of the most
striking mysteries in the Universe. Thanks to the three-year WMAP
observation,\cite{WMAP} the ratio of baryon to photon number
densities can now be determined to a very good precision:
$\eta^{}_{\rm B} \equiv n^{~}_{\rm B}/n^{~}_\gamma = (6.1 \pm 0.2)
\times 10^{-10}$. This tiny quantity measures the observed
matter-antimatter or baryon-antibaryon asymmetry of the Universe,
\begin{equation}
Y^{}_{\rm B} \; \equiv \; \frac{n^{~}_{\rm B} - n^{~}_{\rm\bar
B}}{\bf s} \; \approx \; \frac{\eta^{}_{\rm B}}{7.04} \; \approx
\; (8.66 \pm 0.28) \times 10^{-11} \; ,
\end{equation}
where $\bf s$ denotes the entropy density. To dynamically produce
a net baryon number asymmetry in the framework of the standard
Big-Bang cosmology, three Sakharov necessary conditions have to be
satisfied:\cite{Sakharov} (a) baryon number non-conservation, (b)
C and CP violation, and (c) departure from thermal equilibrium.
Among a number of baryogenesis mechanisms existing in the
literature,\cite{Baryon} the one via leptogenesis\cite{LEP} is
particularly interesting and closely related to neutrino physics.

\subsection{Thermal Leptogenesis}

First of all, let us outline the main points of thermal
leptogenesis in the MSM. The decays of two heavy right-handed
Majorana neutrinos, $N^{}_i \rightarrow l + H^\dagger$ and $N^{}_i
\rightarrow l^c + H$ (for $i=1,2$), are both
lepton-number-violating and CP-violating. The CP asymmetry
$\varepsilon^{}_i$ arises from the interference between the
tree-level and one-loop decay amplitudes. If $N^{}_1$ and $N^{}_2$
have a hierarchical mass spectrum ($ M^{}_1 \ll M^{}_2 $), the
interactions involving $N^{}_1$ can be in thermal equilibrium when
$N^{}_2$ decays. Hence $\varepsilon^{}_2$ is erased before
$N^{}_1$ decays. The CP-violating asymmetry $\varepsilon^{}_1$,
which is produced by the out-of-equilibrium decay of $N^{}_1$, may
finally survive. For simplicity, we assume $ M^{}_1 \ll M^{}_2 $
here and in Sec. 5.2. The possibility of $M^{}_1 \approx M^{}_2$,
which gives rise to the resonant leptogenesis,\cite{DEG} will be
discussed in Sec. 5.3.

In the flavor basis where the mass matrices of charged leptons
($M^{}_l$) and right-handed Majorana neutrinos ($M^{}_{\rm R}$)
are both diagonal, one may calculate the CP-violating asymmetry
$\varepsilon^{}_1$:\cite{Covi}\footnote{For simplicity, we do not
distinguish different lepton flavors in the final states of the
$N^{}_1$ decay. Such flavor effects in leptogenesis may not be
negligible in some cases.\cite{Flavor}}
\begin{eqnarray}
\varepsilon^{}_1 & \equiv & \frac{\Gamma (N^{}_1 \rightarrow l +
H^\dagger) - \Gamma (N^{}_1 \rightarrow l^{\rm c} + H)}
 {\Gamma (N^{}_1 \rightarrow l + H^\dagger)
+ \Gamma (N^{}_1 \rightarrow l^{\rm c} + H)} \cr & \approx & -
\frac {3}{16\pi v^2}  \frac{M^{}_1}{M^{}_2}  \frac{{\rm Im} \left
[ (M^\dagger_{\rm D} M^{~}_{\rm D})^2_{12} \right ]}
{(M^\dagger_{\rm D} M^{~}_{\rm D})^{}_{11}} \; .
\end{eqnarray}
Leptogenesis means that $\varepsilon^{}_1$ gives rise to a net
lepton number asymmetry in the Universe,
\begin{eqnarray}
Y^{}_{\rm  L} \equiv \frac{n^{}_{\rm  L} - n^{}_{\bar{{\rm
L}}}}{\bf s} = \frac{\kappa}{g^{}_*} \; \varepsilon^{}_1 \; ,
\end{eqnarray}
where $g^{}_* = 106.75$ is an effective number characterizing the
relativistic degrees of freedom which contribute to the entropy of
the early Universe, and $\kappa$ accounts for the dilution effects
induced by the lepton-number-violating wash-out processes. The
efficiency factor $\kappa$ can be figured out by solving the full
Boltzmann equations.\cite{Covi} For simplicity, here we take the
following analytical approximation for $\kappa$:\cite{Kappa}
\begin{equation}
\kappa \; \approx \; 0.3 \left (\frac{10^{-3} ~ {\rm
eV}}{\tilde{m}^{}_1} \right ) \left [ \ln \left (
\frac{\tilde{m}^{}_1}{10^{-3} ~ {\rm eV}} \right ) \right ]^{-0.6}
\;
\end{equation}
with $\tilde{m}^{}_1 = (M^\dagger_{\rm D} M^{}_{\rm D})^{}_{11} /
{M^{}_1}$. The lepton number asymmetry $Y^{}_{\rm  L}$ is
eventually converted into a net baryon number asymmetry $Y^{}_{\rm
B}$ via the non-perturbative sphaleron processes,\cite{Kuzmin,HT}
\begin{eqnarray}
Y^{}_{\rm B} = - c\; Y^{}_{\rm L} \; ,
\end{eqnarray}
where $c = 28/79 \approx 0.35$ in the SM. A similar relation
between $Y^{}_{\rm B}$ and $Y^{}_{\rm L}$ can be obtained in the
supersymmetric extension of the MSM.\cite{Covi}

\subsection{Upper Bound of $|\varepsilon^{}_1|$}

In those seesaw models with three right-handed Majorana neutrinos,
the CP-violating asymmetry $\varepsilon^{}_1$ has an upper
bound\cite{BOUND}
\begin{eqnarray}
|\varepsilon^{}_1| & \leq & \frac{3 M^{}_1}{16 \pi v^2}  \left|
\frac{m_2^2 - m_1^2}{m^{}_2} + \frac{m_3^2 - m_1^2}{m^{}_3}
\right| \; .
\end{eqnarray}
Since $m^{}_1$ or $m^{}_3$ must be massless in the MSM, we ought
to obtain more rigorous constraints on
$|\varepsilon^{}_1|$.\cite{ROSS} But it is not proper to directly
substitute $m^{}_1 = 0$ or $m^{}_3 = 0$ into Eq. (5.6). With the
help of Eqs. (2.4) and (2.6), the expression of $\varepsilon^{}_1$
in Eq. (5.2) can be rewritten as
\begin{eqnarray}
\varepsilon^{}_1  & \approx & - \frac {3}{16\pi v^2}
\frac{M^{}_1}{(M^\dagger_{\rm D} M^{}_{\rm D})^{}_{11}}  {\rm Im}
\left [ (M^T_{\rm D} M^*_\nu M^{}_{\rm D})^{}_{11} \right ], \;
\cr & \approx & - \frac {3}{16\pi v^2}
\frac{M^{}_1}{(M^\dagger_{\rm D} M^{}_{\rm D})^{}_{11}}  {\rm Im}
\left\{\left [(V^\dagger M^{}_{\rm D})^T m (V^\dagger M^{}_{\rm
D}) \right ]^{}_{11} \right \} \; ,
\end{eqnarray}
where $m \equiv {\rm Diag} \{m^{}_1, m^{}_2, m^{}_3 \}$ with
either $m^{}_1 =0$ or $m^{}_3 =0$, and $V$ is the MNS matrix. In
the $m^{}_1 = 0$ case, we adopt the Casas-Ibarra-Ross
parametrization of $M^{}_{\rm D}$ and define
\begin{eqnarray}
K \equiv V^\dagger M^{}_{\rm D} = i \sqrt{m}  ~ R  \sqrt{M^{}_{\rm
R}} \;\; .
\end{eqnarray}
Because of $K^{}_{1i} =0$, we obtain $(M^\dagger_{\rm D}
M^{~}_{\rm D})^{}_{11} = (K^\dagger K)^{}_{11} = |K^{}_{21}|^2 +
|K^{}_{31}|^2$. In addition,
\begin{eqnarray}
{\rm Im} \left\{\left [(V^\dagger M^{}_{\rm D})^T m (V^\dagger
M^{}_{\rm D}) \right ]^{}_{11} \right \} & = & {\rm Im}\left[(K^T
m K)^{}_{11} \right] \nonumber \\
& = & m^{}_2 {\rm Im}\left[K^2_{21}\right] + m^{}_3 {\rm
Im}\left[K^2_{31}\right] \;
\end{eqnarray}
and $m^{}_2 {\rm Im} [K^2_{31}] + m^{}_3 {\rm Im} [K^2_{21}] =0$
hold.\cite{ROSS} Then $\varepsilon^{}_1$ in Eq. (5.7) can be
expressed as
\begin{eqnarray}
\varepsilon^{}_1  & \approx & - \frac {3 M^{}_1}{16\pi v^2}
\frac{m^2_3 - m^2_2}{m^{}_3} \frac{{\rm Im}
[K^2_{31}]}{|K^{}_{21}|^2 + |K^{}_{31}|^2} \;.
\end{eqnarray}
The upper bound of $|\varepsilon^{}_1|$ turns out to be
\begin{eqnarray}
|\varepsilon^{}_1|  & \leq &  \frac {3 M^{}_1}{16\pi v^2}
\frac{\Delta m^2_{\rm atm}}{\sqrt{\Delta m^2_{\rm atm} + \Delta
m^2_{\rm sun}}} \;
\end{eqnarray}
in the $m^{}_1 =0$ case. Similarly, one may get
\begin{eqnarray}
|\varepsilon^{}_1|  & \leq &  \frac {3 M^{}_1}{16\pi v^2}
\frac{\Delta m^2_{\rm sun}}{\sqrt{\Delta m^2_{\rm atm}}} \;
~~~~~~~~~~~~
\end{eqnarray}
in the $m^{}_3 =0$ case.

In the MSM, the successful leptogenesis depends on three
parameters: $\varepsilon^{}_1$, $M^{}_1$ and $\tilde{m}^{}_1$.
Because of the washout effects, which are characterized by
$\tilde{m}^{}_1$, the maximal $\varepsilon^{}_1$ does not imply
the minimal $M^{}_1$. Taking $m^{}_1 =0$ for example and making
use of Eqs. (3.2) and (5.8), we obtain
\begin{eqnarray}
\frac{{\rm Im} [K^2_{31}]}{|K^{}_{21}|^2 + |K^{}_{31}|^2}   = -
\frac{ m^{}_3 {\rm Im} [\sin^2 z]}{m^{}_2 |\cos z|^2 + m^{}_3
|\sin z|^2} \; ,
\end{eqnarray}
and
\begin{eqnarray}
\tilde{m}^{}_1 & = &  m^{}_2 |\cos z|^2 + m^{}_3 |\sin z|^2 \; .
\end{eqnarray}
These results indicate that $\tilde{m}^{}_1 \geq m^{}_2$ holds.
Furthermore,
\begin{eqnarray}
m^{}_3 {\rm Im} [ \sin^2 z] \leq m^{}_3 |\sin z|^2 =
\tilde{m}^{}_1 - m^{}_2 |\cos z|^2 \leq  \tilde{m}^{}_1 - m^{}_2
\; .
\end{eqnarray}
With the help of Eqs. (5.10), (5.13), (5.14) and (5.15), we arrive
at a new upper bound on $\varepsilon^{}_1$:\cite{ROSS}
\begin{eqnarray}
|\varepsilon^{}_1|  & \leq &  \frac {3 M^{}_1}{16\pi v^2}
\frac{\Delta m^2_{\rm atm}}{\sqrt{\Delta m^2_{\rm atm} + \Delta
m^2_{\rm sun}}} \left( 1 - \frac{\sqrt{\Delta m^2_{\rm
sun}}}{\tilde{m}^{}_1} \right ) \; ,
\end{eqnarray}
in which the effect of $\tilde{m}^{}_1$ has been taken into
account. For the $m^{}_3 =0$ case, one may similarly obtain
\begin{eqnarray}
|\varepsilon^{}_1|  & \leq &  \frac {3 M^{}_1}{16\pi v^2}
\frac{\Delta m^2_{\rm sun}}{\sqrt{\Delta m^2_{\rm atm}}} \left( 1
- \frac{\sqrt{\Delta m^2_{\rm atm} - \Delta m^2_{\rm
sun}}}{\tilde{m}^{}_1} \right ) \; ,
\end{eqnarray}
where $\tilde{m}^{}_1$ satisfies $\tilde{m}^{}_1 \geq m^{}_1$.

Using the maximal value of $\varepsilon^{}_1$ in Eq. (5.16) or
(5.17), together with the best-fit values of $\Delta m^2_{\rm
sun}$ and $\Delta m^2_{\rm atm}$, we carry out a numerical
analysis of $Y^{}_{\rm B}$ versus $\tilde{m}^{}_1$ and show the
result in Fig. 5.1, where the observationally-allowed range of
$Y^{}_{\rm B}$ is taken to be $8.5 \times 10^{-11} \leq Y^{}_{\rm
B} \leq 9.4 \times 10^{-11}$. We see that the successful
baryogenesis via leptogenesis requires $M^{}_1 \geq 5.9 \times
10^{10} ~{\rm GeV}$ in the $m^{}_1 =0$ case and $M^{}_1 \geq 1.3
\times 10^{13} ~{\rm GeV}$ in the $m^{}_3 =0$ case. Because
$\tilde{m}_1 \geq m_2$ holds for the normal neutrino mass
hierarchy, we have
\begin{eqnarray}
|a^{}_1|^2 + |a^{}_2|^2 + |a^{}_3|^2 \geq  M^{}_1 m^{}_2 \geq 0.53
~ {\rm GeV}^2 \; ,
\end{eqnarray}
where $a^{}_i$ (for $i=1, 2, 3$) are the matrix elements in the
first column of $M^{}_{\rm D}$. Thus the largest $|a^{}_i|$ should
not be smaller than $ 0.42 ~{\rm GeV}$. For the inverted neutrino
mass hierarchy, we can similarly find that the largest $|a^{}_i|$
should be above $ 14.6 ~{\rm GeV}$.
\begin{figure}[tbp]
\begin{center}
\includegraphics[width=7.3cm,height=7.3cm,angle=0]{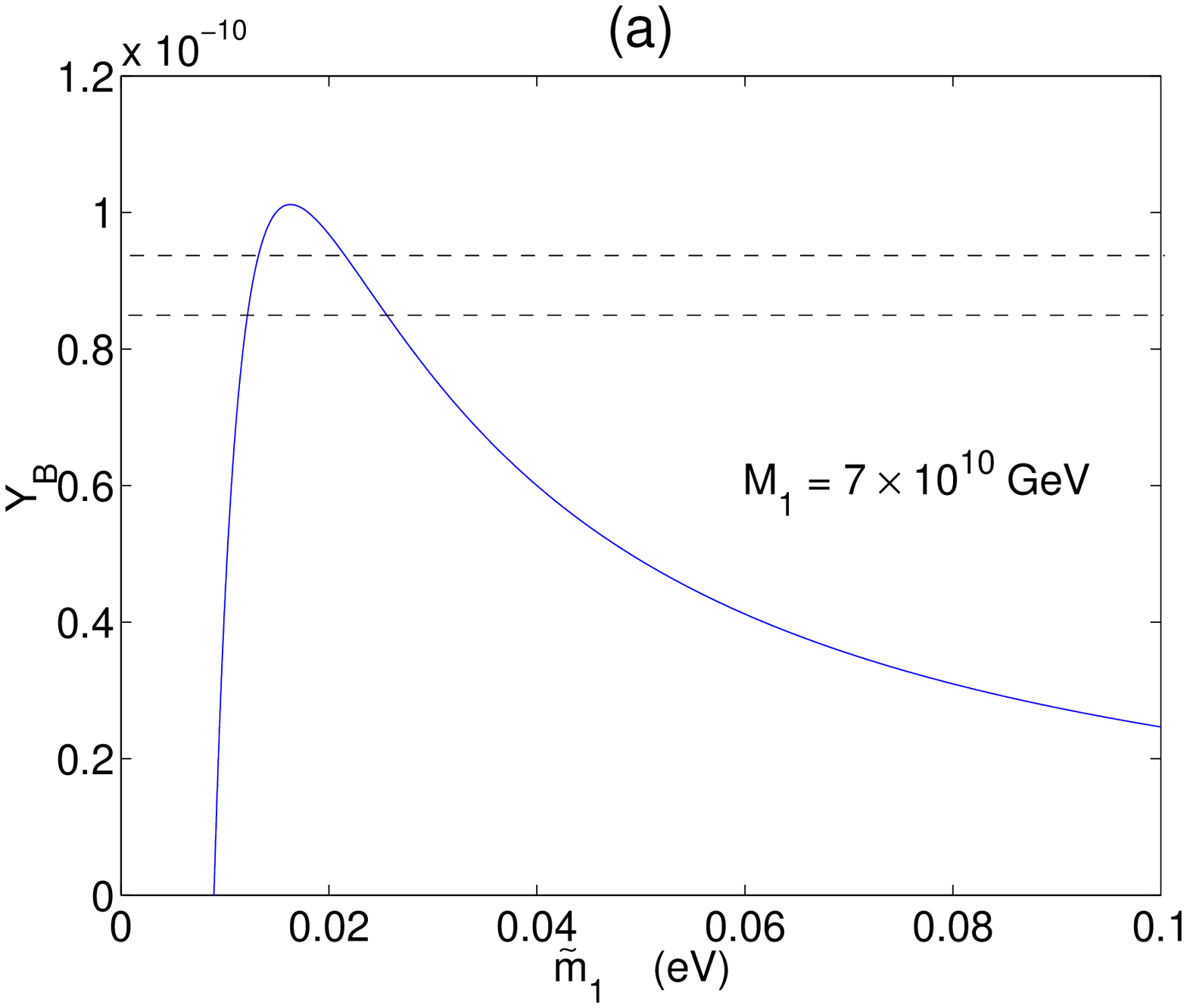}
\includegraphics[width=7.3cm,height=7.3cm,angle=0]{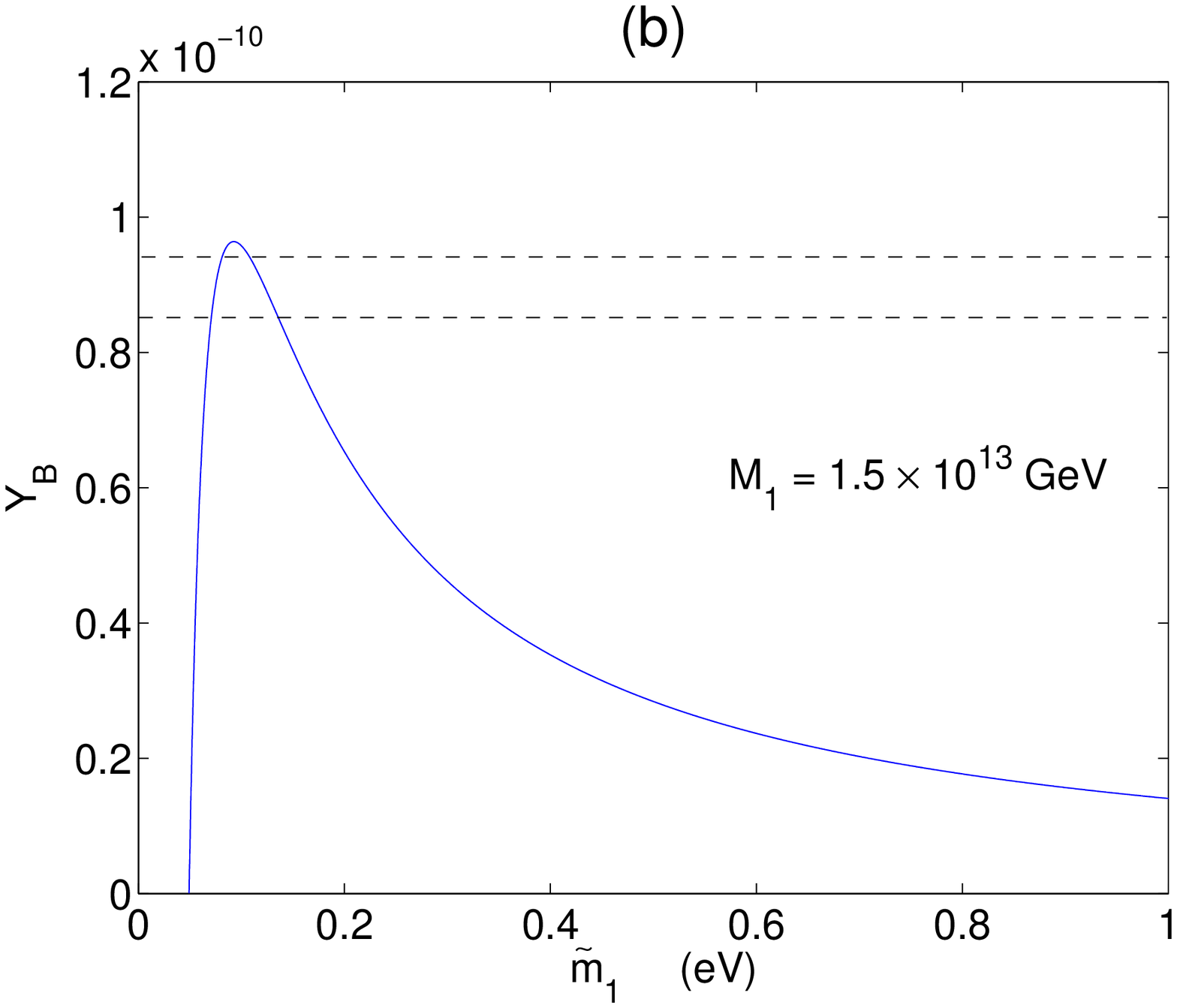}
\end{center}
\vspace{-0.3cm}
\caption{ Numerical illustration of the correlation between
$Y^{}_{\rm B}$ and $\tilde{m}^{}_1$ in the MSM: (a) the $m^{}_1
=0$ case; and (b) the $m^{}_3 =0$ case.}
\end{figure}

\subsection{Resonant Leptogenesis}

In the previous sections, we have discussed the simplest scenario
of thermal leptogenesis with two hierarchical right-handed
Majorana neutrinos. Another interesting scenario is the so-called
resonant leptogenesis.\cite{DEG} When the masses of two heavy
Majorana neutrinos are approximately degenerate (i.e., $M^{}_1
\approx M^{}_2$), the one-loop self-energy effect can be
resonantly enhanced and play the dominant role in
$\varepsilon^{}_1$ and $\varepsilon^{}_2$. It is then possible to
generate the observed baryon number asymmetry $Y^{}_{\rm B}$
through the out-of-equilibrium decays of relatively light and
approximately degenerate $N^{}_1$ and $N^{}_2$. Such a scenario
could allow us to relax the lower bound on the lighter
right-handed Majorana neutrino mass $M^{}_1$ in the MSM and to get
clear of the gravitino overproduction problem in the
supersymmetric version of the MSM.\cite{Gravitino}

When the mass splitting between two heavy Majorana neutrinos is
comparable to their decay widths, the CP-violating asymmetry
$\varepsilon^{}_i$ is dominated by the one-loop self-energy
contribution\cite{Pilaftsis}
\begin{eqnarray}
\varepsilon^{}_i  & = &  \frac{ {\rm Im} \left [ (M^\dagger_{\rm
D} M^{~}_{\rm D})^2_{ij} \right ]}{(M^\dagger_{\rm D} M^{~}_{\rm
D})^{}_{ii} (M^\dagger_{\rm D} M^{~}_{\rm D})^{}_{jj}} \;
\frac{(M^2_i - M^2_j )\; M^{}_i \; \Gamma^{}_j }{(M^2_i  - M^2_j
)^2 + M^2_i \; \Gamma^2_j} \; ,
\end{eqnarray}
where $\Gamma^{}_i \equiv  M^{}_i \left( Y^\dagger_\nu Y^{}_\nu
\right)^{}_{ii}/(8\pi)$ is the tree-level decay width of $N^{}_i$.
If $|M^{}_i - M^{}_j| \sim \Gamma^{}_i / 2$ holds, the factor
$(M^2_i - M^2_j ) M^{}_i \Gamma^{}_j /[(M^2_i  - M^2_j )^2 + M^2_i
\Gamma^2_j]$ may approach its maximal value $1/2$. If the masses
of two heavy Majorana neutrinos are exactly degenerate, however,
$\varepsilon^{}_i$ must vanish as one can see from Eq. (5.19).

A simple scenario of TeV-scale leptogenesis in the MSM has
recently been proposed.\cite{XingZhou} For simplicity, here let us
concentrate on the $m^{}_1 = 0$ case to introduce this
phenomenological scenario. By using the bi-unitary
parametrization, the $3\times 2$ Dirac neutrino mass matrix
$M^{}_{\rm D}$ can be expressed as
\begin{equation}
M^{}_{\rm D} \; = \; V^{}_0 \left(\begin{matrix} 0 & 0 \cr x & 0 \cr
0 & y \cr \end{matrix} \right) U \; ,
\end{equation}
where $V^{}_0$ and $U$ are $3 \times 3$ and $2 \times 2$ unitary
matrices, respectively. Then the seesaw relation $M^{}_\nu =
M^{}_{\rm D} M^{-1}_{\rm R} M^T_{\rm D}$ implies that the flavor
mixing of light neutrinos depends primarily on $V^{}_0$ and the
decays of heavy neutrinos rely mainly on $U$. This observation
motivates us to take $V^{}_0$ to be the tri-bimaximal mixing
pattern\cite{TB}
\begin{equation}
V^{}_0 \; = \; \left( \begin{matrix} 2/\sqrt{6} & 1/\sqrt{3} & 0 \cr
-1/\sqrt{6} & 1/\sqrt{3} & 1/\sqrt{2} \cr 1/\sqrt{6} & ~~
-1/\sqrt{3} ~~ & 1/\sqrt{2} \cr \end{matrix} \right) \; ,
\end{equation}
which is compatible very well with the best fit of current
experimental data on neutrino oscillations\cite{FIT}. On the other
hand, we assume $U$ to be the maximal mixing pattern with a single
CP-violating phase,
\begin{equation}
U = \left ( \begin{matrix} 1/\sqrt{2} & 1/\sqrt{2} \cr -1/\sqrt{2} &
1/\sqrt{2} \cr \end{matrix} \right ) \left (
\begin{matrix} e^{-i\alpha} & 0 \cr 0 & e^{+i\alpha} \cr \end{matrix}\right) \; .
\end{equation}
Since $\alpha$ is the only phase parameter in our model, it should
be responsible both for the CP violation in neutrino oscillations
and for the CP violation in $N^{}_i$ decays. In order to implement
the idea of resonant leptogenesis, we assume that $N^{}_1$ and
$N^{}_2$ are highly degenerate in mass; i.e., the magnitude of $r
\equiv (M^{}_2 - M^{}_1)/M^{}_2$ is strongly suppressed. Indeed
$|r| \sim {\cal O}(10^{-7})$ or smaller has typically been
anticipated in some seesaw models with three right-handed Majorana
neutrinos\cite{Pilaftsis} to gain the successful resonant
leptogenesis.

Given $|r| < {\cal O}(10^{-4})$, the explicit form of $M^{}_\nu$ can
reliably be formulated from the seesaw relation $M^{}_\nu =
M^{}_{\rm D} M^{-1}_{\rm R} M^T_{\rm D}$ by neglecting the tiny mass
splitting between $N^{}_1$ and $N^{}_2$. In such a good
approximation, we obtain
\begin{equation}
M^{}_\nu \; = \; \frac{y^2}{M^{}_2} \left [ V^{}_0 \left (
\begin{matrix} 0 & 0 & 0 \cr 0 & \omega^2 \cos 2\alpha & i \omega \sin
2\alpha \cr 0 & i \omega \sin 2\alpha & \cos 2\alpha \cr
\end{matrix} \right ) V^T_0 \right ] \; ,
\end{equation}
where $\omega = x/y$. The diagonalization $V^\dagger M^{}_\nu V^*
= {\rm Diag} \left \{ 0, m^{}_2, m^{}_3 \right \}$, where $V$ is
just the MNS matrix, yields
\begin{eqnarray}
m^{}_2 & = & \displaystyle \frac{y^2}{2M^{}_2} \left [ \sqrt{\left
(1 + \omega^2 \right )^2 \cos^2 2\alpha + 4 \omega^2 \sin^2
2\alpha} ~ - \left (1 - \omega^2 \right ) \left | \cos 2\alpha
\right | \right ] \; ,
\nonumber \\
m^{}_3 & = & \displaystyle \frac{y^2}{2M^{}_2} \left [ \sqrt{\left
(1 + \omega^2 \right )^2 \cos^2 2\alpha + 4 \omega^2 \sin^2
2\alpha} ~ + \left (1 - \omega^2 \right ) \left | \cos 2\alpha
\right | \right ] \; , ~~
\end{eqnarray}
where $0 < \omega < 1$. Taking account of $m^{}_2 = \sqrt{\Delta
m^2_{21}}$ and $m^{}_3 = \sqrt{\Delta m^2_{21} + |\Delta
m^2_{32}|}$, we obtain $m^{}_2 \approx 8.9 \times 10^{-3}$ eV and
$m^{}_3 \approx 5.1 \times 10^{-2}$ eV by using $\Delta m^2_{21}
\approx 8.0 \times 10^{-5} ~ {\rm eV}^2$ and $|\Delta m^2_{32}|
\approx 2.5 \times 10^{-3} ~ {\rm eV}^2$as the typical
inputs.\cite{FIT} Furthermore,
\begin{equation}
~~~V = \left ( \begin{matrix} 2/\sqrt{6} & \cos\theta/\sqrt{3} & i
\sin\theta/\sqrt{3} \cr -1/\sqrt{6} & \cos\theta/\sqrt{3} + i
\sin\theta/\sqrt{2} & \cos\theta/\sqrt{2} + i \sin\theta/\sqrt{3}
\cr 1/\sqrt{6} & -\cos\theta/\sqrt{3} + i \sin\theta/\sqrt{2} &
\cos\theta/\sqrt{2} - i \sin\theta/\sqrt{3} \cr \end{matrix} \right
) \; , \;\;\;
\end{equation}
where $\theta$ is given by $\tan 2\theta = 2 \omega \tan 2\alpha
/(1+\omega^2)$. Comparing this result with the parameterization of
$V$ in Eq. (2.7), we immediately arrive at
\begin{eqnarray}
\sin^2 \theta^{}_x & = & \frac{1 - \sin^2 \theta}{3 - \sin^2
\theta} \; ,
\nonumber \\
\sin^2 \theta^{}_z & = & \frac{\sin^2 \theta}{3} \; ,
\end{eqnarray}
$\theta^{}_y = \pi/4$, $\delta = - \pi/2$ and vanishing Majorana
phases of CP violation. Eq. (5.26) implies an interesting
correlation between $\theta^{}_x$ and $\theta^{}_z$: $\sin^2
\theta^{}_x = (1 - 2 \tan^2 \theta^{}_z)/3$. When $\theta^{}_z
\rightarrow 10^\circ$, we get $\theta^{}_x \rightarrow 34^\circ$,
which is very close to the present best-fit value of the solar
neutrino mixing angle.\cite{FIT} Note that the smallness of
$\theta^{}_z$ requires the smallness of $\theta$ or equivalently
the smallness of $\alpha$. Eqs. (5.24) and (5.26), together with
$\theta^{}_z < 10^\circ$ and the values of $m^{}_2$ and $m^{}_3$
obtained above, yield $0.39 \lesssim \omega \lesssim 0.42$,
$0^\circ \lesssim \alpha \lesssim 23^\circ$ and $0^\circ \lesssim
\theta \lesssim 18^\circ$. We observe that Eq. (5.24) can reliably
approximate to $m^{}_2 \approx x^2/M^{}_2$ and $m^{}_3 \approx
y^2/M^{}_2$ for $\alpha \lesssim 10^\circ$. The Jarlskog parameter
$J^{}_{\rm CP}$,\cite{J} which determines the strength of CP
violation in neutrino oscillations, is found to be $|J^{}_{\rm
CP}| = \sin 2\theta/(6\sqrt{6}) \lesssim 0.04$ in this scenario.
It is possible to measure $|J^{}_{\rm CP}| \sim {\cal O}(10^{-2})$
in the future long-baseline neutrino oscillation experiments.

We proceed to discuss the baryon number asymmetry via resonant
leptogenesis. Given Eq. (5.20), $M^\dagger_{\rm D} M^{}_{\rm D}$
takes the following form for the $m^{}_1 =0$ case:
\begin{equation}
M^\dagger_{\rm D} M^{}_{\rm D} \; = \; U^\dagger \left (
\begin{matrix} x^2 & 0 \cr 0 & y^2 \cr \end{matrix} \right ) U \; .
\end{equation}
Combining Eqs. (5.27) and (5.19), we obtain the explicit
expression of $\varepsilon^{}_i$:
\begin{equation}
\varepsilon^{}_i = \frac{ - 32 \pi v^2 y^2 \left (1 - \omega^2
\right )^2}{\left (1 + \omega^2 \right ) \left [ 1024 \pi^2 v^4 r^2
+ y^4 \left (1 + \omega^2 \right )^2 \right ]} ~ r \sin 4\alpha \; ,
\end{equation}
in which $r \equiv (M^{}_2 - M^{}_1)/M^{}_2$ has been defined to
describe the mass splitting between $N^{}_1$ and $N^{}_2$. Since
$r$ is extremely tiny, we have some excellent approximations:
$\Gamma^{}_1 = \Gamma^{}_2$, $\varepsilon^{}_1 = \varepsilon^{}_2$
and $\tilde{m}^{}_1 = \tilde{m}^{}_2$, where the effective
neutrino masses are defined as $\tilde{m}^{}_i \equiv
(M^\dagger_{\rm D} M^{}_{\rm D})^{}_{ii}/M^{}_i$. To estimate
$\varepsilon^{}_i$ at the TeV scale, we restrict ourselves to the
interesting $\alpha \lesssim 10^\circ$ region and make use of the
approximate result $y^2 \approx m^{}_3 M^{}_2$ obtained above. We
get $y^2 \approx 5.1 \times 10^{-8} ~ {\rm GeV}^2$ from $m^{}_3
\approx 5.1 \times 10^{-2}$ eV and $M^{}_2 \approx 1$ TeV. In
addition, $\omega \approx 0.42$. Then Eq. (5.28) is approximately
simplified to
\begin{equation}
\varepsilon^{}_i \; \approx \; \left \{ \begin{array}{lll} -9.7
\times 10^{-15} ~ r^{-1} \sin 4\alpha \; , & ~~~~ & {\rm for} ~ r
\gg 2.0 \times 10^{-14} \; , \\
-2.5 \times 10^{13} ~ r \sin 4\alpha \; , & & {\rm for} ~ r \ll 2.0
\times 10^{-14} \; , \end{array} \right . ~
\end{equation}
together with $\varepsilon^{}_i \sim -0.25 \times \sin 4\alpha$
for $r \sim 2.0 \times 10^{-14}$. Note that $|\varepsilon^{}_i|
\sim {\cal O}(10^{-5})$ is in general expected to achieve the
successful leptogenesis. Hence the third possibility $r \sim {\cal
O}(10^{-14})$ requires $\alpha \sim {\cal O}(10^{-4})$, implying
very tiny (unobservable) CP violation in neutrino oscillations. If
$\alpha \sim 5^\circ \cdots 10^\circ$, one may take either $r\sim
10^{-10}$ or $r \sim 10^{-18}$ to obtain $|\varepsilon^{}_i| \sim
{\cal O}(10^{-5})$.

The generated lepton number asymmetry can be partially converted
into the baryon number asymmetry via the $(B-L)$-conserving
sphaleron process\cite{Buch}
\begin{equation}
\eta^{}_{\rm B} \approx - 0.96 \times 10^{-2} \sum^2_{i=1} \left
(\kappa^{}_i \varepsilon^{}_i \right ) \approx -1.92 \times 10^{-2}
~ \kappa^{}_i \varepsilon^{}_i \; .
\end{equation}
To evaluate the efficiency factors $\kappa^{}_i$, we define the
decay parameters $K^{}_i \equiv \tilde{m}^{}_i/m^*$, where $m^*
\simeq 1.08\times 10^{-3}~ {\rm eV}$ is the equilibrium neutrino
mass. When the parameters $\tilde{m}^{}_i$ or $K^{}_i$ lie in the
strong washout region (i.e., $\tilde{m}^{}_i \gg m^{}_*$ or
$K^{}_i \gg 1$), $\kappa^{}_i$ can be estimated by using the
approximate formula\cite{DiBari}
\begin{eqnarray}
\frac{1}{\kappa^{}_i} \; \approx \; \left (2 + 4.38 K^{0.13}_i
e^{-1.25/K^{}_i} \right ) K^{}_i  \; , ~~~~~~
\end{eqnarray}
which is valid when the masses of two heavy right-handed Majorana
neutrinos are nearly degenerate. Given $\alpha \lesssim 10^\circ$,
\begin{equation}
\tilde{m}^{}_i \approx \frac{1}{2} \left (m^{}_2 + m^{}_3 \right )
\approx 2.9 \times 10^{-2} ~ {\rm eV} \;
\end{equation}
holds. Thus we get $K^{}_i \approx 27$ and $\kappa^{}_i \approx
4.4 \times 10^{-3}$. The baryon number asymmetry turns out to be
\begin{equation}
\eta^{}_{\rm B} \; \approx \; \left \{ \begin{array}{lll} 8.2 \times
10^{-19} ~ r^{-1} \sin 4\alpha \; , & ~~~~ & {\rm for} ~ r
\gg 2.0 \times 10^{-14} \;  \\
2.1 \times 10^{9} ~ r \sin 4\alpha \; , & & {\rm for} ~ r \ll 2.0
\times 10^{-14} \;  \end{array} \right .
\end{equation}
Note that these results are obtained by taking $M^{}_2 \approx 1$
TeV. Other results can similarly be achieved by starting from Eq.
(5.28) and allowing $M^{}_2$ to vary, for instance, from 1 TeV to
10 TeV. To illustrate, Fig. 5.2 shows the simple correlation
between $r$ and $\alpha$ to get $\eta^{}_B = 6.1 \times 10^{-10}$,
where $M^{}_2 =1$ TeV, 2 TeV, 3 TeV, 4 TeV and 5 TeV have
typically been input. We see the distinct behaviors of $r$
changing with $\alpha$ in two different regions: $r \gg {\cal
O}(10^{-13})$ and $r \ll {\cal O}(10^{-13})$, in which
$\eta^{}_{\rm B} \propto \varepsilon^{}_i \propto y^2 r^{-1} \sin
4\alpha \propto M^{}_2 r^{-1} \sin 4\alpha$ and $\eta^{}_{\rm B}
\propto \varepsilon^{}_i \propto y^{-2} r \sin 4\alpha \propto
M^{-1}_2 r \sin 4\alpha$ hold respectively as the leading-order
approximations. Thus we have $r \propto \sin 4\alpha$ in the first
region and $r \propto 1/\sin 4\alpha$ in the second region for
given values of $\eta^{}_{\rm B}$ and $M^{}_2$. The leptogenesis
in the $m^{}_3 =0$ case can be discussed in a similar
way.\cite{XingZhou}
\begin{figure}
\vspace{-0.5cm}
\psfig{file=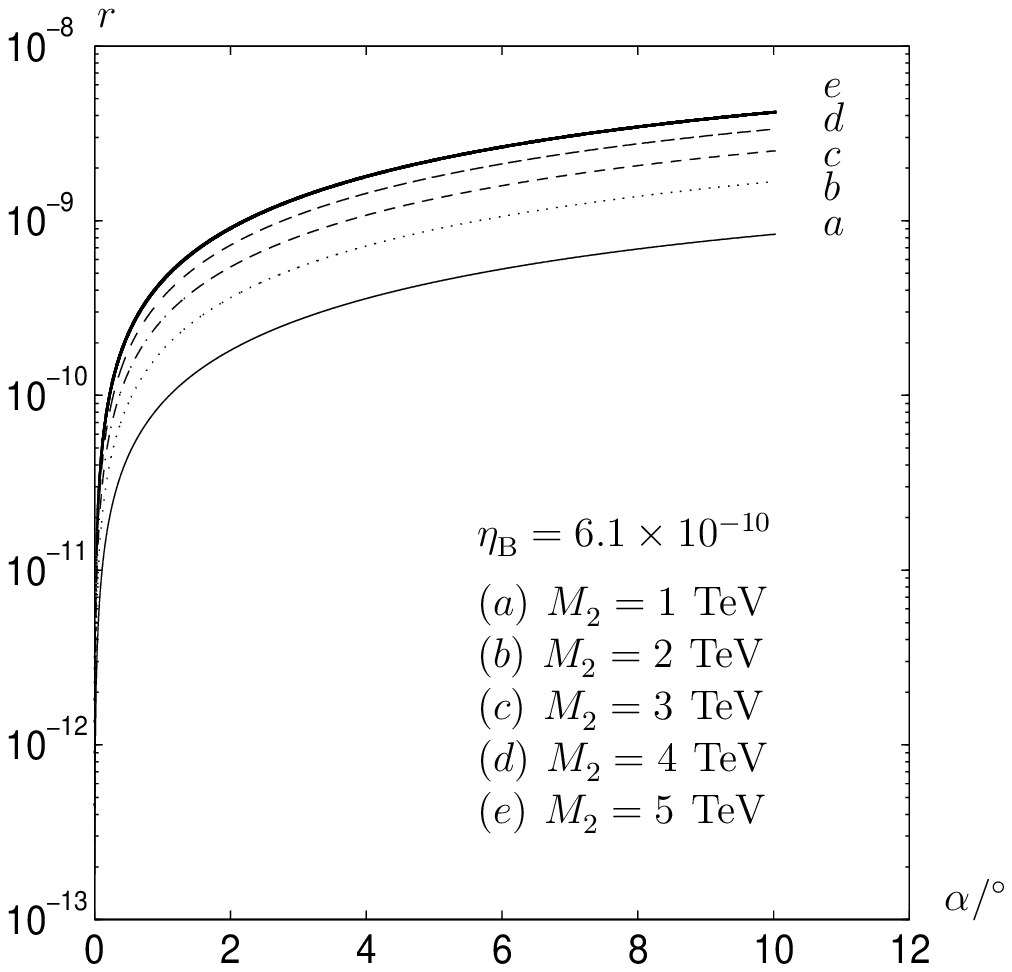, bbllx=2.5cm, bblly=8.8cm, bburx=12.5cm, bbury=19.8cm,%
width=7.5cm, height=7.5cm, angle=0, clip=0}
\psfig{file=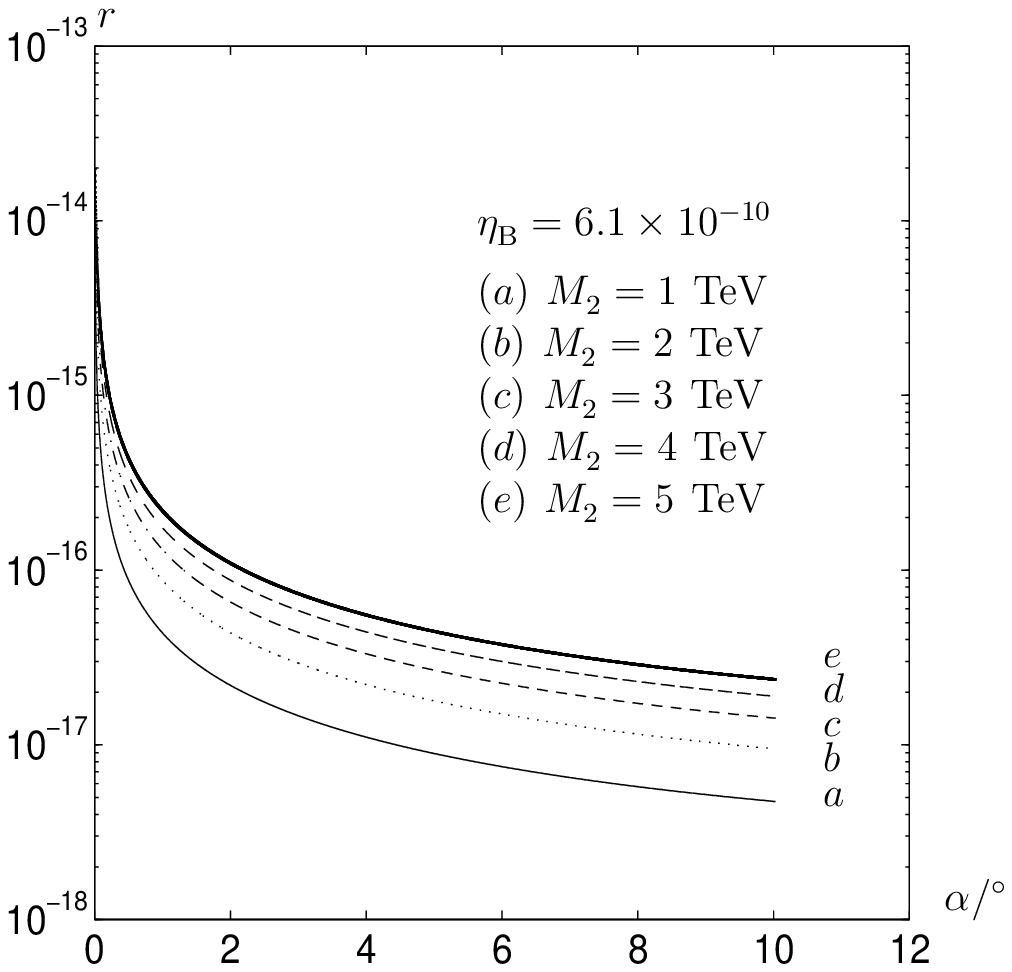, bbllx=2.5cm, bblly=8.5cm, bburx=12.5cm, bbury=19.5cm,%
width=7.5cm, height=7.5cm, angle=0, clip=0}
\vspace{-0.5cm}\caption{ Numerical illustration of the correlation
between the mass splitting parameter $r$ and the CP-violating
phase $\alpha$ in the $m^{}_1 =0$ case to achieve the successful
baryogenesis via leptogenesis. }
\end{figure}

The tiny splitting between $M^{}_1$ and $M^{}_2$ is characterized
by $r$. A natural idea is that $r$ may be zero at a superhigh
energy scale $M^{}_{\rm X}$ and it becomes non-vanishing when the
heavy Majorana neutrino masses run from $M^{}_{\rm X}$ down to the
seesaw scale $M^{}_{\rm S}$.\cite{RL1,Turzynski} Using the
one-loop RGEs,\cite{RGE} one may approximately obtain
\begin{eqnarray}
r \; \approx \; \pm \frac{m^{}_3 M^{}_{\rm S} (1 - r^{}_{23})}{8
\pi^2 v^2} \; {\rm ln}\left( \frac{M^{}_{\rm X}}{M^{}_{\rm S}}
\right )
\end{eqnarray}
in the $m^{}_1 =0$ case, where $r^{}_{23} \equiv m^{}_2 / m^{}_3$
has been defined before. Hence $r$ can be extremely small. When
$M^{}_{\rm X}$ is just the scale of grand unified theories
($M^{}_{\rm X} \sim \Lambda^{}_{\rm GUT} = 10^{16}~{\rm GeV}$),
for example, we have $r \approx 1.6 \times 10^{-12}$ for
$M^{}_{\rm S} = 1~{\rm TeV}$. The possibility that $r$ is
radiatively generated has been studied in detail in the
supersymmetric version of the MSM for both normal and inverted
neutrino mass hierarchies.\cite{Turzynski} In the absence of
supersymmetry and in the presence of one texture zero in
$M^{}_{\rm D}$, however, it is impossible to achieve successful
resonant leptogenesis from the radiative generation of
$r$.\cite{RL1}

Flavor effects in the mechanism of thermal leptogenesis have
recently attracted a lot of attention.\cite{Flavor} Because all
the Yukawa interactions of charged leptons are in thermal
equilibrium at the TeV scale, the flavor issue should be taken
into account in our model. After calculating the CP-violating
asymmetry $\varepsilon^{}_{i\alpha}$ and the corresponding washout
effect for each lepton flavor $\alpha$ ($=e$, $\mu$ or $\tau$) in
the final states of $N^{}_i$ decays, we find that the prediction
for the total baryon number asymmetry $\eta^{}_{\rm B}$ is
enhanced by a factor $\sim 4$ in both $m^{}_1 = 0$ and $m^{}_3 =
0$ cases.\cite{XingZhou} However, such flavor effects may be
negligible when the masses of heavy right-handed Majorana
neutrinos are all of or above ${\cal O}( 10^{12})~{\rm
GeV}$.\cite{Flavor}

\subsection{Leptogenesis in Two-Zero Textures}

We have calculated the CP-violating phases and their RGE running
effects for a typical two-zero texture of $M^{}_{\rm D}$ (i.e.,
the FGY ansatz) in Sec. 4.2. For completeness, here we discuss the
mechanism of thermal leptogenesis in this interesting ansatz by
assuming $M^{}_1 \ll M^{}_2$. Note that we have taken $a^{}_1$,
$b^{}_2$ and $b^{}_3$ of $M^{}_{\rm D}$ to be real and positive,
while $a^{}_2$ is complex and its phase is denoted as $\phi$.
Given $a^{}_3 = b^{}_1 = 0$, the seesaw relation allows us to
get\cite{GUO1}
\begin{eqnarray}
a^2_1 & = & M^{}_1 |(M^{}_\nu)^{}_{11}| \; , ~~~~~~~ |a^{}_2|^2 =
M^{}_1 \frac{|(M^{}_\nu)^{}_{12}|^2}{|(M^{}_\nu)^{}_{11}|} \; ,
\nonumber \\
b^2_3  & = &  M^{}_2 |(M^{}_\nu)^{}_{33}| \; , ~~~~~~~~ b^2_2 ~ ~
= M^{}_2 \frac{|(M^{}_\nu)^{}_{23}|^2}{|(M^{}_\nu)^{}_{33}|} \; .
\end{eqnarray}
With the help of Eqs. (4.5) and (5.2), we obtain
\begin{eqnarray}
\varepsilon^{}_1 & \approx &  \frac{3}{16\pi v^2} \frac{ M^{}_1
|(M^{}_\nu)^{}_{12}|^2 |(M^{}_\nu)^{}_{23}|^2 \sin 2\phi} {\left
\{|(M^{}_\nu)^{}_{11}|^2 + |(M^{}_\nu)^{}_{12}|^2 \right\}
|(M^{}_\nu)^{}_{33}|} \; . ~
\end{eqnarray}
It is clear that $\varepsilon^{}_1$ and $Y^{}_{\rm B}$ only
involve two free parameters: $M^{}_1$ and $\phi$. Because $\phi$
is closely related to the mixing angle $\theta^{}_z$, one may
analyze the dependence of $Y^{}_{\rm B}$ on $\theta^{}_z$ for
given values of $M^{}_1$. For the $m^{}_1=0$ and $m^{}_3=0$ cases,
we plot the numerical results of $Y^{}_{\rm B}$ in Fig. 5.3 (a)
and (b), respectively. Two comments are in order:
\begin{figure}[tbp]
\begin{center}
\includegraphics[width=7.3cm,height=7.3cm,angle=0]{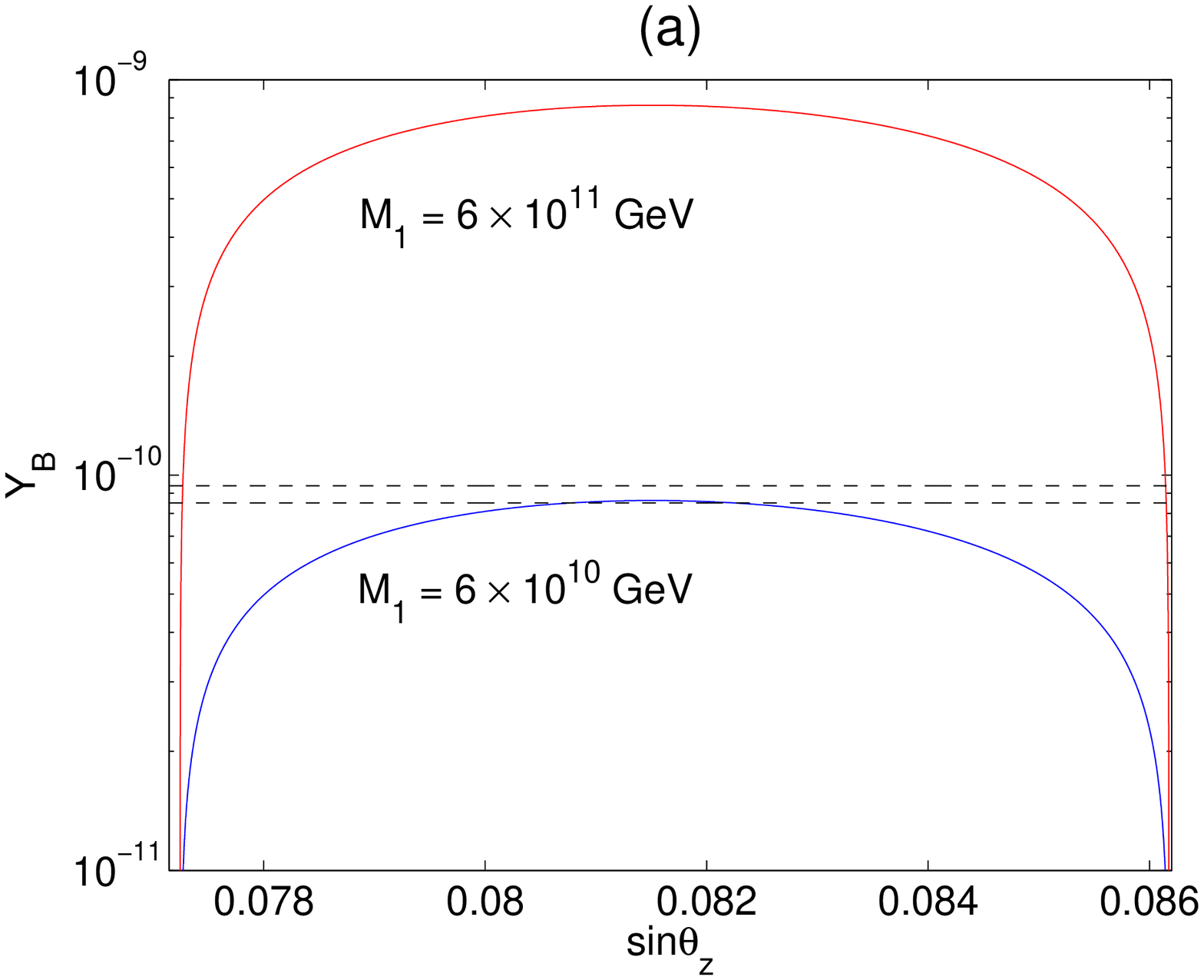}
\includegraphics[width=7.3cm,height=7.3cm,angle=0]{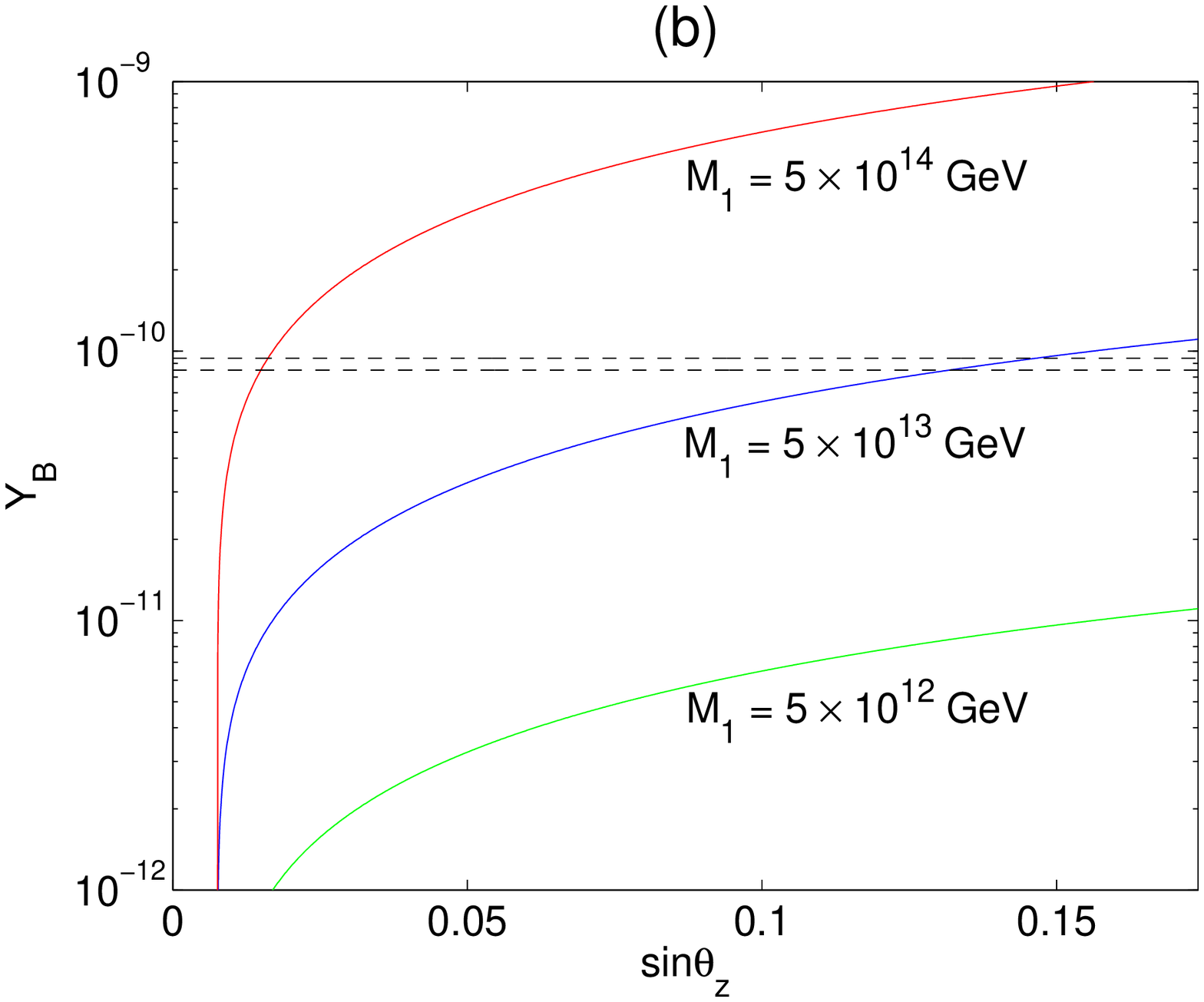}
\end{center} \vspace{-0.3cm}
\caption{Numerical illustration of $Y^{}_{\rm B}$ changing with
$M^{}_1$ and $\sin \theta^{}_z$: (a)  for the $m^{}_1 = 0$ case; (b)
for the $m^{}_3 = 0$ case. The region between two dashed lines in
(a) or (b) corresponds to the range of $Y^{}_{\rm B}$ allowed by
current observational data.}
\end{figure}
\begin{enumerate}
\item In the $m^{}_1=0$ case, current data of $Y^{}_{\rm B}$
require $M^{}_1 \geq 5.9 \times 10^{10} ~ {\rm GeV}$ for the
allowed ranges of $s^{}_z$. Once $s^{}_z$ is precisely measured,
it is possible to fix the value of $M^{}_1$ and then exclude some
possibilities (e.g., the one with $M^{}_1 = 10^{11}~{\rm GeV}$
will be ruled out, if $s_z \approx 0.082$ holds).

\item In the $m^{}_3 =0$ case, the condition $M^{}_1 \geq 3.8
\times 10^{13}~{\rm GeV}$ is imposed by current data of $Y^{}_{\rm
B}$. Although $\theta^{}_z$ is less restricted in this scenario,
it remains possible to pin down the value of $M^{}_1$ once
$\theta^{}_z$ is determined (e.g., $M^{}_1 \approx 5 \times
10^{13}~{\rm GeV}$ is expected, if $s_z \approx 0.14$ holds).
\end{enumerate}
A similar analysis of $Y^{}_{\rm B}$ can be done in the
supersymmetric version of the MSM.\cite{GUO1}

Note that we have simply used the low-energy values of light
neutrino masses and flavor mixing angles in the above calculation
of $\varepsilon^{}_1$ and $Y^{}_{\rm B}$. Now let us take into
account the RGE running effects on these parameters from $\mu =
M^{}_Z$ up to $\mu = M^{}_1$. With the help of Eq. (4.24), we can
obtain an approximate relationship between $\varepsilon^{}_1
(M^{}_1)$ and $\varepsilon^{}_1 (M^{}_Z)$:\cite{MEI}
\begin{eqnarray}
\varepsilon^{}_1 (M^{}_1) \approx I^{-1}_\alpha \varepsilon^{}_1
(M^{}_Z) \; .
\end{eqnarray}
Looking at the running behavior of $I^{}_\alpha$ shown in Fig.
4.2, we conclude that $\varepsilon^{}_1$ is radiatively corrected
by a factor smaller than two. Therefore, $\varepsilon^{}_1
(M^{}_1) \approx \varepsilon^{}_1 (M^{}_Z)$ is actually an
acceptable approximation in the MSM.

In the flavor basis where both $M^{}_l$ and $M^{}_{\rm R}$ are
diagonal, Eq. (5.2) shows that $\varepsilon^{}_1$ only depends on
the nontrivial phases of $M^{}_{\rm D}$. This observation implies
that there might not exist a direct connection between CP
violation in heavy Majorana neutrino decays and that in light
Majorana neutrino oscillations. The former is characterized by
$\varepsilon^{}_i$, while the latter is measured by the phase
parameter $\delta$ of $V$ or more exactly by the Jarlskog
invariant $J^{}_{\rm CP}$. That is to say, $\varepsilon^{}_i$ and
$J^{}_{\rm CP}$ (or $\delta$) seem not to be necessarily
correlated with each other.\cite{Buch2} But their correlation is
certainly possible in the MSM under discussion, in which
$M^{}_\nu$ is linked to $M^{}_{\rm D}$ and $M^{}_{\rm R}$. Taking
account of the flavor effects in leptogenesis,\cite{Flavor}
however, several authors have pointed out that CP violation at low
energies is necessarily related to that at high energies in the
canonical seesaw models.\cite{0609petcov}

The correlation between leptogenesis and CP violation in neutrino
oscillations has been discussed in the
MSM.\cite{DiBari,T1,GUO1,MEI} Here we illustrate how the
cosmological baryon number asymmetry is correlated with the
Jarlskog invariant of CP violation in the FGY ansatz. We plot the
numerical result of $Y^{}_{\rm B}$ versus $J^{}_{\rm CP}$ in Fig.
5.4,\cite{GUO1} where $M^{}_1 = 8 \times 10^{10}~{\rm GeV}$ for
the $m^{}_1 =0$ case and $M^{}_1 = 6 \times 10^{13}~{\rm GeV}$ for
the $m^{}_3 =0$ case have typically been taken. One can see that
the observationally-allowed range of $Y^{}_{\rm B}$ corresponds to
$J^{}_{\rm CP} \sim 1\%$ in the $m^{}_1 =0$ case and $J^{}_{\rm
CP} \sim 2\%$ in the $m^{}_3 =0$ case. The correlation between
$Y^{}_{\rm B}$ and $J^{}_{\rm CP}$ is so strong that it might be
used to test the FGY ansatz after $J^{}_{\rm CP}$ is measured in
the future long-baseline neutrino oscillation experiments.
\begin{figure}[tbp]
\begin{center}
\includegraphics[width=7.3cm,height=7.3cm,angle=0]{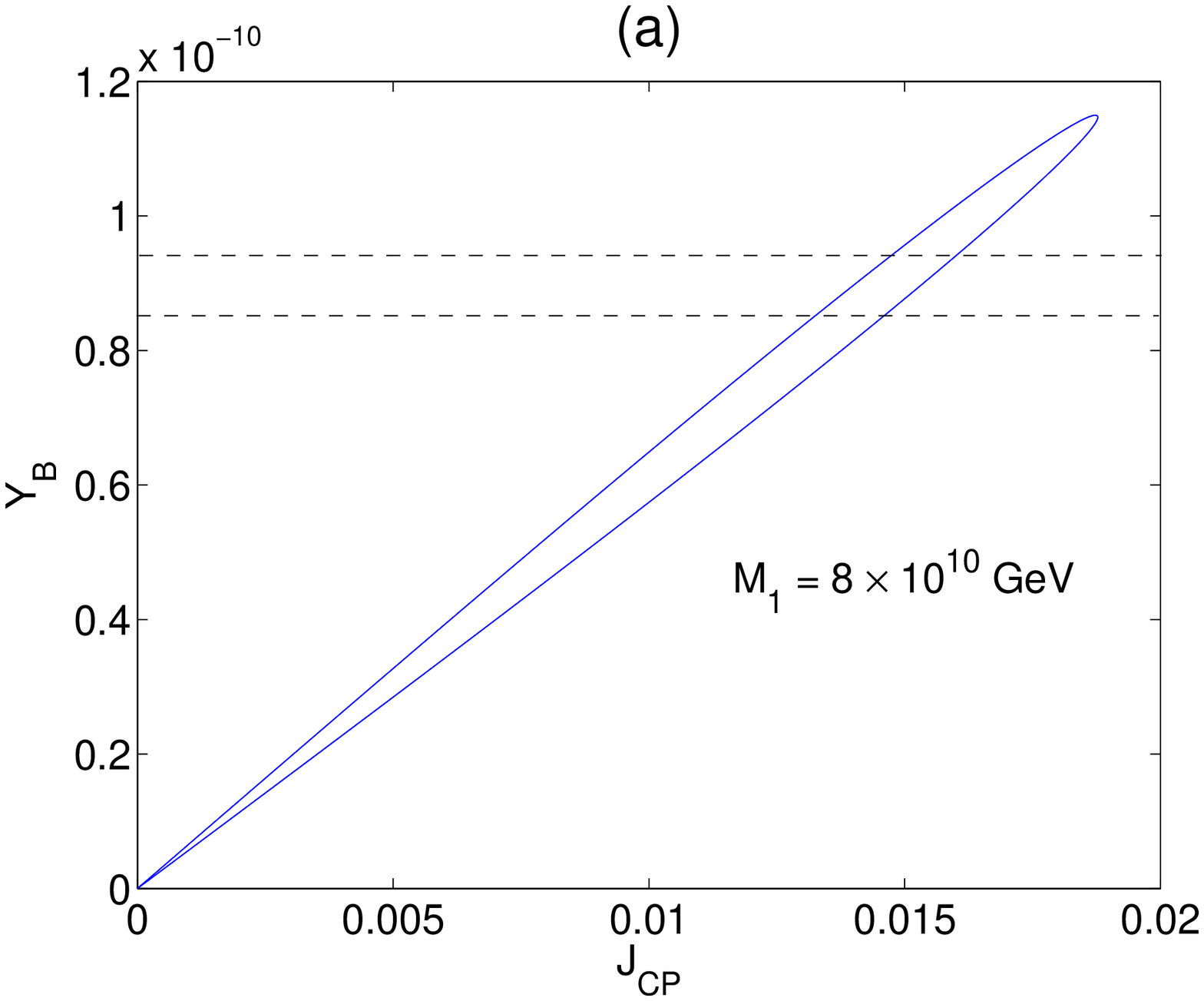}
\includegraphics[width=7.3cm,height=7.3cm,angle=0]{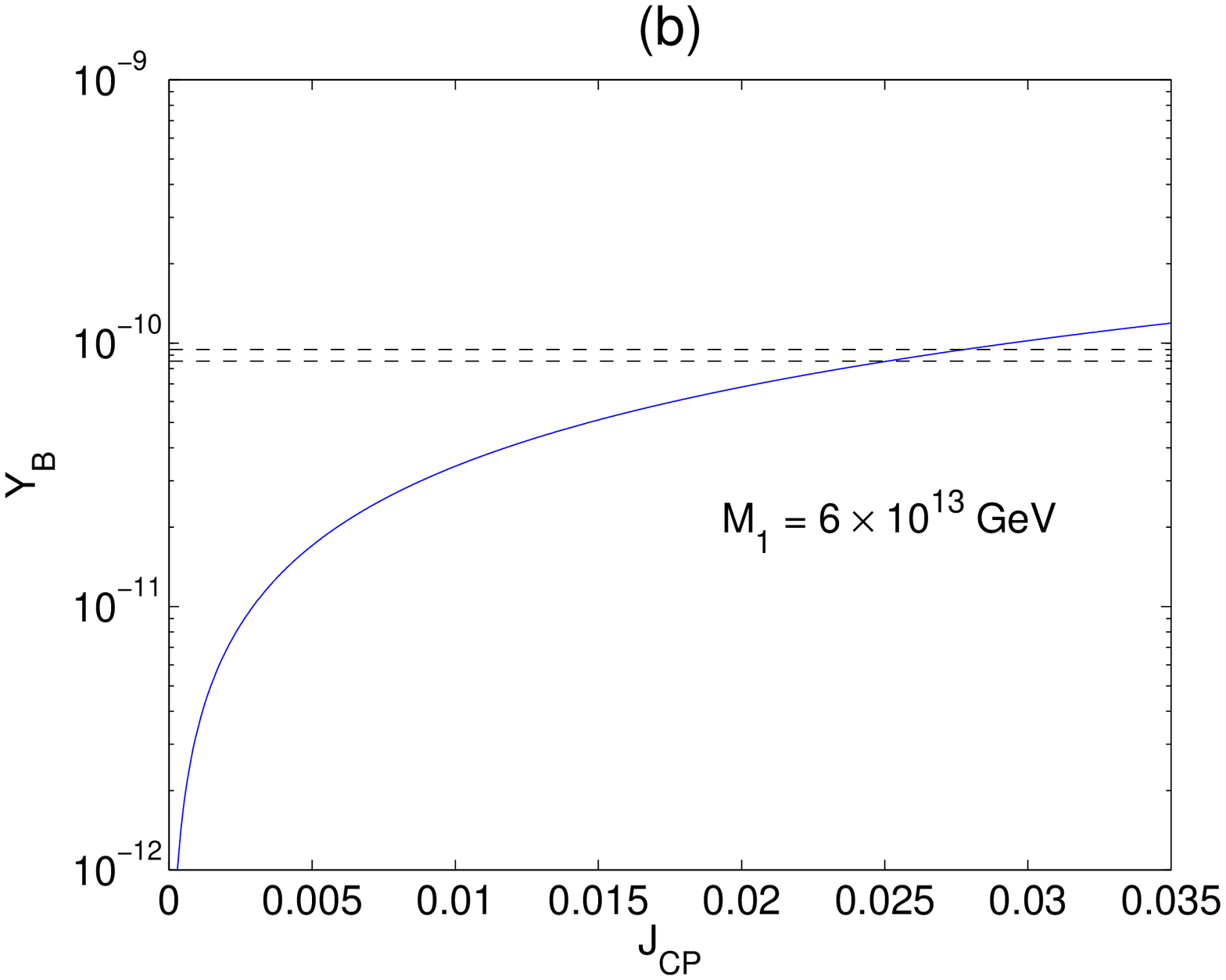}
\end{center} \vspace{-0.3cm}
\caption{ Numerical illustration of the correlation between
$Y^{}_{\rm B}$ and $J^{}_{\rm CP}$: (a) in the $m^{}_1 =0$ case
with $M^{}_1 = 8 \times 10^{10}~{\rm GeV}$; and (b) in the $m^{}_3
=0$ case with $M^{}_1 = 6 \times 10^{13}~{\rm GeV}$.}
\end{figure}

\subsection{Lepton-Flavor-Violating Decays}

The existence of neutrino oscillations implies the violation of
lepton flavors. Hence the lepton-flavor-violating (LFV) decays in
the charged-lepton sector, such as $\mu \to e + \gamma$, should
also take place. They are unobservable in the SM, because their
decay amplitudes are expected to be highly suppressed by the
ratios of neutrino masses ($m^{}_i \lesssim 1$ eV) to the
$W$-boson mass ($M^{}_W \approx 80$ GeV). In the supersymmetric
extension of the SM, however, the branching ratios of such rare
processes can be enormously enlarged. Current experimental bounds
on the LFV decays $\mu \to e + \gamma$, $\tau \to e + \gamma$ and
$\tau \to \mu + \gamma$ are\cite{limit1}
\begin{eqnarray}
{\rm BR}(\mu \rightarrow e \gamma) \; & < & \; 1.2 \times 10^{-11}
\; ,
\nonumber \\
{\rm BR}(\tau \rightarrow e \gamma) \; & < & \;  1.1 \times
10^{-7} \; ,
\nonumber \\
{\rm BR}(\tau \rightarrow \mu \gamma) \; & < & \;  6.8 \times
10^{-8} \; .
\end{eqnarray}
The sensitivities of a few planned experiments\cite{limit2} may
reach ${\rm BR}(\mu \rightarrow e \gamma) \sim 1.3 \times
10^{-13}$, ${\rm BR}(\tau \rightarrow e \gamma) \sim {\cal
O}(10^{-8})$ and ${\rm BR}(\tau \rightarrow \mu \gamma) \sim {\cal
O}(10^{-8})$.

For simplicity, here we restrict ourselves to a very conservative
case in which supersymmetry is broken in a hidden sector and the
breaking is transmitted to the observable sector by a flavor blind
mechanism, such as gravity.\cite{ROSS} Then all the soft breaking
terms are diagonal at high energy scales, and the only source of
lepton flavor violation in the charged-lepton sector is the
radiative correction to the soft terms through the neutrino Yukawa
couplings. In other words, the low-energy LFV processes $l^{}_j
\to l^{}_i + \gamma$ are induced by the RGE effects of the slepton
mixing. The branching ratios of $l^{}_j \to l^{}_i + \gamma$ are
given by \cite{Casas,LFV}
\begin{equation}
{\rm BR}(l^{}_j \rightarrow l^{}_i \gamma) \approx
\frac{\alpha^3}{G^2_{\rm F} m^8_{\rm S}}  \left[ \frac{3 m^2_0 +
A^2_0}{8 \pi^2 v^2 \sin^2\beta} \right]^2 |C^{}_{ij}|^2 \tan^2
\beta \; ,
\end{equation}
where $m^{}_0$ and $A^{}_0$ denote the universal scalar soft mass
and the trilinear term at $\Lambda^{}_{\rm GUT}$, respectively. In
addition,\cite{Petcov}
\begin{eqnarray}
m^8_{\rm S} \; \approx \; 0.5m^2_0 M^2_{1/2} (m^2_0 +
0.6M^2_{1/2})^2 \;
\end{eqnarray}
with $M^{}_{1/2}$ being the gaugino mass; and
\begin{eqnarray}
C^{}_{ij} \; = \; \sum_k (M^{}_{\rm D})^{}_{ik} (M^*_{\rm
D})^{}_{jk} {\rm ln} \frac{\Lambda^{}_{\rm GUT}}{M^{}_k} \;
\end{eqnarray}
with $\Lambda_{\rm GUT} = 2.0 \times 10^{16}$ GeV to be fixed in
our calculations. The LFV decays have been discussed in the
supersymmetric version of the MSM.\cite{ROSS,RAIDAL,MOHAPATRA,Cao}
To illustrate, we are going to compute the LFV processes by taking
account of the FGY ansatz, which only has three unknown parameters
$\theta^{}_z$, $M^{}_1$ and $M^{}_2$.

To calculate the branching ratio of $\mu \rightarrow e + \gamma$,
we need to know the following parameters in the framework of the
minimal supergravity (mSUGRA) model: $M^{}_{1/2}$, $m^{}_0$,
$A^{}_0$, $\tan \beta$ and ${\rm sign } (\mu)$. These parameters
can be constrained from cosmology (by demanding that the proper
supersymmetric particles should give rise to an acceptable dark
matter density) and low-energy measurements (such as the process
$b \rightarrow s + \gamma$ and the anomalous magnetic moment of
muon $g_\mu -2$). Here we adopt the Snowmass Points and Slopes
\cite{SPS} (SPS) listed in Table. 5.1. These points and slopes are
a set of benchmark points and parameter lines in the mSUGRA
parameter space corresponding to different scenarios in the search
for supersymmetry at present and future experiments. Points 1a and
1b are ``typical'' mSUGRA points (with intermediate and large
$\tan \beta$, respectively), and they lie in the ``bulk'' of the
cosmological region where the neutralino is sufficiently light and
no specific suppression mechanism is needed. Point 2 lies in the
``focus point'' region, where a too large relic abundance is
avoided by an enhanced annihilation cross section of the lightest
supersymmetric particle (LSP) due to a sizable higgsino component.
Point 3 is directed towards the co-annihilation region where the
LSP is quasi-degenerate with the next-to-LSP (NLSP). A rapid
co-annihilation between the LSP and the NLSP can give a
sufficiently low relic abundance. Points 4 and 5 are extreme $\tan
\beta$ cases with very large and small values, respectively.
\begin{table}
\begin{center}
\begin{tabular}{|c|c|c|c|c|l|}
 \hline Point & $M^{}_{1/2}$ & $m^{}_0$ & $A^{}_0$  & $\tan \beta$ &
 Slope \\\hline
 1\,a & 250 & 100 & -100 & 10 &  $m^{}_0 = - A^{}_0 = 0.4
 M^{}_{1/2}$, $M^{}_{1/2}$ varies  \\
 1\,b & 400 & 200 & 0 & 30 &    \\
 2 &  300 & 1450 & 0 & 10 &  $m^{}_0 = 2 M^{}_{1/2} + 850$ GeV,
 $M^{}_{1/2}$ varies \\
 3 &  400 & 90 & 0 & 10 &    $m^{}_0 = 0.25 M^{}_{1/2} - 10 $ GeV,
 $M^{}_{1/2}$ varies \\
 4 &  300 & 400 & 0 & 50 &    \\
 5 &  300 & 150 & -1000 & 5 &   \\\hline
\end{tabular}
\end{center}
\caption{Some parameters for the SPS in the mSUGRA. The masses are
given in unit of GeV. $\mu$ appearing in the Higgs mass term has
been taken as $\mu > 0$ for all SPS.}
\end{table}

With the help of Eqs. (4.5) and (5.41), $|C^{}_{ij}|^2$ can
explicitly be written as
\begin{eqnarray}
|C^{}_{12}|^2 & = & |a^{}_1|^2 \; |a^{}_2|^2 \; \left (
\displaystyle {\rm ln} \frac{\Lambda^{}_{\rm GUT}}{M^{}_1} \right
)^2 \; ,
\nonumber \\
|C^{}_{13}|^2 & = & 0 \; ,
\nonumber \\
|C^{}_{23}|^2 & = & |b^{}_2|^2 \; |b^{}_3|^2 \; \left (
\displaystyle {\rm ln} \frac{\Lambda^{}_{\rm GUT}}{M^{}_2} \right
)^2  \; .
\end{eqnarray}
Because of  $|C^{}_{13}|^2=0$, we are left with ${\rm BR}(\tau
\rightarrow e \gamma) = 0$. If ${\rm BR}(\tau \rightarrow e
\gamma) \neq 0$ is established from the future experiments, it
will be possible to exclude the FGY ansatz. Using Eq. (5.35), we
reexpress Eq. (5.42) as
\begin{eqnarray}
|C^{}_{12}|^2 & = & M_1^2 \; |(M^{}_\nu)^{}_{12}|^2 \; \left (
\displaystyle {\rm ln} \frac{\Lambda^{}_{\rm GUT}}{M^{}_1} \right
)^2 \; ,
\nonumber \\
|C^{}_{23}|^2 & = & M_2^2 \; |(M^{}_\nu)^{}_{23}|^2 \; \left (
\displaystyle {\rm ln} \frac{\Lambda^{}_{\rm GUT}}{M^{}_2} \right
)^2 \; .
\end{eqnarray}
As shown in Sec. 5.4, $M^{}_1$ may in principle be constrained by
leptogenesis for given values of $\sin \theta^{}_z$.
\footnote{Note that $g^{}_* =228.75$ in the MSSM. In addition, the
coefficient $3/(16\pi v^2)$ on the right-hand side of Eq. (5.36)
should be replaced by $3/(8\pi v^2 \sin^2\beta)$ in the
supersymmetric version of the MSM.} For simplicity, we choose
$Y^{}_{\rm B} = 9.0 \times 10^{-11}$ as an input parameter, but
$M^{}_2$ is entirely unrestricted from the successful leptogenesis
with $M_2 \gg M_1$.
\begin{figure}[tbp]
\begin{center}
\includegraphics[width=7.3cm,height=7.3cm,angle=0]{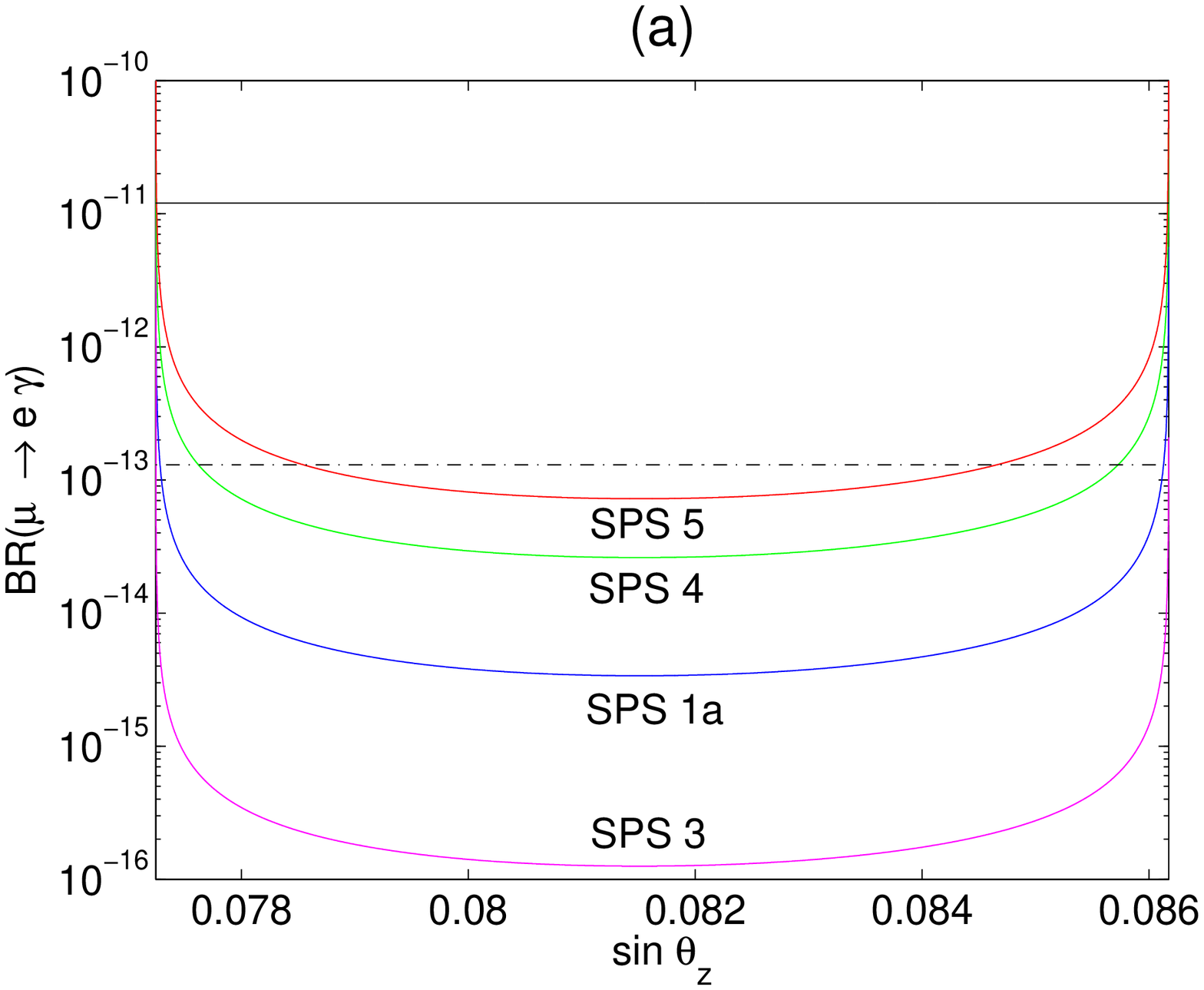}
\includegraphics[width=7.3cm,height=7.3cm,angle=0]{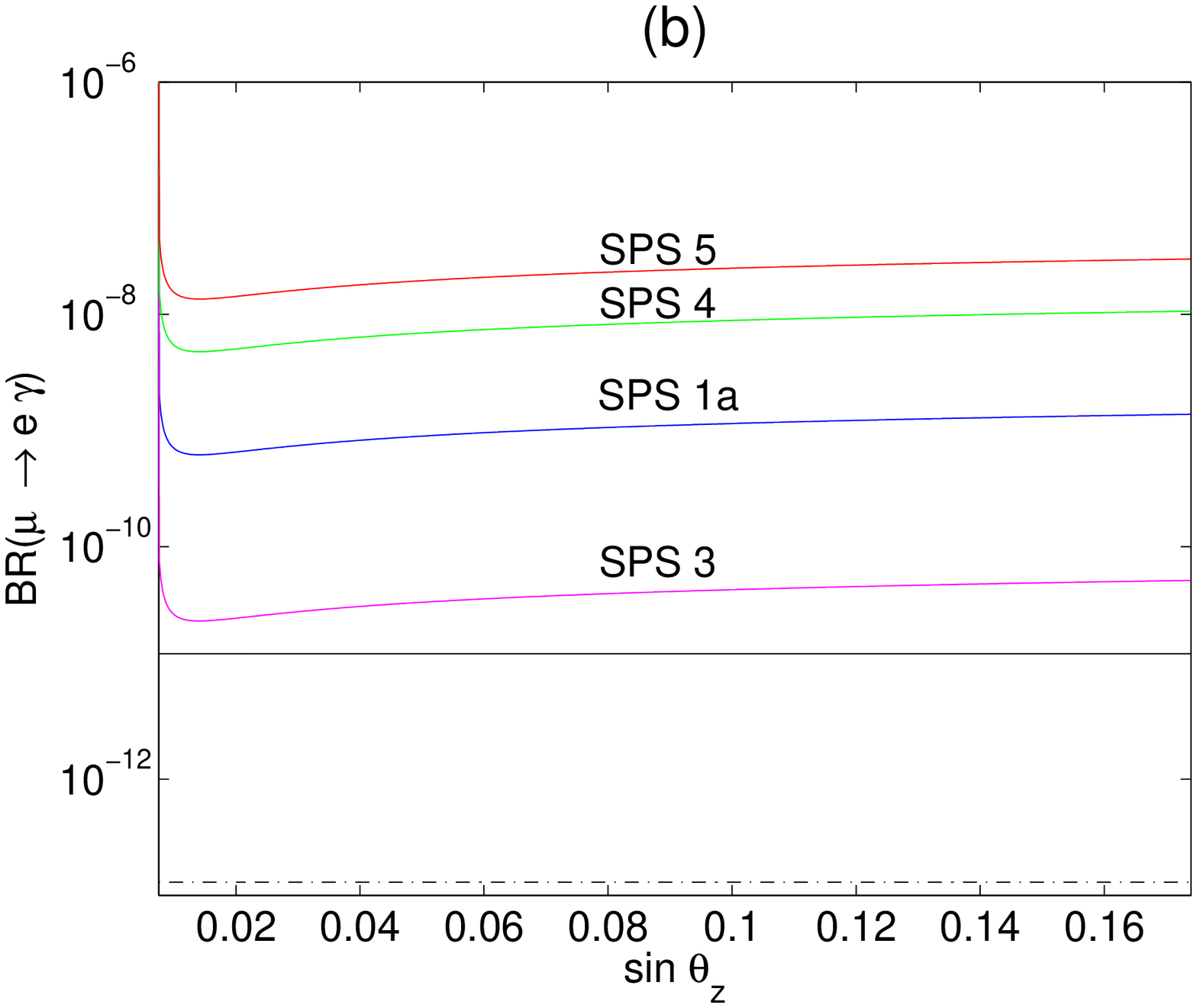}
\end{center} \vspace{-0.3cm}
\caption{Numerical illustration of the dependence of ${\rm BR}(\mu
\rightarrow e \gamma)$ on $\sin \theta^{}_z$: (a) in the $m^{}_1
=0$ case; and (b) in the $m^{}_3 =0$ case. The black solid line
and black dash-dot line denote the present experimental upper
bound on and the future experimental sensitivity to ${\rm BR}(\mu
\rightarrow e \gamma)$, respectively.}
\end{figure}

We numerically calculate BR($\mu \rightarrow e \gamma$) for
different values of $\sin \theta^{}_z$ by using the SPS points.
The results are shown in Fig. 5.5. Since the SPS points 1a and 1b
(or Points 2 and 3) almost have the same consequence in our
scenario, we only focus on Point 1a (or Point 3). When $\sin
\theta^{}_z \rightarrow 0.077$ or $\sin \theta^{}_z \rightarrow
0.086$, the future experiment is likely to probe the branching
ratio of $\mu \rightarrow e + \gamma$ in the $m^{}_1 = 0$ case.
The reason is that $\sin \theta^{}_z \rightarrow 0.077$ (or $\sin
\theta^{}_z \rightarrow 0.086$) implies $\phi \rightarrow - \pi/2$
(or $\phi \rightarrow 0$). Furthermore, the successful
leptogenesis requires a very large $M^{}_1$ due to
$\varepsilon^{}_1 \propto \sin 2 \phi$. It is clear that the SPS
points are all unable to satisfy ${\rm BR}(\mu \rightarrow e
\gamma) \leq 1.2 \times 10^{-11}$ in the $m^{}_3 = 0$ case.
Therefore, we can exclude the $m^{}_3 =0$ case when the SPS points
are taken as the mSUGRA parameters. When $\sin \theta^{}_z \simeq
0.014$, ${\rm BR}(\mu \rightarrow e \gamma)$ arrives at its
minimal value in the $m^{}_3 = 0$ case.  For the SPS slopes,
larger $M^{}_{1/2}$ yields smaller ${\rm BR}(\mu \rightarrow e
\gamma)$. We plot the numerical dependence of BR($\mu \rightarrow
e \gamma$) on $M^{}_{1/2}$ in Fig. 5.6, where we have adopted the
SPS slope 3 and taken $300 ~ {\rm GeV} \leq M^{}_{1/2} \leq 1000 ~
{\rm GeV}$. We find that $M^{}_{1/2} \geq 474 ~ {\rm GeV}$ (or
$M^{}_{1/2} \geq 556 ~ {\rm GeV}$) can result in $ {\rm BR}(\mu
\rightarrow e \gamma) \leq 1.2 \times 10^{-11}$ for $\sin
\theta^{}_z = 0.014$ (or $\sin \theta^{}_z = 0.1$). For all values
of $M^{}_{1/2}$ between 300 GeV and 1000 GeV, ${\rm BR}(\mu
\rightarrow e \gamma)$ is larger than the sensitivity of some
planned experiments, which ought to examine the $m^{}_3 =0$ case
when the SPS slope 3 is adopted. The same conclusion can be drawn
for the SPS slopes 1a and 2. In view of the present experimental
results on muon $g_\mu - 2$, one may get $M^{}_{1/2} \lesssim 430$
GeV for $\tan \beta = 10$ and $A^{}_0 =0$,\cite{0306219} implying
that the $m^{}_3 =0$ case should be disfavored.
\begin{figure}[tbp]
\begin{center}
\vspace{-0.3cm}
\includegraphics[width=7.3cm,height=7.3cm,angle=0]{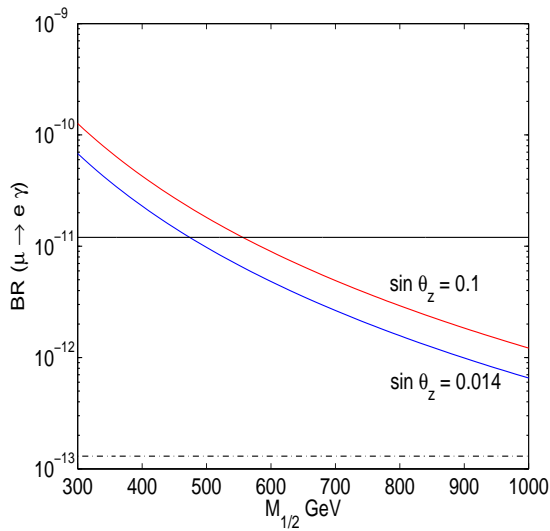}
\end{center} \vspace{-0.3cm}
\caption{ Numerical illustration of the dependence of ${\rm
BR}(\mu \rightarrow e \gamma)$ on $M^{}_{1/2}$ for SPS slope 3 in
the $m^{}_3 =0$ case. The black solid line and black dash-dot line
denote the present experimental upper bound on and the future
experimental sensitivity to ${\rm BR}(\mu \rightarrow e \gamma)$,
respectively.}
\end{figure}

With the help of Eqs. (5.39) and (5.43), one can obtain
\begin{eqnarray}
\frac{{\rm BR}(\tau \rightarrow \mu \gamma)}{{\rm BR}(\mu
\rightarrow e \gamma)} \; = \; \frac{M_2^2 \;
|(M^{}_\nu)^{}_{23}|^2 \; \left [{\rm ln} ({\Lambda^{}_{\rm
GUT}}/{M^{}_2}) \right ]^2 }{M_1^2 \; |(M^{}_\nu)^{}_{12}|^2 \;
\left [ {\rm ln} ({\Lambda^{}_{\rm GUT}}/{M^{}_1}) \right ]^2}
\;\; .
\end{eqnarray}
Since the successful leptogenesis can be used to fix $M^{}_1$, a
measurement of the above ratio will allow us to determine or
constrain $M^{}_2$. It is worth remarking that this ratio is
independent of the mSUGRA parameters.\footnote{Note that
$\varepsilon^{}_1$ is inversely proportional to the mSUGRA
parameter $\sin^2 \beta$. Because $\tan \beta \lesssim 3$ is
disfavored (as indicated by the Higgs exclusion
bounds\cite{LEP2}), here we focus on $\tan \beta \geq 5$ or
equivalently $\sin^2 \beta \geq 0.96$. Hence $\sin^2 \beta \approx
1$ is a reliable approximation in our discussion.} To illustrate,
we show the numerical result of ${\rm BR}(\tau \rightarrow \mu
\gamma)/{\rm BR}(\mu \rightarrow e \gamma)$ in Fig. 5.7 for both
$m^{}_1= 0$ and $m^{}_3= 0$ cases. Below $\Lambda^{}_{\rm GUT}$,
the term $M_2^2 [{\rm ln} (\Lambda^{}_{\rm GUT}/{M^{}_2})]^2$ and
the ratio in Eq. (5.44) reach their maximum values at $M^{}_2 =
\Lambda_{\rm GUT}/e = 7.4 \times 10^{15}$ GeV. Obviously, the
ratio ${\rm BR}(\tau \rightarrow \mu \gamma)/{\rm BR}(\mu
\rightarrow e \gamma)$ is below $2 \times 10^9$ in the $m^{}_1=0$
case and below $8 \times 10^3$ in the $m^{}_3=0$ case. This
conclusion is independent of the mSUGRA parameters.\cite{GuoWL}
\begin{figure}[tbp]
\begin{center}
\includegraphics[width=7.3cm,height=7.3cm,angle=0]{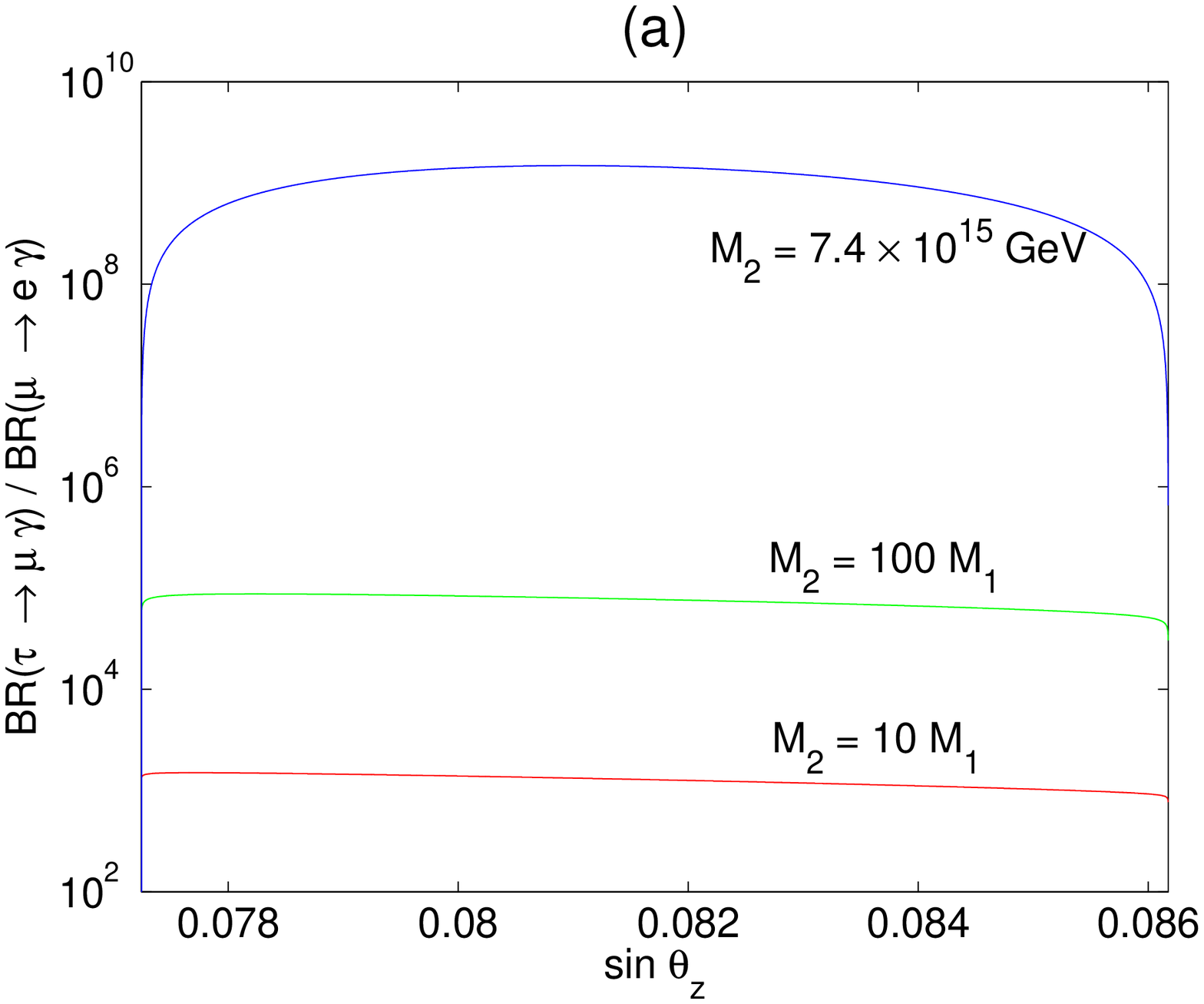}
\includegraphics[width=7.3cm,height=7.3cm,angle=0]{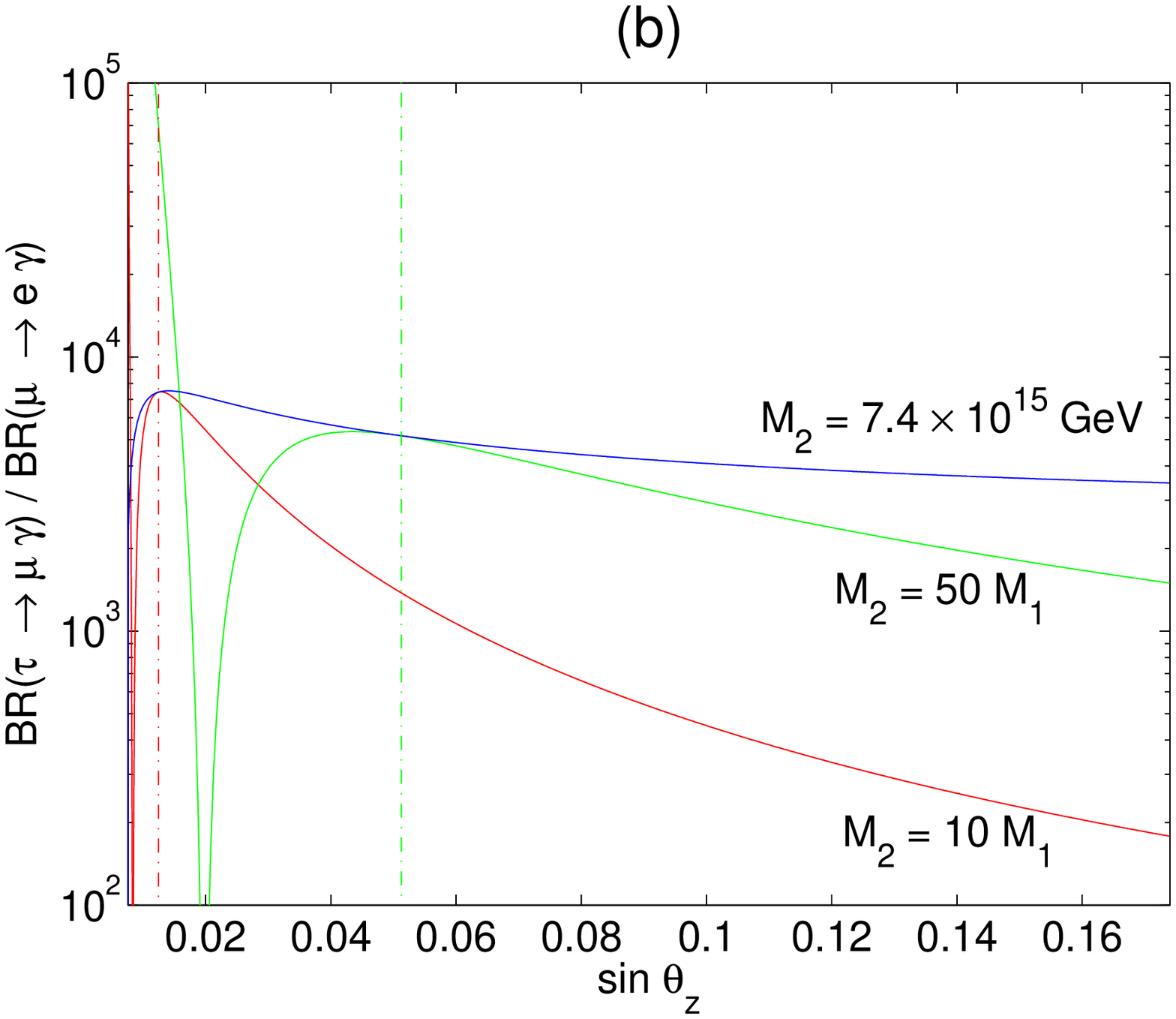}
\end{center}\vspace{-0.3cm}
\caption{Numerical illustration of the dependence of ${\rm
BR}(\tau \rightarrow \mu \gamma)/{\rm BR}(\mu \rightarrow e
\gamma)$ on $\sin \theta^{}_z$: (a) in the $m^{}_1 =0$ case; and
(b) in the $m^{}_3 =0$ case.}
\end{figure}

\section{Concluding Remarks}
\setcounter{equation}{0} \setcounter{figure}{0}

We have presented a review of recent progress in the study of the
MSM, which only contains two heavy right-handed Majorana
neutrinos. The attractiveness of this economical seesaw model is
three-fold:
\begin{itemize}
\item     Its consequences on neutrino phenomenology are almost as rich as
those obtained from the conventional seesaw models with three
heavy right-handed Majorana neutrinos. In particular, the MSM can
simultaneously account for two kinds of new physics beyond the SM:
the cosmological matter-antimatter asymmetry and neutrino
oscillations.

\item     Its predictability and testability are actually guaranteed
by its simplicity. For example, the neutrino mass spectrum in the
MSM is essentially fixed, although current experimental data
remain unable to tell whether $m^{}_1 =0$ or $m^{}_3 =0$ is really
true or close to the truth.

\item     Its supersymmetric version allows us to explore a wealth
of new phenomena at both low- and high-energy scales. On the one
hand, certain flavor symmetries can be embedded in the
supersymmetric MSM; on the other hand, the rare LFV processes can
naturally take place in such interesting scenarios.
\end{itemize}
Therefore, we are well motivated to outline the salient features
of the MSM and summarize its various phenomenological implications
in this article.

In view of current neutrino oscillation data, we have demonstrated
that the MSM can predict the neutrino mass spectrum and constrain
the effective masses of the tritium beta decay and the
neutrinoless double-beta decay. Five distinct parameterization
schemes have been introduced to describe the neutrino
Yukawa-coupling matrix of the MSM. We have investigated neutrino
mixing and baryogenesis via leptogenesis in some detail by taking
account of possible texture zeros of the Dirac neutrino mass
matrix. An upper bound on the CP-violating asymmetry in the decay
of the lighter right-handed Majorana neutrino has been derived.
The RGE running effects on neutrino masses, flavor mixing angles
and CP-violating phases have been analyzed, and the correlation
between the CP-violating phenomena at low and high energies has
been highlighted. It has been shown that the observed
matter-antimatter asymmetry of the Universe can naturally be
interpreted through the resonant leptogenesis mechanism at the TeV
scale. The LFV decays, such as $\mu \to e + \gamma$, have also
been discussed in the supersymmetric extension of the MSM.

Of course, there remain many open questions in neutrino physics.
But we are paving the way to eventually answer them. No matter
whether the MSM can survive the experimental and observational
tests in the near future, we expect that it may provide us with
some valuable hints in looking for the complete theory of massive
neutrinos.

\section*{Acknowledgements}
We are grateful to J. W. Mei for his collaboration in the study of
the MSM. This work is supported in part by the National Natural
Science Foundation of China.


\end{document}